%% file: main.tex
\documentclass{article}

\usepackage[preprint]{froggy_2025}
 
\usepackage{wrapfig}
\usepackage[utf8]{inputenc} 
\usepackage[T1]{fontenc}    
\usepackage[hyphens]{url}
\usepackage{booktabs}       
\usepackage{amsfonts}       
\usepackage{nicefrac}       
\usepackage{microtype}      
\usepackage{xcolor}         

\usepackage{algorithm}
\usepackage[noend]{algpseudocode}
\usepackage{xcolor}
\usepackage{booktabs}
\usepackage{threeparttable}
\definecolor{darkGreen}{rgb}{0.2,0.5,0.2}
\definecolor{mydarkblue}{rgb}{0,0.08,0.45}
\definecolor{darkGreen2}{HTML}{007202}
\usepackage[colorlinks,citecolor=mydarkblue,urlcolor=mydarkblue,linkcolor=mydarkblue]{hyperref}
\usepackage[capitalize,noabbrev]{cleveref}
\usepackage{url}
\usepackage{graphicx}
\usepackage{subcaption}
\usepackage{enumitem}
\usepackage{booktabs,threeparttable}
\usepackage{tabularx}
\newcolumntype{Y}{>{\raggedright\arraybackslash}X}
\usepackage{textcomp}
\usepackage{listings}       
\lstdefinestyle{prompt}{
  basicstyle=\footnotesize\ttfamily,
  breaklines=true,
  breakindent=0pt,
  columns=fullflexible,
  keepspaces=true,
  aboveskip=2pt,
  belowskip=2pt,
  upquote=true,
}
\usepackage{multirow}
\usepackage{mathtools}
\usepackage{booktabs}
\usepackage{siunitx}
\usepackage[normalem]{ulem}
\usepackage{fontawesome5}

\DeclareRobustCommand{\resourceLink}[3]{%
  \href{#1}{\textcolor{froggy-green}{\faIcon{#2}}%
    \nobreak\hspace{0.3em}\textcolor{darkGreen2}{#3}}%
}
\newcommand{\resourceSep}{\hspace{1.6em}}

\title{ProgramDistill: From Interactive Web Apps to Verifiable Reference-Guided SWE Tasks}

\author{
\Authfont{Jeonghye Kim$^1$, Minseon Kim$^2$, Young Jin Kim$^3$, Matheus Pereira$^2$} \\
\vspace{-0.2em}
\Authfont{Marc-Alexandre Côté$^2$, Alessandro Sordoni$^2$, Xingdi Yuan$^2$,  Zhengyan Shi$^2$}\\
\Affilfont{$^1$KAIST ~~ $^2$Microsoft Research Montr\'eal  ~~ $^3$Microsoft AI} \\
\vspace{0.2em}
\Affilfont{jeonghye.kim@kaist.ac.kr ~~ froggy@microsoft.com}\\
\vspace{0.2em}
\Authfont{%
  \resourceLink{https://microsoft.github.io/debug-gym/blog/2026/09/programdistill/}{newspaper}{Blog}%
  \resourceSep
  \resourceLink{https://microsoft.github.io/debug-gym/blog/2026/09/programdistill/dashboard/}{trophy}{Leaderboard}%
  \resourceSep
  \resourceLink{https://huggingface.co/datasets/microsoft/ProgramDistill}{database}{ProgramDistill-Bench}%
  \resourceSep
  \resourceLink{https://aka.ms/froggy}{link}{Froggy}%
}\\
}

\newif\ifrevisionmarks
\revisionmarksfalse

\input{archive/section/browser_related_revision}
\input{archive/section/repair_results_revision}

\input{archive/analysis/generated/canonical-macros.tex}

\begin{document}
\addtocontents{toc}{\protect\setcounter{tocdepth}{-1}}

\maketitle
\begin{abstract}
Coding agents are typically evaluated with desired behavior
specified through issues or instructions.
In practical web development, however, agents may need to infer behavior from working
software and implement it in an incomplete application.
We introduce \textbf{ProgramDistill}, a benchmark evaluating coding agents on features discovered through interaction with fully functional reference applications. We build
ProgramDistill by factorizing applications into features of different granularities, each associated with
replayable behaviors executable via its gold patch.
Our pipeline, \textbf{mine-craft-patch}, discovers 1,975 replay-verified
behaviors across 26 applications and constructs 4,063 tasks without human intervention. Across nine frontier coding agents, GPT-6 Astra and Claude Opus~5 achieve 49.2\% and 28.8\% success
on cumulative workflows in full-application reconstruction.
In partial-application reconstruction, success falls from 100\% to 64.0\%
and from 96\% to 32\% as restoration depth increases from 1 to 8.
ProgramDistill thus provides a scalable benchmark with controlled difficulty for evaluating and diagnosing coding agents, and a natural basis for future curriculum-based training.
\end{abstract}

\input{archive/section/intro}
\input{archive/section/minepatch/setup}
\input{archive/section/minepatch/mining}
\input{archive/section/minepatch/crafting}
\input{archive/section/minepatch/patching}
\input{archive/section/experiments}
\input{archive/section/related_works}
\input{archive/section/conclusion}

\bibliographystyle{abbrv}
\bibliography{main}

\appendix
\input{archive/section/appendix_contents}

\input{archive/section/appendix}

\end{document}

%% file: archive/section/repair_results_revision.tex
\input{archive/analysis/generated/repair-refresh-macros.tex}

%% file: archive/analysis/generated/repair-refresh-macros.tex
\newcommand{\RepairAstraBinary}{84.3}
\newcommand{\RepairAstraChain}{90.2}
\newcommand{\RepairAstraDepthEightBinary}{64.0}
\newcommand{\RepairAstraCumulativeBinary}{81.2}
\newcommand{\RepairAstraCumulativeChain}{88.2}
\newcommand{\RepairAstraScopeChainLogic}{96.2}
\newcommand{\RepairAstraScopeChainUI}{84.9}
\newcommand{\RepairAstraScopeChainGap}{11.3}
\newcommand{\RepairAstraScopeBinaryLogic}{92.9}
\newcommand{\RepairAstraScopeBinaryUI}{76.9}
\newcommand{\RepairAstraScopeBinaryGap}{16.0}

\newcommand{\RepairAstraProbePercent}{0.3}
\newcommand{\RepairPooledProbePercent}{7.2}

\newcommand{\RepairDepthOneReference}{34.60}
\newcommand{\RepairDepthOneCurrent}{27.69}
\newcommand{\RepairDepthOneUnchanged}{12.4}
\newcommand{\RepairDepthOneStubs}{11.1}

\newcommand{\RepairDepthEightReference}{8.46}
\newcommand{\RepairDepthEightCurrent}{6.81}
\newcommand{\RepairDepthEightUnchanged}{23.3}
\newcommand{\RepairDepthEightStubs}{24.6}

%% file: archive/analysis/generated/canonical-macros.tex
\newcommand{\CanonicalAstraAtomicWeighted}{58.98}

\newcommand{\CanonicalAstraBinaryWeighted}{49.15}

\newcommand{\CanonicalAstraChainWeighted}{64.58}

\newcommand{\CanonicalAstraStepsMean}{719.4}

\newcommand{\CanonicalOpusBinaryWeighted}{28.81}

\newcommand{\CanonicalOpusChainWeighted}{47.42}

\newcommand{\CanonicalOpusStepsMean}{627.5}

\newcommand{\CanonicalSolBinaryWeighted}{21.07}

\newcommand{\CanonicalSolChainWeighted}{34.22}

\newcommand{\CanonicalSolStepsMean}{719.0}

%% file: archive/section/intro.tex


\section{Introduction}

\definecolor{observeref}{HTML}{3C83D5}
\definecolor{editcode}{HTML}{7551B7}
\definecolor{observecur}{HTML}{22A66A}

\begin{figure}[!h]
\centering
\vspace{-0.3cm}
\includegraphics[width=0.882\linewidth]{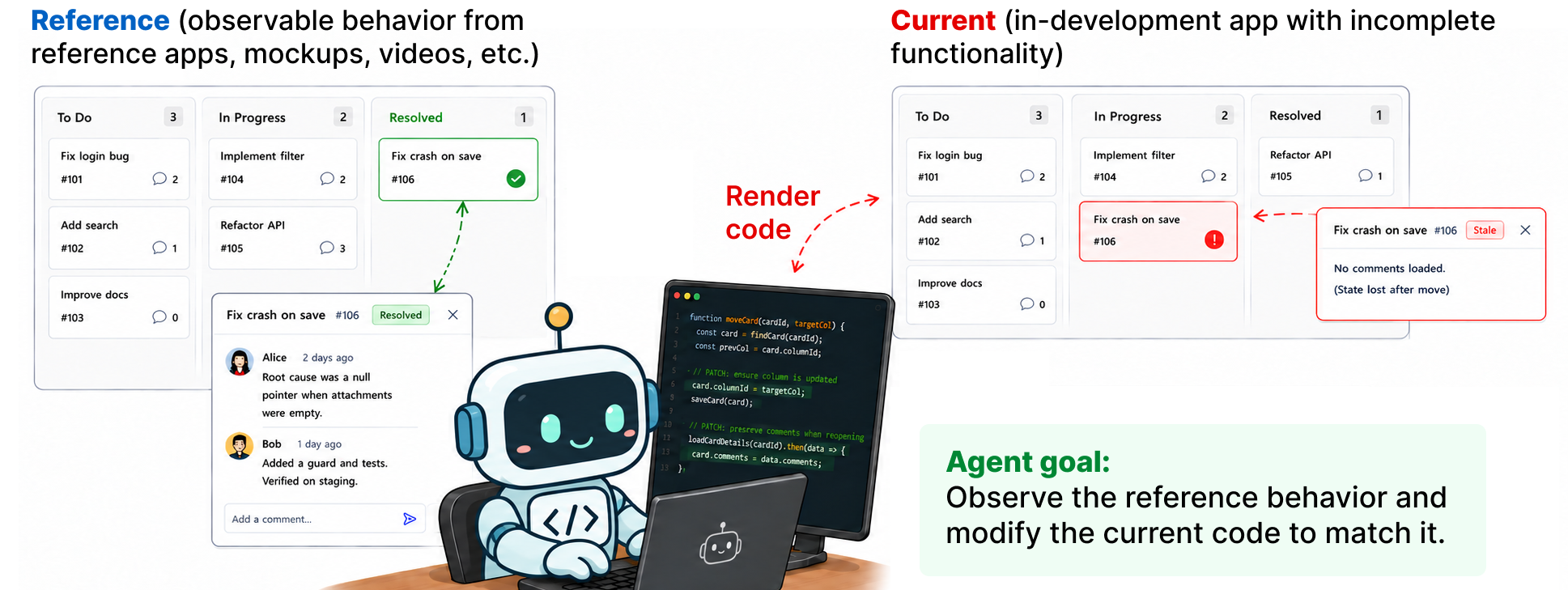}
\caption{\textbf{ProgramDistill evaluates reference-guided software engineering for
interactive web applications.}
\\ A coding agent is given a working
\textcolor{observeref}{\textbf{reference}} application whose source implementation is hidden
and an editable
\textcolor{observecur}{\textbf{current}} application with missing functionality.
The agent interacts with the reference to infer the intended behavior, restores that
behavior by modifying the current implementation, and validates the result against the
reference.}
\label{fig:minepatch_concept}
\end{figure}

Large language models (LLMs) increasingly serve as coding agents capable of inspecting repositories and modifying source code in response to issues, instructions, or tests~\citep{swebench,yang2024sweagent,bugpilot,r2egym,xu2026swe}. These settings, however, typically assume that the desired behavior has already been specified. In practical web development, developers may instead need to infer intended behavior directly from a working reference, such as an earlier product version, an interactive prototype, a comparable application, or a demonstration video. As illustrated in Figure~\ref{fig:minepatch_concept}, an agent must interact with the reference, modify the current implementation, and validate the resulting behavior against it. Even a simple interaction such as dragging a card may require inferring both its visible effect and the application state that persists afterward.

Recent work has begun to treat running software itself as part of the specification. ProgramBench~\citep{programbench}, for example, asks agents to probe compiled programs in C/C++, Go, or Rust and reconstruct them from observable execution behavior. This establishes a behavior-to-code setting, but treats the program as a whole-program reconstruction target. Interactive web applications expose additional structure. Their functionality unfolds through user interface (UI) actions and application states, often with explicit behavioral prerequisites. For instance, a user may need to log in before creating an object and create that object before modifying it later. Such stateful dependencies naturally organize behaviors into \emph{prerequisite lineages}.

This raises a different question. \textbf{Can we uncover this latent behavioral structure and use it to turn a working interactive web application into a scalable source of verifiable software-engineering tasks?} Rather than treating an application as a single reconstruction problem, we identify features at varying levels of granularity and construct a SWE task for each, together with replayable interactions and a gold patch. These tasks can be composed along prerequisite lineages into increasingly cumulative repairs, yielding a controllable restoration-depth axis. We call this transformation \emph{application-to-task factorization}.

The resulting tasks evaluate a complementary form of behavioral distillation. Given a working \emph{reference} and an incomplete \emph{current} implementation, the agent must infer the required behavior from the reference and realize it through source-code changes in the current application. We call this \emph{reference-to-current distillation}, where success is measured by whether the replayable interaction executes identically after applying the submitted patch. Together, application-to-task factorization and reference-to-current distillation define \textbf{ProgramDistill}. Here, \emph{distillation} refers to extracting and transferring executable behavior across these transformations.

To realize this framework at scale, ProgramDistill introduces a fully automated \textbf{\texttt{mine--craft--patch}} pipeline orchestrated by multiple LLM agents. It \textbf{\textcolor{darkGreen2}{\emph{mines}}} reproducible behaviors as replayable browser-interaction traces, \textbf{\textcolor{darkGreen2}{\emph{crafts}}} tasks by masking the source implementations responsible for those behaviors, and \textbf{\textcolor{darkGreen2}{\emph{patches}}} the resulting applications by asking coding agents to recover missing functionality from a working reference. Mining and crafting together realize application-to-task factorization, while patching realizes reference-to-current distillation. Each mined behavior serves as both a \emph{task unit} and a \emph{behavioral verifier}, with the same trace succeeding on the intact application, failing after masking, and being replayed after repair. These verified task units can then be composed along prerequisite lineages, increasing restoration depth from atomic repair to full-application reconstruction. Across 26 applications, the pipeline mines 1,975 replay-verified behaviors and constructs 4,063 tasks without human intervention.

We evaluate nine frontier coding agents. In full-application reconstruction, GPT-6 Astra~\citep{openai2026gpt6astra} and Claude Opus~5~\citep{anthropic2026claude} recover 49.2\% and 28.8\% of evaluated workflows. In partial-application reconstruction, their success falls from 100\% to \RepairAstraDepthEightBinary\% and from 96\% to 32\%, respectively, as restoration depth increases from 1 to 8. Trajectory analysis reveals a growing mismatch between reconstruction burden and agent effort, with observation effort per required behavior declining especially sharply as tasks deepen. Astra, meanwhile, achieves the strongest repair performance while exhibiting the highest observation activity and the fewest edit/write steps. Together, these results identify effort allocation across observation, validation, and editing as an important dimension of reference-guided software engineering alongside implementation capability. ProgramDistill thus provides a scalable, controlled benchmark for evaluating reconstruction performance and the interactive information-seeking strategies that contribute to it across varying restoration depths. We make the following contributions:

\begin{itemize}[itemsep=0pt, topsep=0pt, parsep=4pt, partopsep=0pt, leftmargin=*]

\item
\textbf{A benchmark for reference-guided software engineering.}
We introduce ProgramDistill, comprising 4,063 replay-verified SWE tasks derived from 1,975 behaviors across 26 interactive web applications, spanning atomic repair, cumulative repair, and full-application reconstruction.

\item
\textbf{Program distillation from executable behavior.}
We formulate program distillation along two axes. Application-to-task factorization converts working applications into structured SWE tasks, while reference-to-current distillation requires agents to recover behavior from an executable reference. We operationalize both with a fully automated mine--craft--patch pipeline.

\item
\textbf{Restoration depth for evaluation and curriculum construction.}
Prerequisite lineages provide a controlled progression from atomic repair to full reconstruction, exposing systematic failures as restoration depth increases and providing a natural curriculum for future training through trajectory distillation or reinforcement learning.
\end{itemize}

%% file: archive/section/minepatch/setup.tex
\section{Application-to-Task Factorization via the Mine--Craft--Patch Pipeline}
\label{sec:minepatch}

\begin{figure}[h!]
\centering
\includegraphics[width=\linewidth]{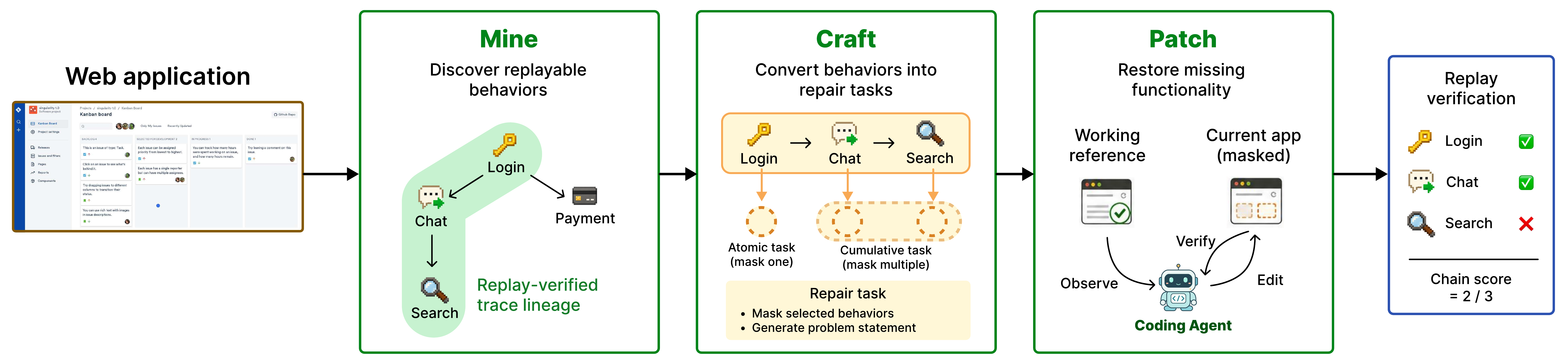}
\caption{\textbf{Overview of the mine--craft--patch pipeline.}
\textbf{\textcolor{darkGreen2}{\emph{Mining}}} discovers replayable behaviors
and organizes them into a prerequisite trace tree by extending previously
verified traces.
\textbf{\textcolor{darkGreen2}{\emph{Crafting}}} constructs atomic repair
tasks by masking one behavior, or combines masks along a lineage to construct
progressively deeper cumulative repair tasks.
\textbf{\textcolor{darkGreen2}{\emph{Patching}}} asks a coding agent to restore
the masked functionality from a working reference, then replays the corresponding
lineage traces to measure how much of the behavior chain has been recovered.}
\label{fig:minepatch_pipeline}
\end{figure}

Figure~\ref{fig:minepatch_pipeline} illustrates
\textbf{\texttt{mine--craft--patch}}, our new synthetic task generation pipeline that orchestrates multiple LLM
agents to synthesize verifiable tasks from web applications without relying on human-written issues, tests, or behavioral
annotations. The pipeline is model-agnostic, and we use GPT-5.6 Sol \citep{openai2026gpt56sol} for all stages in our experiments.

\subsection{Setup: Deterministic and Observable Application Instances}

To support the full mine–craft–patch pipeline, we set up each application as a live, self-contained execution environment accessible through a common browser interface. The environment provides a stable reference for behavior observation, an editable application for task construction and repair, and a reproducible runtime for replay-based verification.

\subsubsection{Applications and Runtime Instances}

\paragraph{Application corpus.}
The \texttt{mine--craft--patch} pipeline operates on 26 web applications drawn from two complementary sources. The corpus includes self-contained applications adapted from the OSWorld web-application suite~\citep{xie2024osworld}, as well as real-world open-source projects and SaaS clones drawn from public repositories. The former provide a controlled and diverse interaction substrate, while the latter contribute larger codebases, deeper workflows, nontrivial state, and heterogeneous architectures. Details of the corpus are provided in Appendix~\ref{app:corpus}.

\paragraph{Reference and editable instances.}
For each web application, we serve a fixed production build as the \emph{reference instance} and a development-server version as the \emph{editable instance}. The reference remains unchanged throughout task construction and repair, while masking and agent edits are applied to the source code and reflected in the editable instance through hot reload. Both instances expose the same application interface, allowing behavior observed in the reference to be reproduced and evaluated in the editable instance.

\paragraph{Deterministic execution.}
Reliable replay requires a reproducible starting state. The pipeline therefore resets application state before collection and replay and uses a shared deterministic clock across the database, backend, and frontend. This reduces variation from time-dependent behavior while preserving normal ordering and timeout semantics. Details of the deterministic execution setup are provided in Appendix~\ref{app:deterministic}.

\subsubsection{Browser Helper}

\definecolor{fcframe}{HTML}{00841A}

\lstdefinestyle{fcjson}{
basicstyle=\ttfamily\scriptsize,
breaklines=true,
breakatwhitespace=false,
columns=fullflexible,
showstringspaces=false,
keepspaces=true,
}

A shared Playwright-based browser helper~\citep{playwright} provides a
common interaction interface for mining, replay verification, and repair.
It executes high-level browser actions and returns structured observations
of visible text, accessibility information, and interactive elements.
For reliable replay, the helper resolves elements using stable observable
attributes rather than volatile DOM identifiers and waits for application
state to settle before recording observations.
The same interaction and observation semantics are used throughout the
pipeline.
The helper also records screenshots for analysis and behavior relabeling,
while the partial-application and full-application reconstruction agents evaluated in this work receive
only structured browser observations.
Appendix~\ref{app:browser-helper} provides implementation details.

%% file: archive/section/minepatch/mining.tex
\subsection{\textcolor{darkGreen2}{Mining}: Discovering and Verifying Interactive Behaviors}
\label{subsec:mining}

Mining discovers what an application can do and records those behaviors as
replayable specifications for task construction and evaluation. By exploring
the live application with access to its source, the pipeline builds a bank
$\mathcal{B}$ of verified traces, each containing browser actions, expected
outcome signals, and an optional parent trace. Parent links preserve the
prerequisite context needed to reproduce dependent behaviors, allowing later
stages to mask and evaluate them along the same lineage.

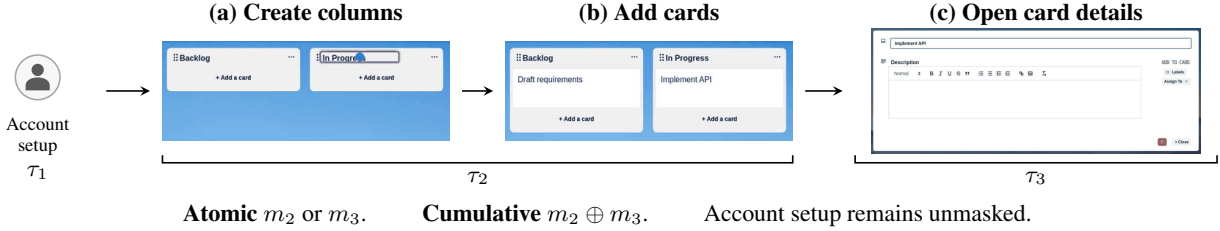
\begin{figure*}[h!]
\centering
\input{archive/figures/trello_running_example}
\caption{\textbf{A behavioral dependency in \texttt{Trello (Knowankit)}.}
Account setup (\(\tau_1\), represented by the icon) precedes board/card
creation (\(\tau_2\)) and the editor interaction (\(\tau_3\)).
The brackets group panels (a)--(b) into one trace and
panel (c) into its dependent trace.}
\label{fig:running-example}
\end{figure*}

\paragraph{From a behavior goal to a verified trace.}
Given the current trace bank, the pipeline first proposes a set of new
behavior goals grounded in source and UI evidence, each with an optional
parent trace. When a parent is specified, the application is reset and its
lineage is replayed to establish the prerequisite state. An LLM agent then
explores the live application to pursue the proposed goal, choosing each
action from the current observation while the environment records its
interactions to form an exploratory trace \(\tilde{\tau}_d\).
In Figure~\ref{fig:running-example}, for instance, the editor behavior
\(\tau_3\) is explored after replaying account setup (\(\tau_1\)) and
board/card creation (\(\tau_2\)). The resulting trace activates the column
title editor and opens the card details, with expected signals checking the
title field values and the presence of the \texttt{Description} label.

Because this initial exploration may include detours, the pipeline re-collects the behavior from a clean reset, providing \(\tilde{\tau}_d\) to the LLM agent as additional prompt context about a successful route. The agent still chooses each action from live observations, but can now omit unnecessary exploratory steps. This parallels findings in self-distillation that conditioning on
additional privileged information can elicit more concise reasoning
trajectories~\citep{kim2026why}. During this clean re-collection, the
environment records replay-stable actions and exposes candidate signals
derived from the observed state. The agent selects the signals that support
the achieved outcome. The resulting trace \(\tau_d\) therefore contains the
actions, expected signals, and parent reference needed for deterministic
replay.

Only traces that pass the replay verifier described below are admitted.
Each admitted trace is labeled according to the behavior actually achieved
and extends its parent lineage, yielding
\(L_d=(\tau_1,\ldots,\tau_d)\); for a root trace,
\(L_1=(\tau_1)\). Mining repeats this process to expand coverage and discover
deeper dependent behaviors. We implement these stages with specialized
Planner, Collector, Relabeler, and Reflector roles.
Appendix~\ref{app:mining-orchestration} details their control flow,
Appendix~\ref{app:prompts} provides the corresponding prompts, and
Appendix~\ref{app:trace-evolution} illustrates how a trace evolves through
the mining pipeline.

\paragraph{Replay-based verification.}
For the intact application $A$ and a trace lineage
$
L_d = (\tau_1,\ldots,\tau_d),
$
we define the replay verifier $V(A,L_d)\in\{0,1\}$.
Starting from a reset application and browser state, it replays
$\tau_1,\ldots,\tau_d$ in order without LLM intervention and returns
1 only if all recorded actions complete and all expected behavioral
signals are satisfied.

A candidate trace $\tau_d$ is admitted when $V(A,L_d)=1$.
The same verifier is used for mask validation and repair evaluation
in Sections~\ref{subsec:crafting} and~\ref{subsec:patching}.

%% file: archive/figures/trello_running_example.tex
\begingroup
\color{black}
\pgfmathsetlengthmacro{\trelloboardwidth}{0.23*\linewidth}
\pgfmathsetlengthmacro{\trellocardwidth}{0.29*\linewidth}
\begin{tikzpicture}[x=\linewidth,y=1cm,
    every node/.style={inner sep=0pt,font=\small},
    line width=0.6pt]
    \node[font=\small\bfseries] at (0.265,1.02) {(a) Create columns};
    \node[font=\small\bfseries] at (0.540,1.02) {(b) Add cards};
    \node[font=\small\bfseries] at (0.850,1.02) {(c) Open card details};

    \begin{scope}[shift={(0.05,0.18)},x=1pt,y=1pt,scale=0.70]
        \draw[draw=black!35,fill=black!3] (0,0) circle[radius=12pt];
        \fill[black!65] (0,4) circle[radius=3.3pt];
        \fill[black!65] (-6.5,-7) -- (-6.5,-4.5)
            .. controls (-6.5,1.5) and (6.5,1.5) .. (6.5,-4.5)
            -- (6.5,-7) -- cycle;
    \end{scope}
    \node[align=center,font=\scriptsize] at (0.05,-0.60) {Account\\setup};

    \node at (0.265,0) {\includegraphics[width=\trelloboardwidth,
        height=0.65in,keepaspectratio]{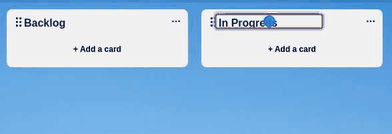}};
    \node at (0.540,0) {\includegraphics[width=\trelloboardwidth,
        height=0.65in,keepaspectratio]{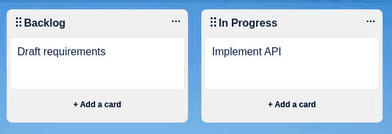}};
    \node at (0.850,0) {\includegraphics[width=\trellocardwidth,
        height=0.65in,keepaspectratio]{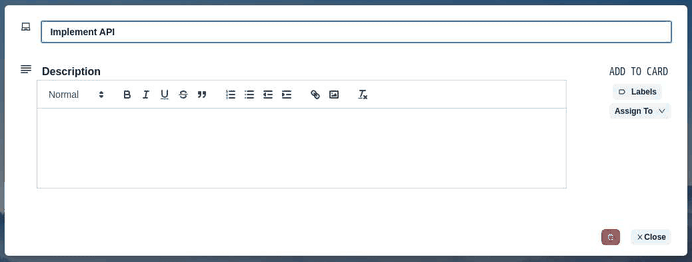}};

    \draw[->,>=stealth] (0.105,0) -- (0.140,0);
    \draw[->,>=stealth] (0.390,0) -- (0.415,0);
    \draw[->,>=stealth] (0.665,0) -- (0.695,0);

    \node at (0.05,-1.08) {\(\tau_1\)};
    \draw (0.150,-0.85) -- (0.150,-0.95)
        -- (0.655,-0.95) -- (0.655,-0.85);
    \node at (0.403,-1.15) {\(\tau_2\)};
    \draw (0.705,-0.85) -- (0.705,-0.95)
        -- (0.995,-0.95) -- (0.995,-0.85);
    \node at (0.850,-1.15) {\(\tau_3\)};
\end{tikzpicture}

\smallskip
{\small \textbf{Atomic} \(m_2\) or \(m_3\).
\qquad \textbf{Cumulative} \(m_2\oplus m_3\).
\qquad Account setup remains unmasked.}
\endgroup

%% file: archive/section/minepatch/crafting.tex
\subsection{\textcolor{darkGreen2}{Crafting}: From Verified Behaviors to Repair Tasks}
\label{subsec:crafting}

Crafting turns replay-verified behaviors into repair tasks by
removing their source implementations and checking that the resulting
failures are confined to the intended targets rather than their
prerequisites. Validated masks can then be combined along a lineage to vary
how many dependent behaviors must be restored together.

\begin{figure}[h!]
    \centering

    \begin{subfigure}[t]{0.27\linewidth}
        \centering
        \includegraphics[width=\linewidth]{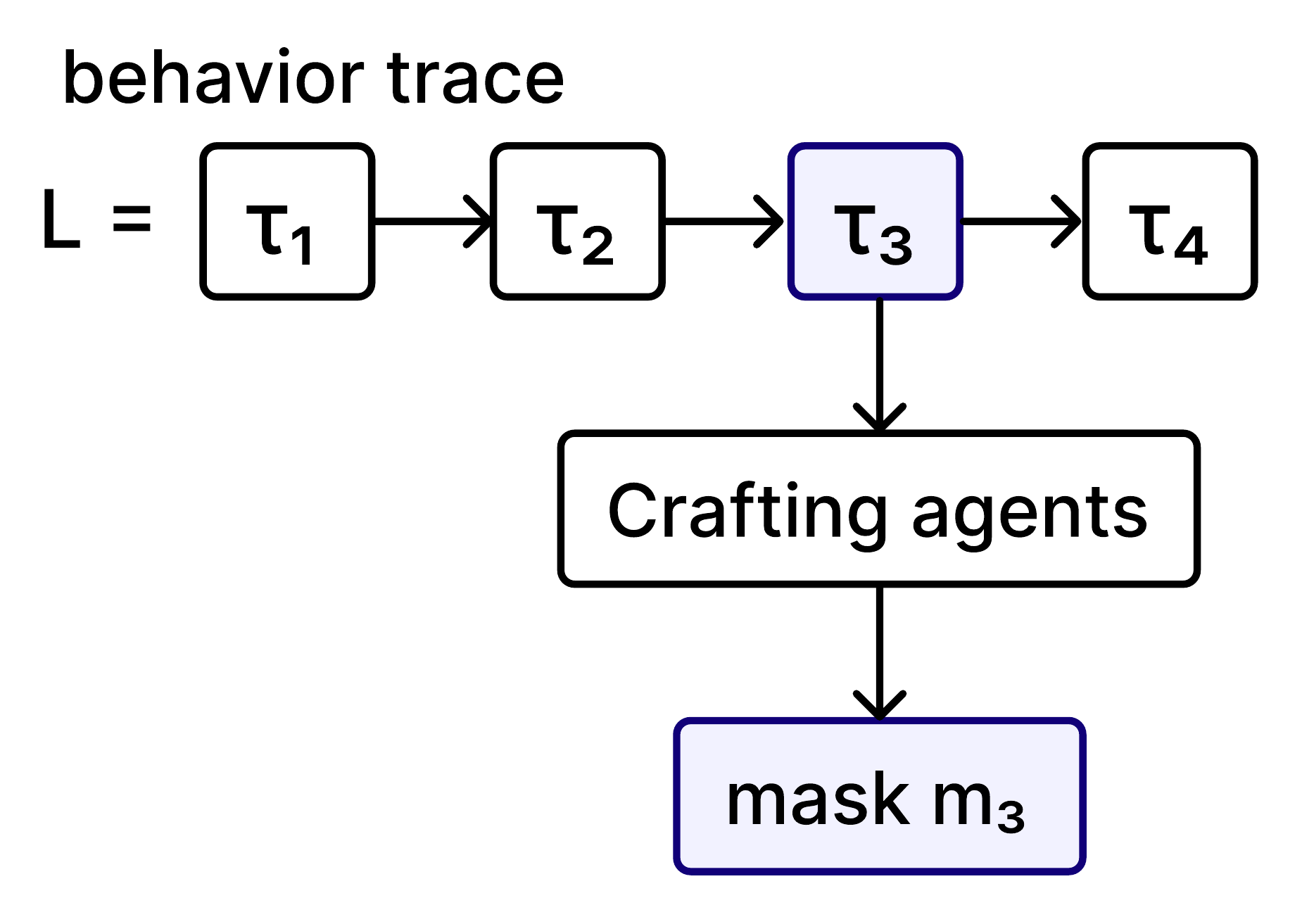}
        \caption{Trace-conditioned mask}
    \end{subfigure}
    \hfill
    \begin{subfigure}[t]{0.27\linewidth}
        \centering
        \includegraphics[width=\linewidth]{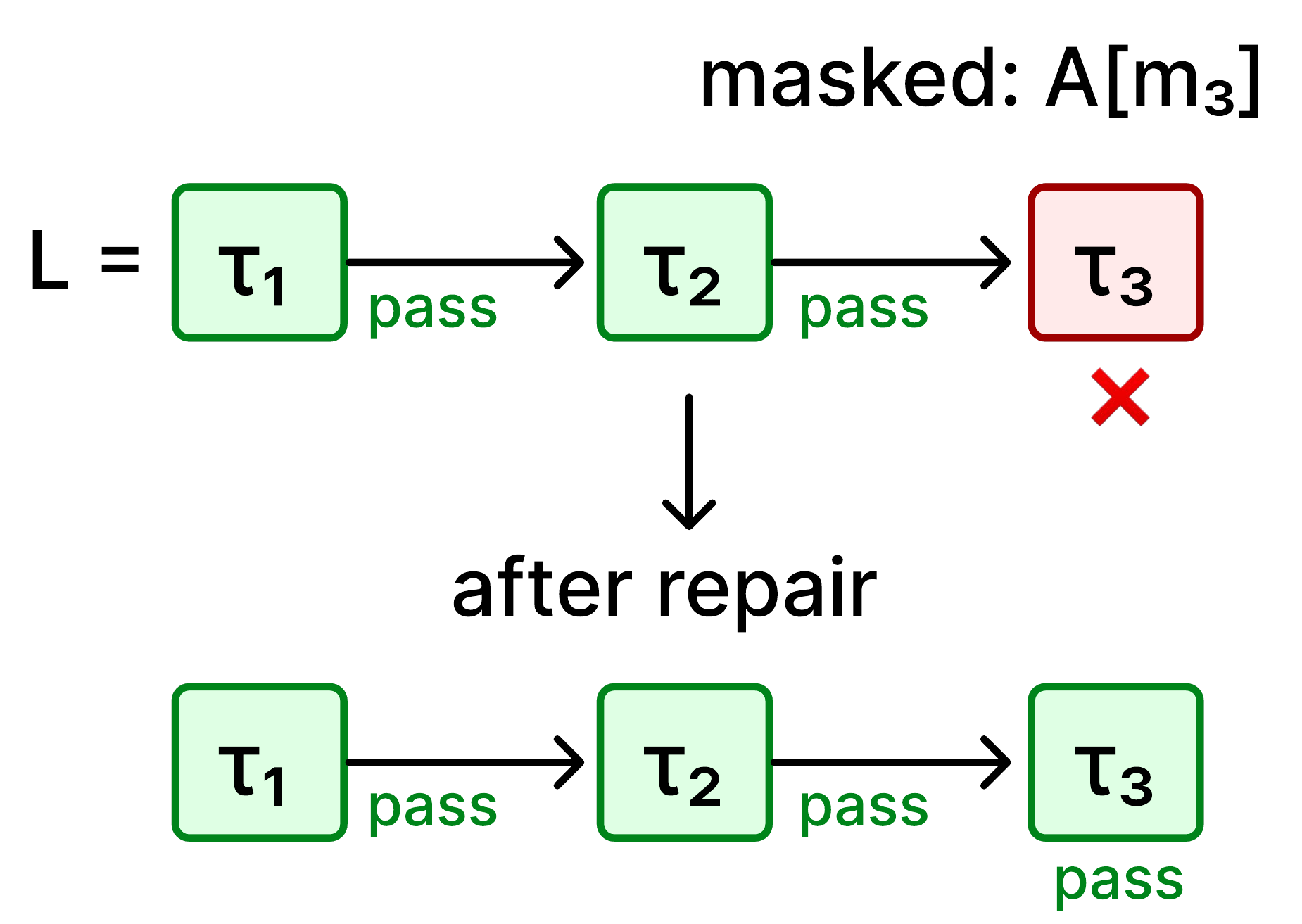}
        \caption{Atomic mask}
    \end{subfigure}
    \hfill
    \begin{subfigure}[t]{0.27\linewidth}
        \centering
        \includegraphics[width=\linewidth]{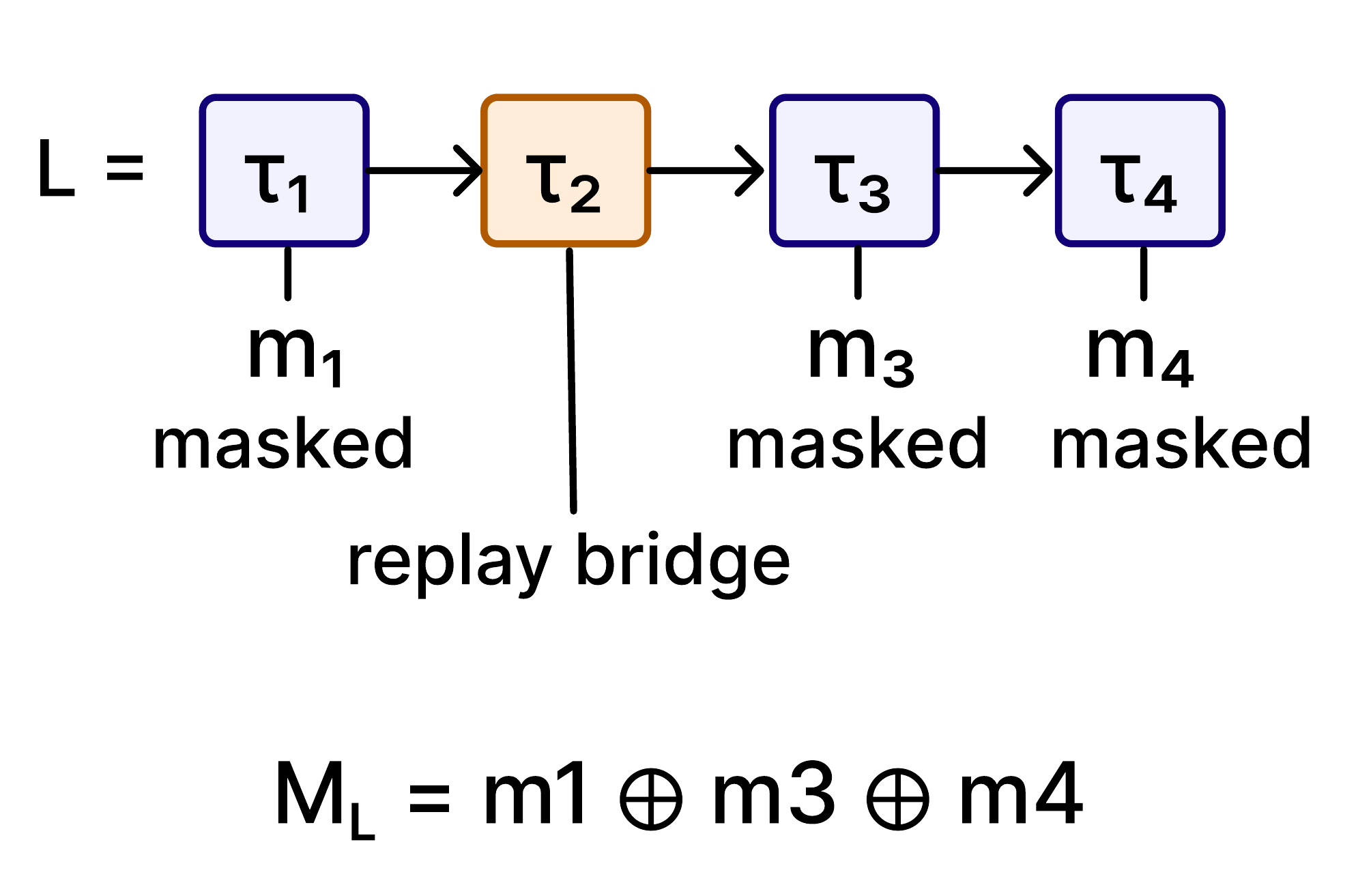}
        \caption{Cumulative mask}
    \end{subfigure}

    \caption{\textbf{Crafting verified behaviors into repair tasks.}
    (a) A behavior-specific mask is generated from a verified trace.
    (b) Atomic validation requires all prerequisite traces to pass and
    the target trace to fail after masking. Gold-patch replay verifies
    recovery of the complete lineage.
    (c) Validated masks are composed along a prerequisite lineage,
    with unmasked traces serving as replay bridges and masked behaviors
    forming joint restoration targets.}
    \label{fig:crafting}
\end{figure}

We control the
resulting tasks along two dimensions.
\emph{Mask scope} determines how much of a feature is removed.
A logic-only mask leaves the existing user interface (UI) in place but
removes the implementation that makes it work, while a logic-and-UI mask
removes both the feature's behavior and its UI.
\emph{Task composition} determines whether the task targets a single
behavior or multiple dependent behaviors along a prerequisite lineage.
Each behavior is first considered for an atomic task that masks only
that behavior. Validated atomic masks are then combined along a lineage
to form cumulative tasks that require several behaviors to be restored
together.
Figure~\ref{fig:crafting} summarizes this process, from
trace-conditioned masking through atomic validation to cumulative
composition.

\paragraph{Trace-conditioned masking.}
As illustrated in Figure~\ref{fig:crafting}(a), consider a verified lineage
$
L_d=(\tau_1,\ldots,\tau_d)
$
with target trace $\tau_d$.
The Crafting agents identify the source implementation responsible
for the observed behavior of $\tau_d$ and propose a mask $m_d$.
We write $A[m_d]$ for the application obtained by applying $m_d$ to the
intact application $A$.
The prompts for the Crafting agents are provided in
Appendix~\ref{app:taskgen-prompts}.

For each logic-only or logic-and-UI mask, the pipeline generates a concise
behavioral problem statement describing the feature's user-facing purpose
and workflow while withholding the original implementation, repair procedure,
and exact verification signals.
The coding agent must infer the missing behavior from the working reference.

\paragraph{Atomic task validation.}
An \emph{atomic repair task} applies only the target mask $m_d$,
leaving the prerequisite implementations unmasked.
Since $L_d$ was replay-verified on the intact application,
$
V(A,L_d)=1
$
by construction.
A mask is accepted only if the masked application builds and launches
successfully, all prerequisite traces
$\tau_1,\ldots,\tau_{d-1}$ remain replayable, and the target behavior fails.
For $d>1$, with
$
L_{d-1}=(\tau_1,\ldots,\tau_{d-1}),
$
these conditions are
\[
\mathrm{Build}(A[m_d])=1,
\qquad
V(A[m_d],L_{d-1})=1,
\qquad
V(A[m_d],L_d)=0.
\]
For a root trace with $d=1$, only the build and target-failure
conditions apply. These conditions yield an SWE-bench-style
fail-to-pass (F2P) objective for the target and pass-to-pass (P2P)
checks for its unmasked prerequisites~\citep{swebench}.
Figure~\ref{fig:crafting}(b) illustrates this counterfactual validation,
where the prerequisite traces continue to pass after masking while the target trace fails and must be recovered by the repair.

In Figure~\ref{fig:running-example}, the atomic mask \(m_3\)
must preserve creation of the board and cards in panels (a)--(b), but break
the editor interaction whose outcome is shown in panel (c).
Composing \(m_2\) and \(m_3\) instead makes both board/card creation and the
editor interaction repair targets, while account setup remains an unmasked
prerequisite.

A separate LLM-based mask-depth critic rejects superficial changes,
such as toggling feature flags or removing call sites while leaving
the substantive implementation intact.
Failed proposals may be revised through bounded feedback rounds.
Only masks that pass both counterfactual validation and the critic
are retained.
If no proposed mask is accepted, the trace remains available as a
prerequisite but does not become a repair target.

\paragraph{Cumulative task composition.}
For a verified lineage
$
L_d
$
ending at a repair target, let
$
j_1<\cdots<j_{r_L}=d
$
index the traces with validated masks.
The \emph{restoration depth} $r_L$ is the number of repair targets
along the lineage.

Combining the corresponding masks yields the cumulative mask
\[
M_L
=
m_{j_1}\oplus\cdots\oplus m_{j_{r_L}},
\]
where $\oplus$ combines the accepted source modifications.
The resulting masked application is $A[M_L]$.
As shown in Figure~\ref{fig:crafting}(c), unmasked traces serve as
\emph{replay bridges} that establish prerequisite state without becoming
repair targets themselves.
Thus, lineage depth $d$ may exceed restoration depth $r_L$.
Atomic tasks have $r_L=1$, whereas cumulative tasks have $r_L>1$.

To construct a cumulative task, the pipeline first attempts to compose
the validated atomic masks deterministically.
Because all component masks are defined against the same intact baseline
$A$, non-overlapping edits are combined directly and overlapping deletions
are merged by union.
A \texttt{git} three-way merge provides an independent consistency check.
If the two results disagree, an LLM-based merge agent resolves the
remaining overlaps while preserving the non-conflicting masks.
Its prompt is provided in Appendix~\ref{app:prompt-merger}.

Of the 1,201 cumulative tasks in our experiments, 629 were composed
deterministically and 572 required the merge agent.
The fraction requiring the merge agent increases with restoration depth,
from 25.6\% at $r_L=2$ to 89.7\% at $r_L=8$ and 100\% at $r_L>8$.
All 572 agent-assisted merges succeeded without omitting any intended
mask component.

\paragraph{Gold patches.}
For each validated atomic or cumulative task, reversing its masking diff
yields a \emph{gold patch}~\citep{swebench} that recovers the original
implementation.
We validate the gold patch by applying it to the masked application and
requiring the complete task lineage to pass replay verification.
Atomic counterfactual checks provide target-specific negative controls,
while gold-patch replay provides an end-to-end positive control that the
restoration and preservation requirements can be satisfied jointly.

%% file: archive/section/minepatch/patching.tex
\subsection{\textcolor{darkGreen2}{Patching}: Restoring Behavior from a Working Reference}
\label{subsec:patching}

Patching evaluates whether a coding agent can recover application behavior
by observing a working reference and implementing it in an incomplete
application. We consider two settings. Partial-application reconstruction
starts from the masked repository produced by crafting, whereas
full-application reconstruction starts from a minimal executable scaffold.
In both settings, the agent interacts with the reference without access to
its source, and the traces collected during mining are replayed to evaluate
the resulting implementation. In Figure~\ref{fig:running-example},
partial-application reconstruction requires restoring the masked board/card
creation behavior, the editor behavior, or both, so that replay reproduces
the corresponding reference behavior.

\paragraph{Partial-application reconstruction.}
The agent receives the masked repository and a generated
behavioral problem statement. It can edit the writable, hot-reloading
\texttt{current} application while comparing it through the browser with the
working \texttt{reference}. The masking diff, gold patch, grading traces, and
reference source remain hidden, and public-network egress is disabled.
Appendix~\ref{app:harness-architecture} describes the resulting isolation,
Appendix~\ref{app:solver-instruction} provides the agent instruction, and
Section~\ref{sec:analysis} analyzes observed shortcut attempts.

\paragraph{Trace-level verification and lineage scoring.}
Let $\Delta$ denote the agent-submitted patch.
We write $A[M_L,\Delta]$ for the repaired application obtained by
applying $\Delta$ to the masked application $A[M_L]$.
Evaluation uses the verifier defined in
Section~\ref{subsec:mining}, with the recorded actions, selectors,
and expected signals unchanged.
Both repair targets and replay bridges must satisfy the recorded
action and outcome checks.
The submitted patch need not match the gold patch at the source level.
Function names, variable names, and code structure may differ as long as
the repaired application reproduces the reference behavior.

For repair targets indexed by
$
j_1<\cdots<j_{r_L}=d,
$
we define
\begin{equation}
\label{eq:binary-score}
\mathrm{BinaryScore}(L_d,\Delta)
=
V(A[M_L,\Delta],L_d),
\qquad
\mathrm{ChainScore}(L_d,\Delta)
=
\frac{1}{r_L}
\sum_{k=1}^{r_L}
V(A[M_L,\Delta],L_{j_k}).
\end{equation}
The binary score is 1 only when the complete task lineage passes,
while the chain score gives partial credit for the recovered prefix.
For an atomic task, $r_L=1$, so the two scores coincide.

For the two-target cumulative task in
Figure~\ref{fig:running-example}, suppose the repaired application passes
through board/card creation \(\tau_2\) but fails at the editor interaction
\(\tau_3\). Then \(\mathrm{BinaryScore}=0\), whereas
\(\mathrm{ChainScore}=1/2\), giving partial credit for the recovered behavior.

\paragraph{Full-application reconstruction.}
In full-application reconstruction, the agent starts from a minimal
executable scaffold and reconstructs the application by interacting
with the working reference. We evaluate the reconstructed application using replay-verified behavior
traces $\mathcal{B}$ collected during mining.
Because each trace contains a replayable interaction sequence and expected
behavioral signals, sampled traces can serve directly as behavioral tests.
Atomic recovery measures the fraction of atomic behavior tests that pass.
Each test scores one target behavior while replaying its prerequisites to
establish the required state.
For cumulative workflows, binary recovery requires the complete lineage
to pass, while the chain score gives partial credit for recovered prefixes. Because reconstruction starts from a scaffold rather than an existing
implementation, prerequisite replay checks workflow correctness rather
than preservation of previously working behavior.

%% file: archive/section/experiments.tex
\section{Experiments} \label{sec:exp}

\subsection{Benchmark Generation Statistics}
\label{sec:exp-construction}

\begin{figure*}[ht!]
\centering
\includegraphics[width=\textwidth]{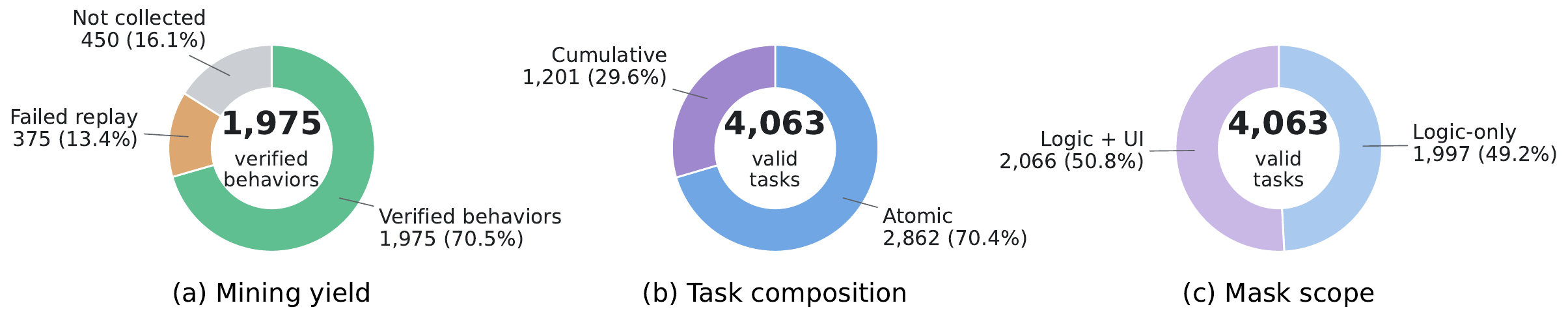}
\caption{\textbf{ProgramDistill benchmark generation.}
(a) Mining yields 1,975 replay-verified behaviors, with remaining goals either
uncollected or failing replay verification.
(b) The resulting 4,063 repair tasks comprise atomic and cumulative tasks.
(c) Tasks are distributed across logic-only and logic-and-UI mask scopes.}
\label{fig:benchmark_construction}
\end{figure*}

We run the full \textbf{\texttt{mine--craft--patch}} pipeline over 26 web applications using
GPT-5.6 Sol as the construction model.
Figure~\ref{fig:benchmark_construction} summarizes the resulting mining yield, task
composition, and mask scopes.

\paragraph{Mining.}
Across the 26 applications, mining proposes 2,800 candidate
goals, of which 2,350 are collected as interaction traces.
After clean-state replay, 2,165 traces reproduce successfully, and
1,975 survive the final replay check to become verified behavior traces.
The mined prerequisite trees reach a maximum lineage depth $d$ of 17
(mean 3.40), a maximum width of 60, and a mean branching factor of 1.78.
Lineage depth is the number of traces from the root to a target, including
replay bridges used to establish prerequisite state.

\paragraph{Lineage.}

\begin{figure}[h!]
\centering
\includegraphics[width=\linewidth]{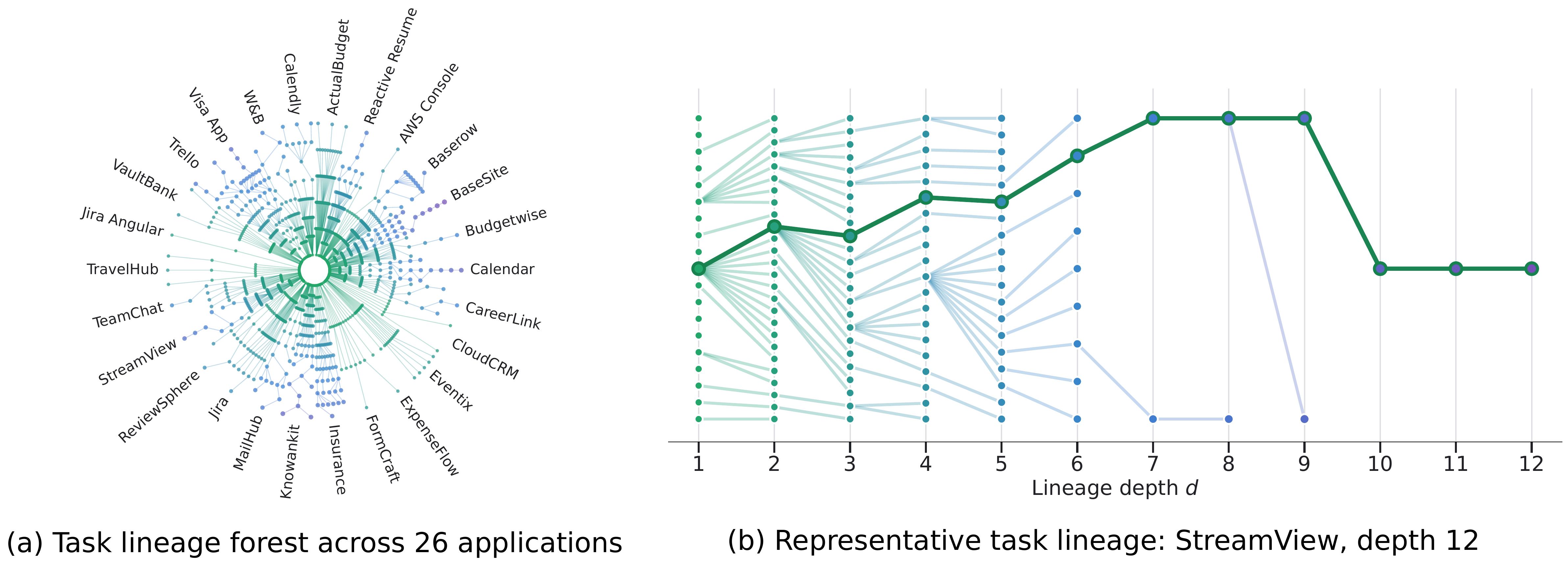}
\caption{\textbf{Prerequisite lineages across applications.}
(a) Mined lineage forests across all 26 applications.
(b) A representative \texttt{StreamView} lineage reaching depth 12, with one
root-to-leaf prerequisite chain highlighted.}
\label{fig:minepatch_summary}
\end{figure}

As described in Section~\ref{subsec:crafting}, cumulative tasks are constructed
by composing masks along mined prerequisite lineages.
Figure~\ref{fig:minepatch_summary} visualizes this structure at two scales.
Panel~(a) shows variation in lineage depth and branching across applications,
reflecting the range of dependent workflows discovered during mining.
Panel~(b) shows a representative \texttt{StreamView} lineage reaching depth 12, with
columns indexing lineage depth $d$ and one root-to-leaf chain highlighted.

\paragraph{Crafting.}
The pipeline constructs 4,063 repair tasks from the verified traces:
2,862 atomic tasks and 1,201 cumulative tasks.
Across both task types, 1,997 use logic-only masks and 2,066
use logic-and-UI masks. Not every trace in a mined lineage becomes an independent repair target.
Traces that do not satisfy the task-admission criteria in
Section~\ref{subsec:crafting} are retained as replay bridges to establish
prerequisite state.
Accordingly, as defined in Section \ref{subsec:crafting}, the \emph{restoration depth} $r_L$ counts the lineage behaviors
that become repair targets and whose code must be restored.

\paragraph{ProgramDistill-300.}
Evaluating all 4,063 tasks for every model would be prohibitively expensive,
so we define a fixed 300-task evaluation suite.
We stratify tasks by restoration depth $r_L$ using quotas of 50, 45, 45, 40,
35, 30, 30, and 25 for depths 1 through 8.
Within each depth, tasks are selected round-robin across applications to preserve
coverage of all 26 apps and prevent applications with larger task pools from
dominating. The resulting suite covers 269 distinct lineages and contains 50 atomic and 250 cumulative tasks, with 140 logic-only and 160 logic-and-UI masks.

\subsection{Partial-Application Reconstruction}
\label{sec:exp-repair}

We evaluate nine frontier models---GPT-6 Astra \citep{openai2026gpt6astra}, GPT-5.6 Sol~\citep{openai2026gpt56sol}, Claude Opus~5~\citep{anthropic2026claude}, Claude Sonnet~5~\citep{anthropic2026sonnet5},
Gemini~3.7 Flash~\citep{google2026gemini37flash}, Gemini~3.6 Flash~\citep{google2026gemini36flash}, Gemini~3.1 Pro Preview~\citep{google2026gemini31pro}, Grok~4.6~\citep{xai2026grok46} and GPT-5.3 Codex~\citep{openai2026gpt53codex}---on ProgramDistill-300,
with reasoning effort set to high. Each model receives a masked application and a problem statement and must
recover the missing behavior by inspecting a reference, editing the current implementation, and validating its changes against the live application. Depth-1 tasks are atomic repairs, while depths 2 through 8 combine multiple repair targets from a prerequisite lineage. 
\paragraph{Agent harness.}
All models use the same modified R2E-Gym \citep{r2egym} agent harness. The harness exposes four coding tools:
\texttt{execute\_bash}, \texttt{file\_editor}, \texttt{search}, and
\texttt{finish}. Through \texttt{execute\_bash}, the agent can also invoke the
\texttt{browser} CLI to observe and interact with both the masked current
application and the working reference.
Appendix~\ref{app:harness} describes the harness design and execution settings.

\subsubsection{Model Performance Comparison}

\begin{figure*}[h!]
  \centering
  \includegraphics[width=\linewidth]{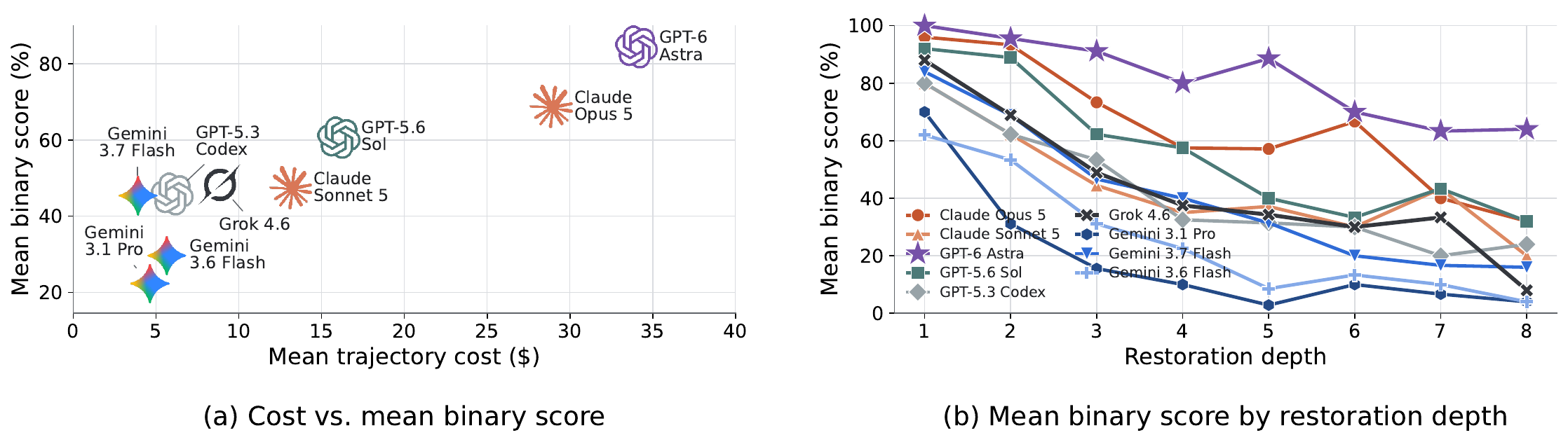}
  \caption{\textbf{Partial-application reconstruction performance.}
  \textbf{(a)} Mean binary score versus mean cost per trajectory (USD).
  \textbf{(b)} Mean binary score by restoration depth $r_L$.
  Performance generally declines as depth increases.
  Figure~\ref{fig:eval-repair-chain} in Appendix~\ref{app:chain-score}
  shows the same panels under the chain score.}
  \label{fig:eval-repair}
\end{figure*}

We evaluate nine models on \texttt{ProgramDistill-300}, comparing
repair performance with model cost. Figure~\ref{fig:eval-repair}(a) plots mean binary score against mean trajectory cost in USD, while
Appendix~\ref{app:chain-score} reports the corresponding chain-score
results. GPT-6 Astra achieves the highest mean binary score at \RepairAstraBinary\%,
followed by Claude Opus~5 at 68.7\% and GPT-5.6 Sol at 60.7\%.
Their mean trajectory costs are \$33.99, \$28.86, and \$16.01,
respectively. Astra gains 15.7 percentage points over Opus~5 at about
18\% higher mean trajectory cost.
Grok~4.6, Claude Sonnet~5, GPT-5.3 Codex, and Gemini~3.7 Flash form
a middle tier, with scores between 45.3\% and 48.3\%.

Higher expenditure does not always yield better repair performance.
Gemini~3.7 Flash nearly matches GPT-5.3 Codex (45.3\% vs.\ 45.7\%)
at a mean trajectory cost of \$3.89 versus \$5.99, while Grok~4.6
slightly exceeds Claude Sonnet~5 (48.3\% vs.\ 47.3\%) at \$8.83
versus \$13.12 per trajectory.

Mask scope provides a complementary view of repair difficulty.
Astra's chain score is \RepairAstraScopeChainLogic\% on \emph{logic-only}
tasks and \RepairAstraScopeChainUI\% on \emph{logic-and-UI} tasks.
The \RepairAstraScopeChainGap{}-point gap shows that interface reconstruction
remains harder even when partial restoration receives credit.
Appendix~\ref{app:mask-scope} reports the full scope comparison.

\paragraph{Degradation with restoration depth.}
Astra is the only model to solve all depth-1 tasks,
achieving 100\% binary success, compared with 96\% for Opus~5 and
92\% for Sol (Figure~\ref{fig:eval-repair}(b)).
Because cumulative tasks are composed of individually validated
atomic repairs and are themselves verified with gold patches, this
depth-1 result establishes a strong empirical baseline for task
solvability.
Performance nevertheless drops as multiple dependent repairs must be
recovered together.
Astra scores \RepairAstraDepthEightBinary\% at depth~8, a 36-point
decline, while Opus~5 and Sol both score 32.0\%.
Every other model retains less than half of its depth-1 performance.
These results suggest that restoration depth exposes a compositional
challenge, as agents can often solve individual repairs but struggle to
combine multiple repairs while preserving their dependencies.
Appendix~\ref{app:depth-context} examines this pattern at similar mean
prompt sizes per step.

\begin{figure}[!h]
\centering
\includegraphics[width=\linewidth]{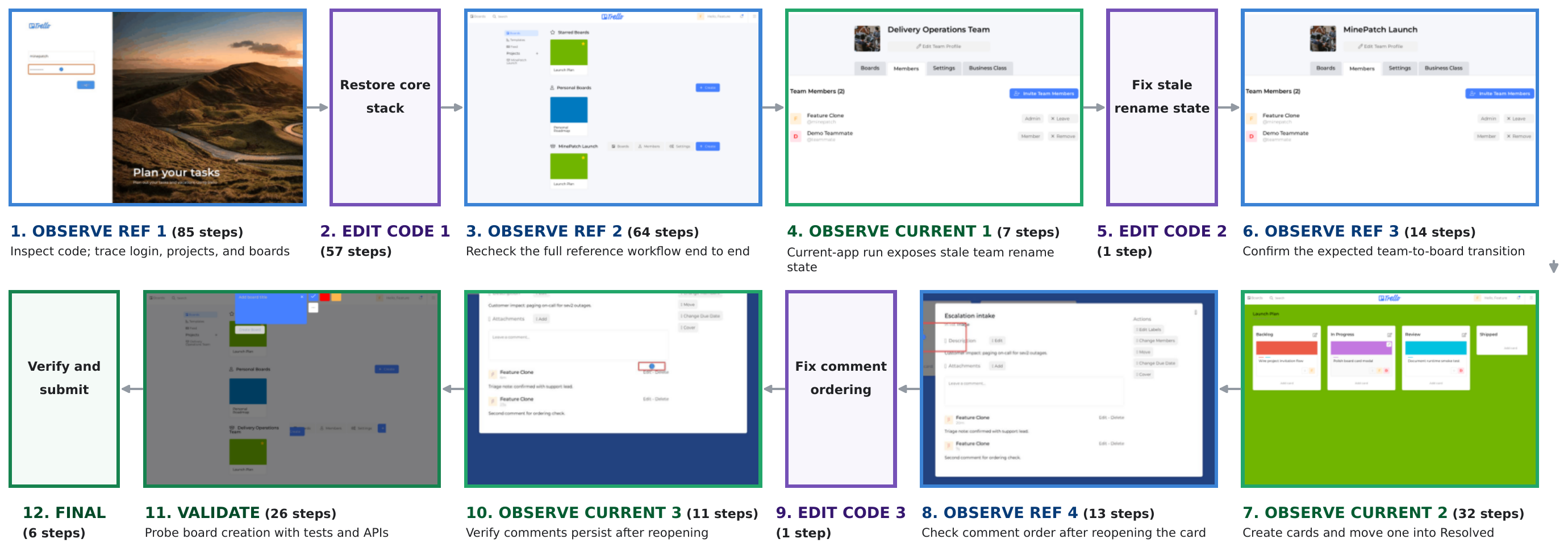}
\caption{\textbf{Partial-application reconstruction on a depth-8 \texttt{Trello (Vdevired)} task.}
The depth-8 task spans eight dependent workflow stages---login, project setup, team and board setup, list and card creation, permissions, card details, cross-list movement, and comment editing---all of which must remain functional together.
Across 317 steps, Claude Opus~5 repeatedly alternates between
\textcolor{observeref}{\textbf{OBSERVE REF}} to infer intended behavior,
\textcolor{editcode}{\textbf{EDIT CODE}} to modify the implementation, and
\textcolor{observecur}{\textbf{OBSERVE CURRENT}} to validate the repair
before final submission.}
\label{fig:trello-repair}
\end{figure}

\paragraph{Example of a successful deep repair trajectory.}
Although performance declines with restoration depth, some deep tasks
are successfully repaired through sustained iterative refinement.
Figure~\ref{fig:trello-repair} shows one such Claude Opus~5 trajectory
on a depth-8 cumulative repair task in \texttt{Trello (Vdevired)}. The agent
first inspects the reference and repository to recover the core
functionality, then repeatedly revisits the reference and current
application to identify residual mismatches, apply targeted fixes, and
validate them. The trajectory illustrates how a successful deep repair
can require repeated observe--edit--validate cycles across multiple
dependent behaviors.

\subsubsection{Analysis}
\label{sec:analysis}

To characterize repair workflows, we compare the recorded trajectories of all
nine models on \texttt{ProgramDistill-300}.

\paragraph{Model-level behavior.}

\begin{figure*}[h!]
\centering
\includegraphics[width=\linewidth]{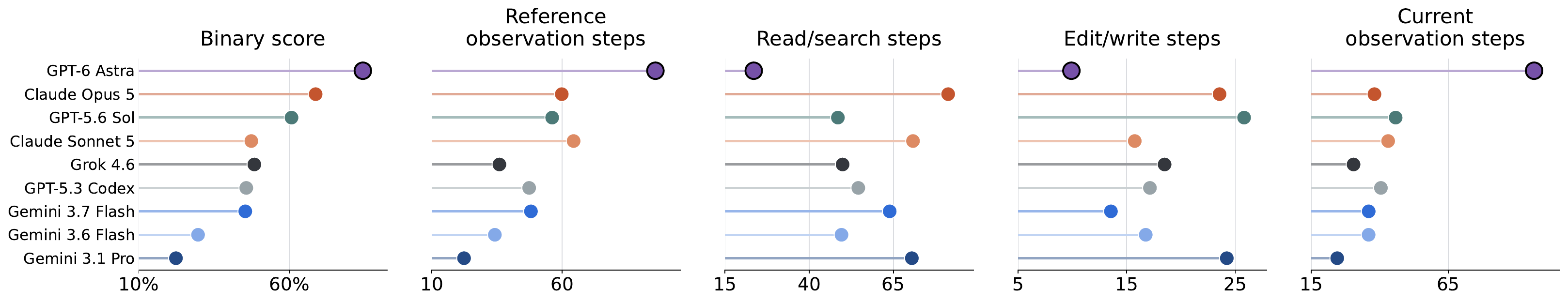}
\caption{\textbf{Model-level performance and interaction behavior.}
Mean binary score alongside mean reference observation, read/search,
edit/write, and current-app observation steps per trajectory.}
\label{fig:model-behavior}
\end{figure*}

Models exhibit markedly different interaction patterns during repair.
Figure~\ref{fig:model-behavior} compares reference and current-app
observation, repository read/search, and edit/write activity alongside
repair performance.

Astra stands out for an observation-intensive, edit-light workflow.
It records the most reference and current-app observation steps, with
current-app observation averaging 96.3 steps per trajectory, approximately
$2.1\times$ the next-highest mean of 45.8 for Sol.
At the same time, Astra makes the fewest edit/write steps, averaging only
9.9 per trajectory, while achieving the strongest repair performance.
Opus~5 and Sonnet~5 are also relatively observation-intensive, but combine
this with substantially more editing, averaging 23.5 and 15.7 edit/write
steps, respectively.

Astra's interaction pattern is therefore characterized by extensive
behavioral checking, particularly of its own implementation, followed by
comparatively selective code changes.
More broadly, these results suggest that repair performance may depend
not only on implementation capability but also on \textbf{how effectively
agents allocate effort across observation, validation, and editing.} Table~\ref{tab:behavior} in Appendix~\ref{app:interaction-stats}
reports interaction statistics across models and restoration depths.

\begin{figure*}[t!]
\centering
\includegraphics[width=\linewidth]{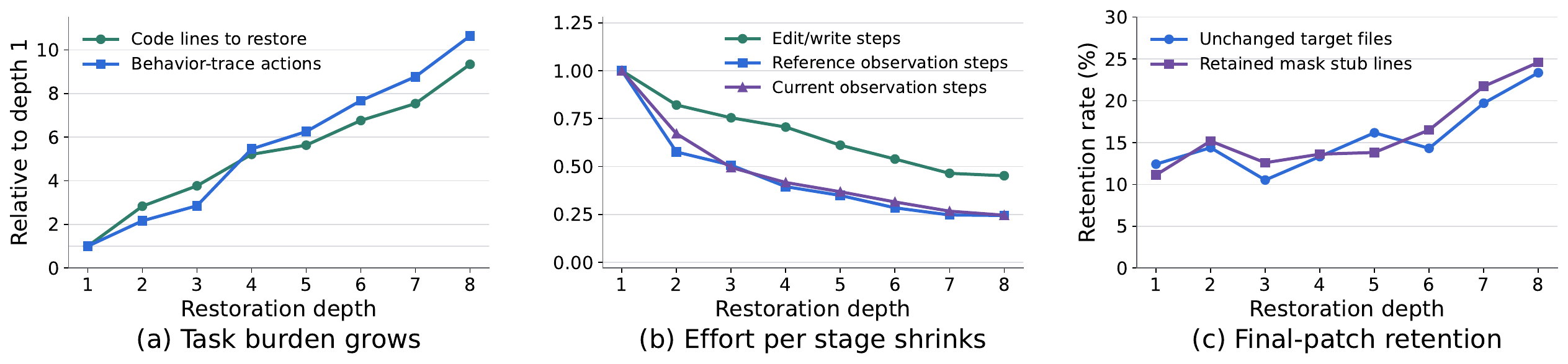}
\caption{\textbf{Scaling of repair burden and agent effort with restoration depth $r_L$.}
(a) Repair burden increases with restoration depth.
(b) Agent effort per repair target, measured by edit/write and
reference/current-app observation steps.
(c) Final-patch retention of unchanged target files and mask-added stub lines.
Panels (a) and (b) are normalized to their depth-1 values, while
panel (c) reports raw percentages.}
\label{fig:depth-behavior}
\end{figure*}

\paragraph{Scaling with restoration depth.}
To better understand why repair performance declines with restoration
depth, we examine how task burden and agent effort scale as more
behaviors must be restored. Figure~\ref{fig:depth-behavior} shows a
growing mismatch between the two. From depth~1 to depth~8, the mean number of code lines to restore grows
by 9.3$\times$, while the mean total number of browser actions across
target behavior traces grows by 10.7$\times$.

In contrast, agent effort per repair target decreases, with the sharpest
contraction in behavioral observation. Reference observation
steps per target fall from \RepairDepthOneReference{} to
\RepairDepthEightReference{}, and current-app observations from
\RepairDepthOneCurrent{} to \RepairDepthEightCurrent{}, reductions of
roughly 75\% in both cases. Edit/write steps also decline, but more
moderately, by roughly 55\%.
\textbf{Thus, as restoration depth grows, agents devote disproportionately less
effort to understanding and validating each required behavior.}

Final patches also leave more of the masked implementation unrestored
as restoration depth increases. Between depths~1 and~8, the mean
fraction of unchanged target files changes from
\RepairDepthOneUnchanged\% to \RepairDepthEightUnchanged\%, while
the fraction of mask-added stub lines retained in the final patch
changes from \RepairDepthOneStubs\% to \RepairDepthEightStubs\%.
Although intermediate depths fluctuate, the
overall pattern suggests that deeper tasks leave more required
implementation unrecovered. Together with the decline in binary score
and the reduced observation and editing effort per target, this indicates
that agents struggle to keep pace as reconstruction burden grows.

\paragraph{Shortcut attempts.}

Beyond the intended repair workflow, agents may attempt to recover
implementation information from sources outside the supplied code and
permitted reference interface. We therefore identify access attempts
involving Git history, public app-source retrieval, protected harness
or evaluator files, and direct reference access.

\begin{figure*}[h!]
\centering
\includegraphics[width=\linewidth]{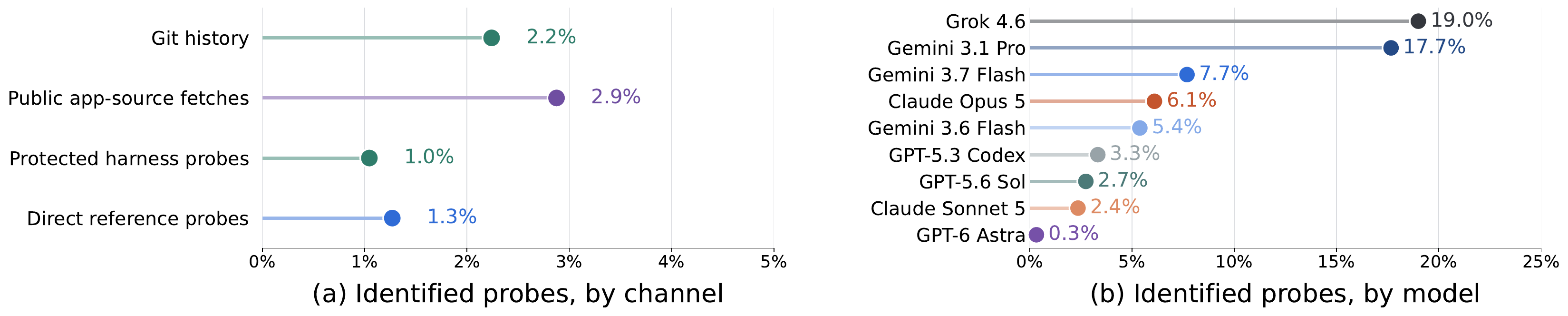}
\caption{\textbf{Identified shortcut attempts.}
(a) Share of trajectories with an identified attempt in each channel.
(b) Within-model share with at least one identified attempt.}
\label{fig:cheating-attempts}
\end{figure*}

Figure~\ref{fig:cheating-attempts}(a) shows the frequency of public app-source
retrieval, Git history, and other identified access channels.
At least one such attempt appears in \RepairPooledProbePercent\% of trajectories, with
substantial variation across models. Grok~4.6 has the highest rate at
19.0\%, followed by Gemini~3.1~Pro at 17.7\%, compared with
\RepairAstraProbePercent\% for Astra.

These channels are blocked by design through several safeguards.
Workspaces contain no pre-mask Git history, stale artifacts are
regenerated from the masked source, public egress is disabled, harness
and grading files are hidden or protected, and the reference is
accessible only through the browser helper. Details of the runtime architecture
and isolation safeguards are provided in Appendix~\ref{app:harness-architecture}.

\input{archive/section/reconstruction_results}

\subsubsection{Analysis} \label{sec:full-app-analysis}

\input{archive/section/reconstruction_failure_analysis}

%% file: archive/section/reconstruction_results.tex
\subsection{Full-Application Reconstruction}
\label{sec:full-app}

We next evaluate the full-application reconstruction setting introduced
in Section~\ref{subsec:patching}. Starting from a minimal executable
scaffold, the agent must rebuild the application by using a product-level
capability description and interacting with a live reference.
We evaluate this setting on twelve stateful web applications using
GPT-6 Astra, Claude Opus~5, and GPT-5.6 Sol at max reasoning effort.
The harness configuration is described in Appendix~\ref{app:harness}.

\paragraph{Evaluation suite and scoring.}
Following the protocol in Section~\ref{subsec:patching}, we evaluate
full-application reconstruction using two types of behavioral tests.
Atomic behavior tests check whether one replay-verified target behavior
is reproduced, with its prerequisites replayed to establish the required
state. Cumulative workflow tests assess
whether multiple dependent behaviors are recovered together.
The evaluation suite contains 590 atomic behavior tests and 413
cumulative workflow tests across twelve applications.

\subsubsection{Model Performance Comparison}

\begin{table*}[h!]
\centering
\caption{\textbf{Full-application reconstruction performance.}
Atomic recovery measures the fraction of atomic behavior tests passed,
while cumulative recovery requires all target behaviors in a workflow
to pass. Bold indicates the
best result among three models.
Appendix~\ref{app:chain-score} reports the
corresponding cumulative chain scores.}
\label{tab:full-app-reconstruction}
\begingroup
\setlength{\tabcolsep}{3pt}
\newcommand{\appheader}[1]{%
  \textbf{\begin{tabular}[c]{@{}c@{}}#1\end{tabular}}}
\resizebox{\textwidth}{!}{%
\begin{tabular}{lccccccccccccc}
\toprule
\appheader{Model} & \appheader{Reactive\\Resume} & \appheader{AWS\\Console} & \appheader{Baserow} & \appheader{BudgetWise} & \appheader{CareerLink} & \appheader{MailHub} & \appheader{Jira\\(Oldboyxx)} & \appheader{Trello\\(Knowankit)} & \appheader{StreamView} & \appheader{TeamChat} & \appheader{Trello\\(Vdevired)} & \appheader{Weights \&\\Biases} & \appheader{Mean} \\
\midrule
\multicolumn{14}{l}{\textit{Atomic behavior recovery}} \\
\addlinespace[2pt]
GPT-6 Astra (max) & \textbf{50.0} & \textbf{56.0} & \textbf{14.0} & \textbf{60.0} & \textbf{62.0} & 88.0 & 64.0 & 53.3 & \textbf{82.0} & \textbf{68.0} & \textbf{28.9} & \textbf{78.0} & \textbf{58.98} \\
Opus 5 (max) & 20.0 & 16.0 & 10.0 & 48.0 & 42.0 & \textbf{92.0} & \textbf{82.0} & \textbf{55.6} & 32.0 & 58.0 & 6.7 & 40.0 & 42.03 \\
GPT-5.6 Sol (max) & 8.0 & 8.0 & 8.0 & 30.0 & 18.0 & 78.0 & 72.0 & \textbf{55.6} & 32.0 & 50.0 & 26.7 & 16.0 & 33.39 \\
\addlinespace[4pt]
\midrule
\multicolumn{14}{l}{\textit{Cumulative binary score}} \\
\addlinespace[2pt]
GPT-6 Astra (max) & \textbf{43.8} & \textbf{45.8} & \textbf{5.9} & \textbf{42.6} & \textbf{55.6} & 81.8 & 75.0 & 38.1 & \textbf{75.9} & \textbf{45.5} & \textbf{23.5} & \textbf{72.7} & \textbf{49.15} \\
Opus 5 (max) & 17.2 & 25.0 & 2.9 & 21.3 & 27.8 & \textbf{86.4} & \textbf{86.1} & 38.1 & 13.8 & 33.3 & 0.0 & 24.2 & 28.81 \\
GPT-5.6 Sol (max) & 4.7 & 0.0 & 2.9 & 12.8 & 8.3 & 59.1 & 75.0 & \textbf{47.6} & 24.1 & 27.3 & 20.6 & 3.0 & 21.07 \\
\bottomrule
\end{tabular}}
\endgroup
\end{table*}

Table~\ref{tab:full-app-reconstruction} shows that full-application
reconstruction remains challenging even for the strongest models.
GPT-6 Astra performs best on both atomic behaviors and cumulative
workflows, reaching \CanonicalAstraAtomicWeighted\% atomic recovery and
\CanonicalAstraBinaryWeighted\% cumulative recovery. Claude Opus~5
achieves 42.03\% atomic and 28.81\% cumulative recovery, while
GPT-5.6 Sol achieves 33.39\% and 21.07\%, respectively.

Notably, this ordering matches the masked-repair results, with Astra
outperforming Opus~5 and Opus~5 outperforming Sol in both settings.
The consistency across substantially different task regimes suggests
that ProgramDistill captures stable differences in reference-guided
software engineering capability rather than effects specific to one
task formulation.

Across models, cumulative recovery is consistently lower than atomic recovery, showing that reproducing atomic behaviors does not necessarily translate into recovering complete multi-step workflows.
Performance also varies sharply across applications. For Astra,
cumulative recovery ranges from 81.8\% on \texttt{MailHub} to 5.9\% on \texttt{Baserow}.
One possible explanation is that some applications expose more regular
and repetitive interaction structures, while others require coordinating
more heterogeneous state and workflows. \texttt{MailHub}, for example, is largely
organized around recurring message and thread operations. In contrast,
lower-scoring applications involve more varied representations, such as
user-defined table schemas in \texttt{Baserow}, resume layouts and JSON
import/export in \texttt{Reactive Resume}, service-specific resource
configurations in \texttt{AWS Console}, and plan and SIM selections across cart
and checkout in \texttt{BudgetWise}. Maintaining consistency across these
heterogeneous representations may make full-application reconstruction
substantially harder.

\begin{table*}[htbp]
\centering
\scriptsize
\setlength{\tabcolsep}{3.5pt}
\caption{\textbf{Agent activity during full-application reconstruction.}
Min, mean, and max per-run counts of agent steps, observation steps,
and browser state returns. Observation steps use the same request-based
definition as in the repair analysis.
Browser state returns count recorded helper responses, including those
returned by interactions and failed actions, and multiple responses may occur
within one step.}
\label{tab:rebuild-behavior}

\begin{tabular}{
    >{\raggedright\arraybackslash}p{3.25cm}
    S[table-format=4.2, table-column-width=1.1cm]
    S[table-format=4.2, table-column-width=1.3cm]
    S[table-format=4.2, table-column-width=1.2cm]
    @{\hspace{7pt}}
    S[table-format=4.2, table-column-width=1.1cm]
    S[table-format=4.2, table-column-width=1.3cm]
    S[table-format=4.2, table-column-width=1.2cm]
    @{\hspace{7pt}}
    S[table-format=4.2, table-column-width=1.1cm]
    S[table-format=4.2, table-column-width=1.3cm]
    S[table-format=5.0, table-column-width=1.2cm]
}
\toprule
& \multicolumn{3}{c}{\textbf{Steps}}
& \multicolumn{3}{c}{\textbf{Observation steps}}
& \multicolumn{3}{c}{\textbf{Browser state returns}} \\
\cmidrule(lr){2-4}
\cmidrule(lr){5-7}
\cmidrule(lr){8-10}
\textbf{Model}
& {Min} & {Mean} & {Max}
& {Min} & {Mean} & {Max}
& {Min} & {Mean} & {Max} \\
\midrule
GPT-6 Astra (max) & 292 & 719.4 & 1921 & 221 & 643.3 & 1796 & 406 & 1154.1 & 3843 \\
Opus 5 (max) & 371 & 627.5 & 940 & 59 & 258.2 & 557 & 1427 & 4932.1 & 12821 \\
GPT-5.6 Sol (max) & 348 & 719.0 & 1165 & 204 & 419.3 & 705 & 277 & 854.7 & 1441 \\
\bottomrule
\end{tabular}

\end{table*}

\paragraph{Observation strategies in reconstruction.}
Unlike partial-application reconstruction, where trajectories average roughly 200 agent
steps, full-application reconstruction produces much longer runs,
averaging around 700 steps and reaching up to 1,921
(Table~\ref{tab:rebuild-behavior}). In this setting, Opus~5 often uses
shell \texttt{for} loops to issue multiple browser CLI calls within a
single agent step, effectively collecting several browser states at
once. This yields 7.9 browser state returns per agent step on average,
compared with 1.6 for Astra and 1.2 for Sol.

Distinguishing observation steps from browser state returns exposes
this difference in interaction style. Despite the much larger volume of
browser feedback, Opus~5 still recovers fewer behaviors than Astra,
suggesting that observation volume alone does not explain reconstruction
performance.

%% file: archive/section/reconstruction_failure_analysis.tex
\input{archive/analysis/generated/final-failure-macros.tex}

Full-application reconstruction remains difficult even for the strongest agents.
We therefore ask why agents fail to recover behaviors that are available to
inspect in a working reference.
We analyze \FailureCount{} failed atomic behaviors across the
\FailureRuns{} reconstruction runs in Table~\ref{tab:full-app-reconstruction},
using the submitted source, and recorded reference and current
observations with GPT-6 Astra. 
Each failure is assigned to one of four categories: not observed, observed but
not implemented, implemented with the wrong observable form, or implemented
with the wrong state, route, or result. We also examine why the agent's own
validation did not expose and correct the remaining difference.

\begin{figure*}[htbp]
\centering
\includegraphics[width=\linewidth]{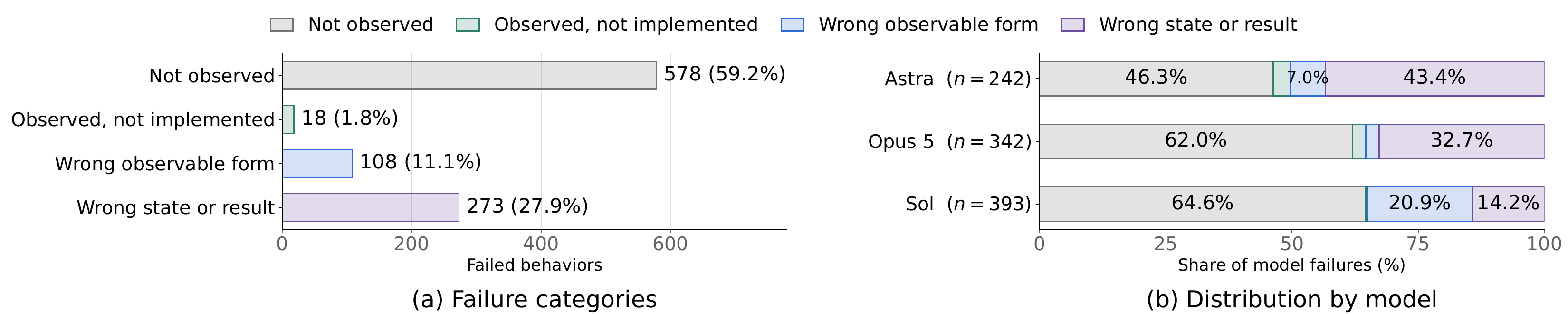}
\caption{\textbf{Full-application reconstruction failures.}
Across \FailureRuns{} runs on \FailureApps{} applications,
\textbf{(a)} counts failures in the four categories and
\textbf{(b)} shows their distribution within each model's failures.}
\label{fig:full-app-analysis}
\end{figure*}

\paragraph{Behavior-level reconstruction outcomes.}
As shown in Figure~\ref{fig:full-app-analysis}, the largest failure
category is behavior that was not observed in the reference, accounting
for \FailureNotObservedPercent\% of failures. The remaining failures occur
after the relevant behavior has been observed. Of these,
\FailureNotImplementedPercent\% are observed but not implemented,
\FailureWrongFormPercent\% are implemented with the wrong observable form,
and \FailureWrongStatePercent\% produce the wrong state, route, or result.
These outcomes reveal difficulties both in exploring the reference and in
faithfully reproducing behaviors that have already been observed.

\paragraph{What validation missed.}
Agents checked that their reconstructed applications worked, but these
checks did not always establish faithful reproduction of the reference
behavior. In the failure cases in
Figure~\ref{fig:reconstruction-example-cases}, agents often validated
related functionality or a limited set of workflows without rechecking
the failing workflow after the final relevant source edit. For
example, validating native resume creation and export did not reveal a
failure in external JSON Resume import, while syntax-only checks did not
reveal that report edits in W\&B were not persisted.

Checks such as syntax validation, successful backend requests, or
selected text matches can establish that an application runs without
showing that its behavior matches the reference. As a result, omitted
behaviors, incorrect observable forms, and wrong state transitions can
remain undetected even when the relevant reference behavior has already
been observed. A related pattern appears in partial-application reconstruction, where current-app
observation effort per target decreases as restoration depth increases
(Figure~\ref{fig:depth-behavior}).

\input{archive/section/reconstruction_examples}

%% file: archive/analysis/generated/final-failure-macros.tex
\newcommand{\FailureRuns}{36}
\newcommand{\FailureApps}{12}
\newcommand{\FailureTargets}{1770}
\newcommand{\FailurePassed}{793}
\newcommand{\FailureCount}{977}
\newcommand{\FailureReviewed}{977}
\newcommand{\FailureClassified}{977}
\newcommand{\FailureWrongForm}{108}
\newcommand{\FailureWrongFormPercent}{11.1}
\newcommand{\FailureWrongState}{273}
\newcommand{\FailureWrongStatePercent}{27.9}
\newcommand{\FailureNotImplemented}{18}
\newcommand{\FailureNotImplementedPercent}{1.8}
\newcommand{\FailureNotObserved}{578}
\newcommand{\FailureNotObservedPercent}{59.2}

%% file: archive/section/reconstruction_examples.tex
\begin{figure*}[htbp]
\centering

\begin{subfigure}[t]{0.49\linewidth}
\centering
\includegraphics[width=\linewidth]{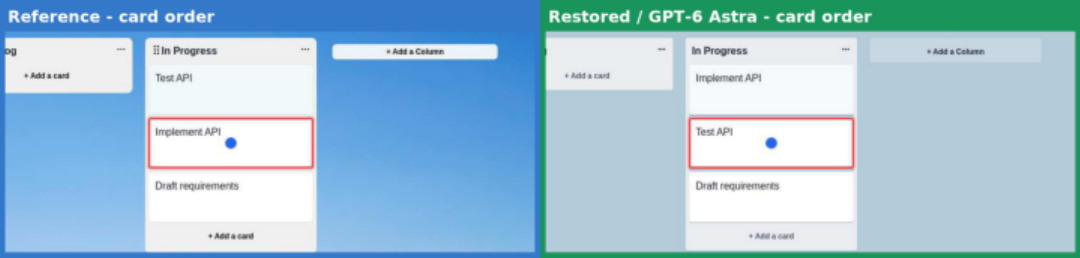}
\caption{\texttt{Trello (Knowankit)}, GPT-6 Astra}
\label{fig:reconstruction-trello-order}
\end{subfigure}
\hfill
\begin{subfigure}[t]{0.49\linewidth}
\centering
\includegraphics[width=\linewidth]{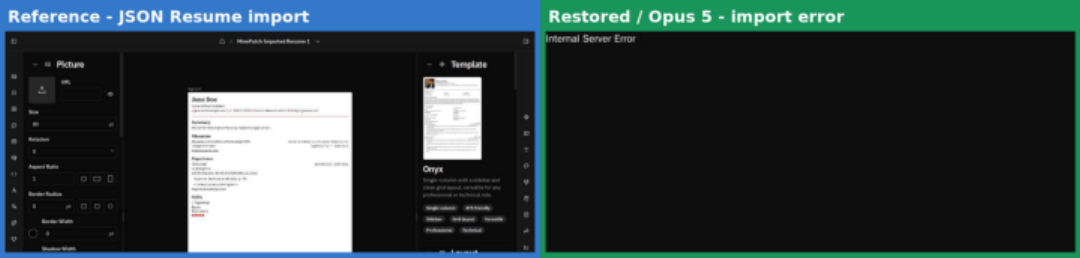}
\caption{\texttt{Reactive Resume}, Opus~5}
\label{fig:reconstruction-resume-import}
\end{subfigure}

\medskip

\begin{subfigure}[t]{0.49\linewidth}
\centering
\includegraphics[width=\linewidth]{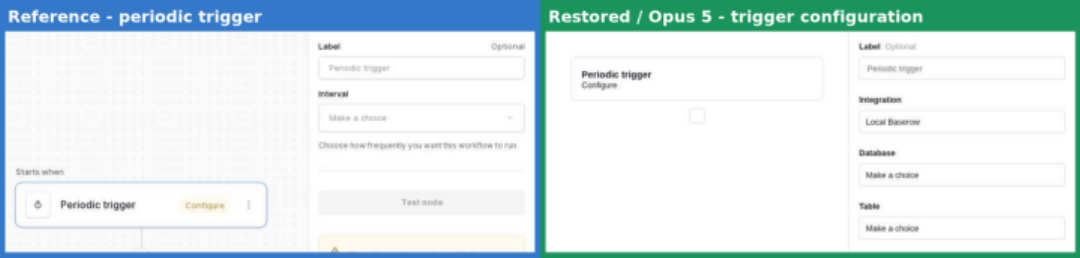}
\caption{\texttt{Baserow}, Opus~5}
\label{fig:reconstruction-periodic-trigger}
\end{subfigure}
\hfill
\begin{subfigure}[t]{0.49\linewidth}
\centering
\includegraphics[width=\linewidth]{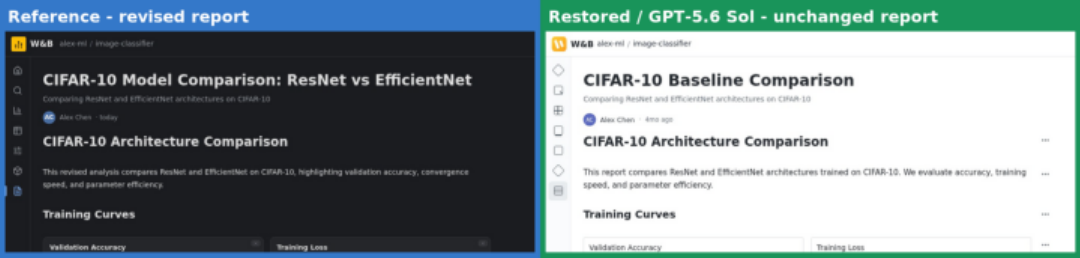}
\caption{W\&B, GPT-5.6 Sol}
\label{fig:reconstruction-report-save}
\end{subfigure}

\caption{\textbf{Reconstruction errors missed during validation.}
In each panel, the reference application appears on the left and the
final submission on the right.
\textbf{(a)} Dragging a card leaves the cards in the wrong order.
\textbf{(b)} Importing a JSON Resume produces a rendering failure.
\textbf{(c)} A periodic trigger shows database settings instead of an
interval setting.
\textbf{(d)} Changes to a report's title and summary are not saved.}
\label{fig:reconstruction-example-cases}
\end{figure*}

\paragraph{Example failure cases.}
Figure~\ref{fig:reconstruction-example-cases} illustrates four
reconstruction errors that remain after agent validation. For each
case, we replay the same action sequence on the reference and the final
submission to expose the behavioral mismatch. Across the four examples,
the agent's validation does not recheck the exact
workflow that reveals the remaining error after its final relevant edit.

In \texttt{Trello (Knowankit)}, Astra reconstructs card dragging, but a center
drop leaves the cards in the opposite order from the reference. After
the final relevant edit, the agent validates other board interactions
rather than this drag workflow, so the ordering error remains
undetected.

In \texttt{Reactive Resume}, Opus~5 implements JSON Resume import without
converting the uploaded schema into the representation expected by the
application, causing the imported resume to fail during rendering.
Its final validation covers native resume creation, editing, and export,
but does not exercise import of the external JSON Resume schema.

In \texttt{Baserow}, a trigger determines how an automated workflow starts. A
periodic trigger should expose an \texttt{Interval} setting that
specifies how often the workflow runs. Opus~5 changes the trigger title
to \texttt{Periodic trigger}, but leaves the underlying database-trigger
form in place. The reconstructed application therefore shows
\texttt{Integration}, \texttt{Database}, and \texttt{Table} instead of
the expected \texttt{Interval} setting. The agent validates a
database-trigger workflow but does not recheck the periodic-trigger
configuration.

Finally, in W\&B, Sol renders a report editor but does not persist changes
to the title and summary. After the same editing sequence, the reference
retains the revised report while the reconstructed application still
shows the original text. The agent's final checks cover syntax without
reopening the edited report, so the failed state update remains undetected.

%% file: archive/section/related_works.tex
\section{Related Works}

\paragraph{Repository-level software-engineering benchmarks.}
A large body of work evaluates coding agents on repository-level software
engineering. SWE-bench~\citep{swebench} asks agents to resolve real GitHub
issues, while R2E-Gym~\citep{r2egym}, SWE-smith~\citep{swesmith}, and related
pipelines scale executable task construction from commits, tests, and source
repositories, and open agent platforms such as
OpenHands~\citep{wang2025openhands} standardize how such agents are built and
run. DeepSWE~\citep{deepswe} pushes the same setting toward
long-horizon engineering work, where a single task spans many dependent
changes. SWE-bench Multimodal~\citep{swebench_multimodal} augments issue
descriptions with visual evidence, and SWE-Together~\citep{swetogether}
extends this setting to multi-turn sessions in which users progressively
clarify or correct requirements. These benchmarks increasingly capture
realistic repository editing and execution, but the desired behavior is still
supplied through an issue, test-derived task, visual artifact, or user
interaction. ProgramDistill instead requires the agent to recover the
specification by interacting with a working application, and uses the
application's own executable behaviors as both repair targets and replayable
verifiers.

\paragraph{Visual and interactive web development.}
Web2Code~\citep{web2code} and Design2Code~\citep{design2code} study
reconstructing webpages from visual references, while
Interaction2Code~\citep{interaction2code} extends this setting to interactive
behavior. Vision2Web~\citep{vision2web} and VISTA~\citep{vista} further broaden
the scope toward interactive and full-stack web development from text, image,
or design specifications. ProgramDistill differs in that the target behavior
is not provided as a fixed artifact. The agent must decide what to inspect,
exercise a live reference to uncover stateful behavior, relate those
observations to an existing codebase, implement the missing functionality, and
validate the result against the running application. The distinction is thus a
shift from \emph{specification consumption} to \emph{specification discovery}.

\paragraph{Behavioral reconstruction and reverse engineering.}
Recovering an unstated specification has been studied from the
source side, where SpecRover~\citep{specrover} infers program intent to guide
repair, and from the verification side, where oracle
automation~\citep{molina2024oracle} derives the assertions a test needs. In
ProgramDistill both come from the running application itself, since a mined
trace is at once the intent to restore and the oracle that checks it.
ProgramBench~\citep{programbench} and MirrorCode~\citep{mirrorcode} provide
the closest precedent for treating executable software as a behavioral
specification, where agents probe a program without its source and reconstruct an
implementation that matches its externally visible behavior. Their primary
unit, however, is holistic program reimplementation. ProgramDistill instead
factorizes an interactive application into replayable behaviors, preserves
their prerequisite relationships, and recomposes them into tasks of increasing
depth. The same behavioral substrate supports atomic repair, cumulative repair
across dependent workflows, and full-application reconstruction under a common
verifier.

\paragraph{Browser-use agents and environments.}
A broad line of work studies agents that complete user-specified tasks
through interaction with existing websites, including
WebArena~\citep{webarena}, VisualWebArena~\citep{visualwebarena},
WorkArena~\citep{browsergym}, Mind2Web~\citep{deng2023mind2web},
WebVoyager~\citep{webvoyager}, and BU Bench~\citep{browser_use2024}.
OSWorld~\citep{xie2024osworld} evaluates agents on computer-use
tasks involving existing web and desktop applications. A complementary line of work develops browser interfaces and execution
infrastructure for such agents.
BrowserGym~\citep{browsergym} provides a unified environment with
common observation and action spaces across web-agent benchmarks.
Browser Use~\citep{browser_use2024} integrates page observations and
browser-control tools into an agent framework.
Anthropic's browser use tool~\citep{claudebrowseruse} provides tools
for reading page structure and interacting with referenced elements.
Playwright MCP~\citep{playwrightmcp} exposes Playwright browser
automation through accessibility snapshots and element-level actions.

ProgramDistill similarly couples coding with iterative browser-based
observation and execution, but the code serves a different purpose.
Rather than producing an automation program that operates a website,
the agent modifies the application itself to recover behavior observed
from a working reference.
It must infer intended behavior through interaction, repair or
reconstruct the source implementation, and validate the result against
the reference.
Moreover, ProgramDistill turns discovered interactions into both
software-engineering tasks and replayable verifiers, using browser
interaction not only for execution but also for task construction and
evaluation.

Taken together, prior benchmarks isolate important parts of the software
development process. ProgramDistill brings them into a single closed-loop
setting in which an agent must \emph{observe, infer, implement, and validate}
behavior against a live reference. This closely mirrors reference-guided web
development, where a developer studies an existing product or prototype,
modifies the implementation, and repeatedly checks the result. At the same
time, ProgramDistill turns the behaviors uncovered from the application into
the tasks and verifiers themselves, providing one substrate for studying
localized repair, compositional workflows, and full-application reconstruction.

%% file: archive/section/conclusion.tex
\section{Discussion}

\paragraph{Limitations and future work.}
Agents in both partial- and full-application reconstruction receive
structured browser observations of visible text, accessibility information,
and interactive elements, without accompanying multimodal screenshot inputs.
Extending the observation interface with screenshots would enable visually
grounded interaction, while evaluating visual fidelity would require criteria
beyond the current replay-based behavioral checks.

Our experiments also focus on self-contained web applications.
Extending ProgramDistill to more open-ended settings involving external
services, nondeterministic state, and desktop or mobile applications remains
an important direction for future work.

Benchmark construction in our experiments uses GPT-5.6 Sol throughout the
mine--craft--patch pipeline.
The coverage and quality of mined behaviors and generated repair tasks may
therefore depend on the capability of the construction model.
More capable models could improve behavior discovery, masking, and task
construction, while also increasing generation cost.
Characterizing this quality--cost trade-off across construction models is an
important direction for future work.

ProgramDistill also depends on the browser and agent infrastructure used to
expose, record, and replay application behavior.
The current browser helper relies on replay-stable selectors and settled
structured observations, while the agent harness determines how models access
and revisit those observations.
Improvements in replay robustness, observation quality, or agent tooling could
expand the set of behaviors that can be reliably mined and may affect measured
agent performance.
Characterizing the sensitivity of ProgramDistill to these infrastructure
choices is another direction for future work.

Beyond evaluation, the replay-verifiable tasks and agent trajectories produced
by ProgramDistill provide a natural basis for training coding agents.
Trajectories from frontier agents can be distilled into training data for
smaller models, while replay-verifiable tasks can serve directly as
reinforcement learning environments.
Restoration depth further provides a controllable curriculum axis from atomic
repairs to increasingly compositional workflows.
We leave the use of ProgramDistill for training coding agents to future work.

\paragraph{Potential misuse.} Reference-guided software engineering could be misused to clone proprietary applications or reproduce their functionality without authorization. Our experiments are limited to self-contained applications derived from public or open-source repositories, and the framework is intended only for research and authorized software development.

\paragraph{Conclusion.}
We introduced ProgramDistill, a framework that turns the end-to-end workflow of reference-guided web development into verifiable software-engineering tasks.
Agents must inspect a working application, infer its behavior, recover that behavior through source-code changes, and validate the result against the reference.
By factorizing applications into replayable behaviors while preserving their prerequisite relationships, ProgramDistill provides a common substrate spanning localized repair, compositional workflows, and full-application reconstruction.
Across frontier coding agents, performance declines with restoration
depth, while full-application reconstruction leaves substantial behavior
unrecovered. These complementary settings expose gaps between observing
reference behavior, implementing it, and validating the resulting application,
making ProgramDistill a diagnostic benchmark for current coding agents.

More broadly, ProgramDistill shows that executable software can serve not only as a target for evaluation, but as a scalable source of specifications, tasks, verifiers, and learning signals. This opens a path toward a continual cycle in which working software generates increasingly challenging tasks that both reveal the limits of current coding agents and provide the experience needed to push beyond them.

\section*{Acknowledgment}

We thank Emiliano Penaloza, Christopher Cui, Jonathan Light, and Roger Creus Castanyer for their valuable discussions and insightful feedback.

%% file: archive/section/appendix_contents.tex
\clearpage
\addtocontents{toc}{\protect\setcounter{tocdepth}{2}}
\begingroup
\definecolor{appendixgreen}{HTML}{004D00}
\hypersetup{colorlinks=true,linkcolor=appendixgreen}
\renewcommand{\contentsname}{Appendix Contents}
\tableofcontents
\endgroup
\clearpage

%% file: archive/section/appendix.tex
\providecommand{\workshopbuild}{0}

\newtcolorbox{promptbox}[1]{enhanced, breakable, colback=white,
  colframe=fcframe, colbacktitle=green!10, coltitle=black,
  boxrule=1.2pt, arc=2pt, left=3pt, right=3pt, top=3pt, bottom=3pt,
  fonttitle=\small\bfseries, title={#1}}

\newpage
\section{Application Corpus}
\label{app:corpus}

The corpus contains 26 self-contained interactive web applications drawn from
two sources. Eighteen are adapted from the OSWorld web-application
suite~\citep{xie2024osworld}, which provides a controlled and diverse
interaction substrate. The remaining eight are real-world open-source projects
and SaaS clones taken from public repositories. These contribute larger
codebases, deeper workflows, nontrivial state, and heterogeneous architectures,
spanning Django, Angular, React, and Next.js stacks with SQL and document
back ends.

\begin{figure}[h!]
    \centering
    \begin{subfigure}{0.23\linewidth}
        \includegraphics[width=\linewidth]{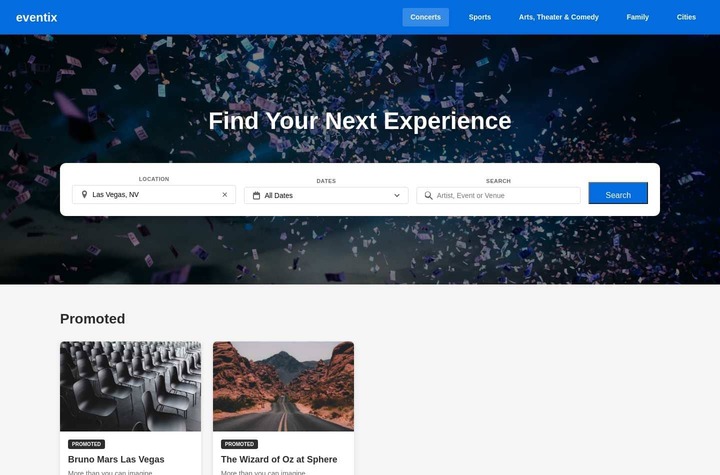}
        \caption{\texttt{Eventix} (events)}
    \end{subfigure}\hfill
    \begin{subfigure}{0.23\linewidth}
        \includegraphics[width=\linewidth]{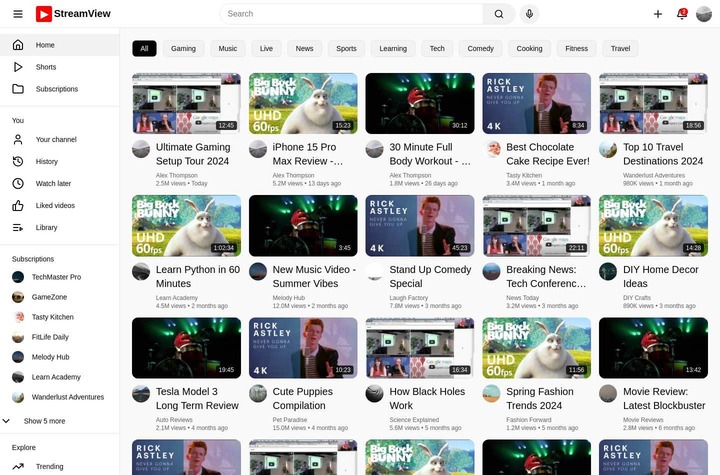}
        \caption{\texttt{StreamView} (video)}
    \end{subfigure}\hfill
    \begin{subfigure}{0.23\linewidth}
        \includegraphics[width=\linewidth]{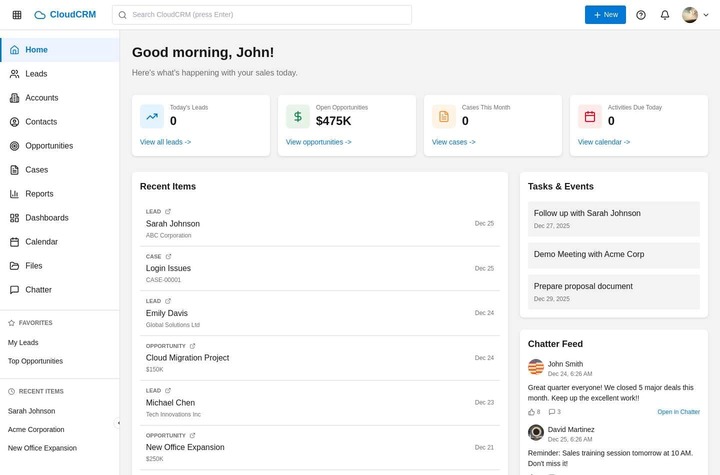}
        \caption{\texttt{CloudCRM} (CRM)}
    \end{subfigure}\hfill
    \begin{subfigure}{0.23\linewidth}
        \includegraphics[width=\linewidth]{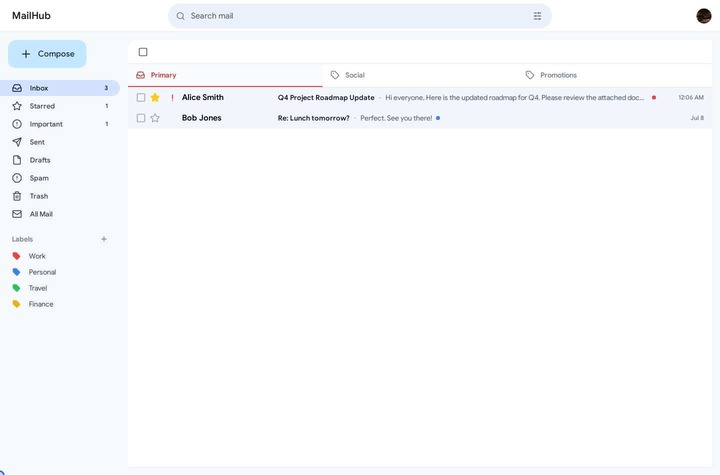}
        \caption{\texttt{MailHub} (email)}
    \end{subfigure}

    \vspace{6pt}

    \begin{subfigure}{0.23\linewidth}
        \includegraphics[width=\linewidth]{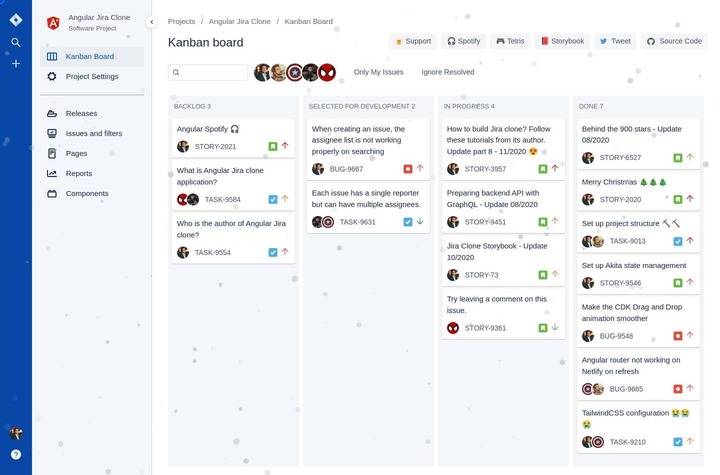}
        \caption{\texttt{Jira clone} (kanban)}
    \end{subfigure}\hfill
    \begin{subfigure}{0.23\linewidth}
        \includegraphics[width=\linewidth]{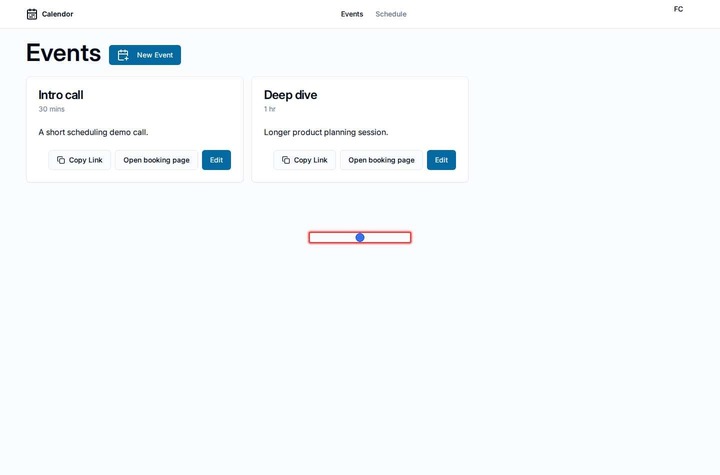}
        \caption{\texttt{Calendly} (scheduling)}
    \end{subfigure}\hfill
    \begin{subfigure}{0.23\linewidth}
        \includegraphics[width=\linewidth]{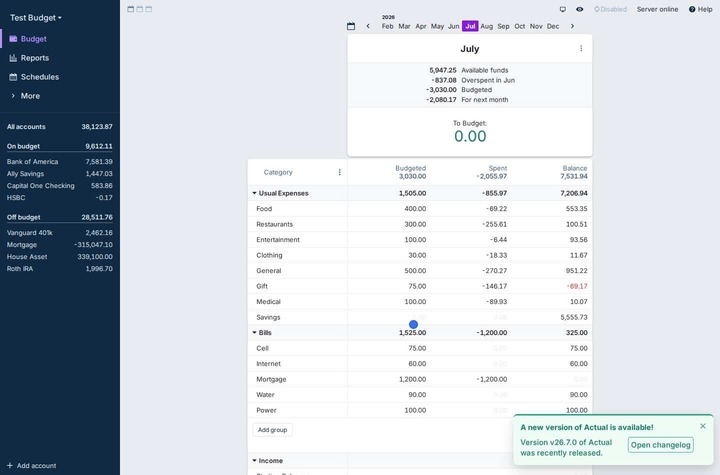}
        \caption{\texttt{Actual Budget} (finance)}
    \end{subfigure}\hfill
    \begin{subfigure}{0.23\linewidth}
        \includegraphics[width=\linewidth]{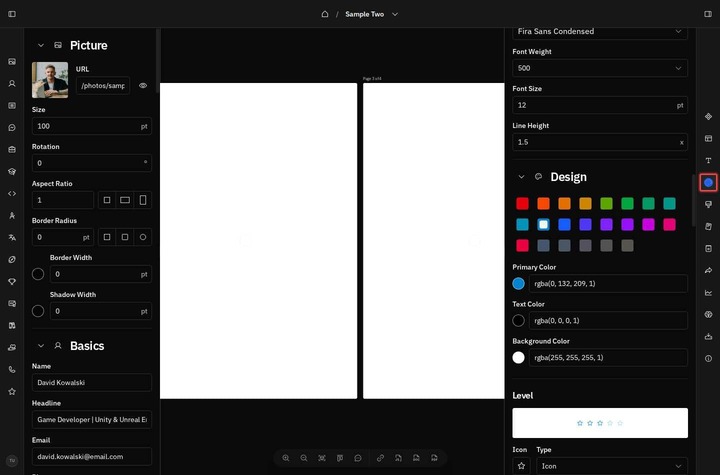}
        \caption{\texttt{Reactive Resume} (resume)}
    \end{subfigure}

    \caption{Example applications in the ProgramDistill corpus. Top row (a--d)
    shows OSWorld-derived apps. Bottom row (e--h) shows real-world open-source
    clones.}
    \label{fig:corpus}
\end{figure}

Each application is pinned to a single upstream commit. Its runtime image is
rebuilt from that commit and checked to reproduce the vendored source tree, so
mining, masking, repair, and replay all run against a byte-identical
application. Table~\ref{tab:corpus} lists each application with its source
repository, pinned commit, and numbers of atomic and cumulative tasks.
The number of mined tasks varies across applications, reflecting differences in
how many behaviors can be reliably discovered and replay-verified under each
application's state and interaction dynamics. Figure~\ref{fig:corpus} shows representative interfaces from both sources.

%

\begin{table}[htbp]
\centering
\small
\setlength{\tabcolsep}{4pt}
\caption{\textbf{The 26 applications in the ProgramDistill corpus.} Task counts are the atomic
and cumulative partial-application reconstruction tasks mined from each application.}
\label{tab:corpus}
\newlength{\corpusappcol}\newlength{\corpusrepocol}
\ifnum\workshopbuild=1\footnotesize
  \setlength{\corpusappcol}{0.30\linewidth}\setlength{\corpusrepocol}{0.325\linewidth}
\else
  \setlength{\corpusappcol}{0.33\linewidth}\setlength{\corpusrepocol}{0.355\linewidth}
\fi
\begin{tabular}{@{}p{\corpusappcol}p{\corpusrepocol}lrrr@{}}
\toprule
\textbf{Application} & \textbf{Source repository} & \textbf{Commit} &
\textbf{Atomic} & \textbf{Cum.} & \textbf{Total} \\
\midrule
\multicolumn{6}{@{}l}{\textit{OSWorld-derived applications}}\\
\addlinespace[2pt]
AWS Console & \texttt{Task-Web/awsconsole\_web} & \texttt{794c1cc1} & 155 & 42 & 197 \\
BaseSite & \texttt{Task-Web/basesite} & \texttt{4c06e3d7} & 147 & 70 & 217 \\
BudgetWise & \texttt{Task-Web/budgetwise\_web} & \texttt{bf3903f6} & 198 & 92 & 290 \\
Calendar & \texttt{Task-Web/calendar\_web} & \texttt{6a7a8d1d} & 136 & 45 & 181 \\
CareerLink & \texttt{Task-Web/careerlink\_web} & \texttt{33352a1c} & 197 & 70 & 267 \\
CloudCRM & \texttt{Task-Web/cloudcrm\_web} & \texttt{1c7a1653} & 53 & 3 & 56 \\
Eventix & \texttt{Task-Web/eventix\_web} & \texttt{f60c3463} & 20 & 3 & 23 \\
ExpenseFlow & \texttt{Task-Web/expenseflow\_web} & \texttt{487fd168} & 2 & 0 & 2 \\
FormCraft & \texttt{Task-Web/formcraft\_web} & \texttt{5c20ecdf} & 25 & 4 & 29 \\
Insurance & \texttt{Task-Web/insurance\_claim\_web} & \texttt{720120c7} & 92 & 49 & 141 \\
MailHub & \texttt{Task-Web/mailhub\_web} & \texttt{4603ec38} & 198 & 49 & 247 \\
ReviewSphere & \texttt{Task-Web/reviewsphere\_web} & \texttt{8e4f9706} & 69 & 21 & 90 \\
StreamView & \texttt{Task-Web/streamview\_web} & \texttt{ea965045} & 105 & 53 & 158 \\
TeamChat & \texttt{Task-Web/teamchat\_web} & \texttt{ed077b4d} & 161 & 68 & 229 \\
TravelHub & \texttt{Task-Web/travelhub\_ad\_web} & \texttt{b45dae8e} & 34 & 11 & 45 \\
VaultBank & \texttt{Task-Web/vaultbank\_web} & \texttt{e9ebb7e4} & 131 & 33 & 164 \\
Visa Application & \texttt{Task-Web/visaapplication\_web} & \texttt{c0e14ed0} & 52 & 42 & 94 \\
Weights \& Biases & \texttt{Task-Web/wandb\_web} & \texttt{1c781220} & 126 & 52 & 178 \\
\addlinespace[3pt]
\midrule
\multicolumn{6}{@{}l}{\textit{Real-world open-source projects and SaaS clones}}\\
\addlinespace[2pt]
Actual Budget & \texttt{actualbudget/actual} & \texttt{e0880f39} & 36 & 18 & 54 \\
Reactive Resume & \texttt{amruthpillai/reactive-resume} & \texttt{dcf1b28c} & 246 & 126 & 372 \\
Baserow & \texttt{baserow/baserow} & \texttt{8d7c0111} & 207 & 105 & 312 \\
Trello (Knowankit) & \texttt{knowankit/trello-clone} & \texttt{19e9d64a} & 93 & 46 & 139 \\
Jira (Oldboyxx) & \texttt{oldboyxx/jira\_clone} & \texttt{26a9e77b} & 122 & 62 & 184 \\
Jira (Trungvose) & \texttt{trungvose/jira-clone-angular} & \texttt{20b50e9a} & 24 & 2 & 26 \\
Trello (Vdevired) & \texttt{vdevired/trello-clone} & \texttt{1a40f9e5} & 86 & 60 & 146 \\
Calendly & \texttt{WebDevSimplified/calendly-clone} & \texttt{730c5f19} & 147 & 75 & 222 \\
\midrule
\textbf{Total} & & & \textbf{2862} & \textbf{1201} & \textbf{4063} \\
\bottomrule
\end{tabular}
\end{table}

\ifnum\workshopbuild=1
\subsection{Mined Lineage Structure}
\label{app:construction}

Figure~\ref{fig:minepatch_summary} visualizes the mined prerequisite structure at two scales, showing a corpus-wide view across all 26 applications and a representative \texttt{StreamView} lineage reaching depth 12.

\begin{figure}[h!]
\centering
\includegraphics[width=\linewidth]{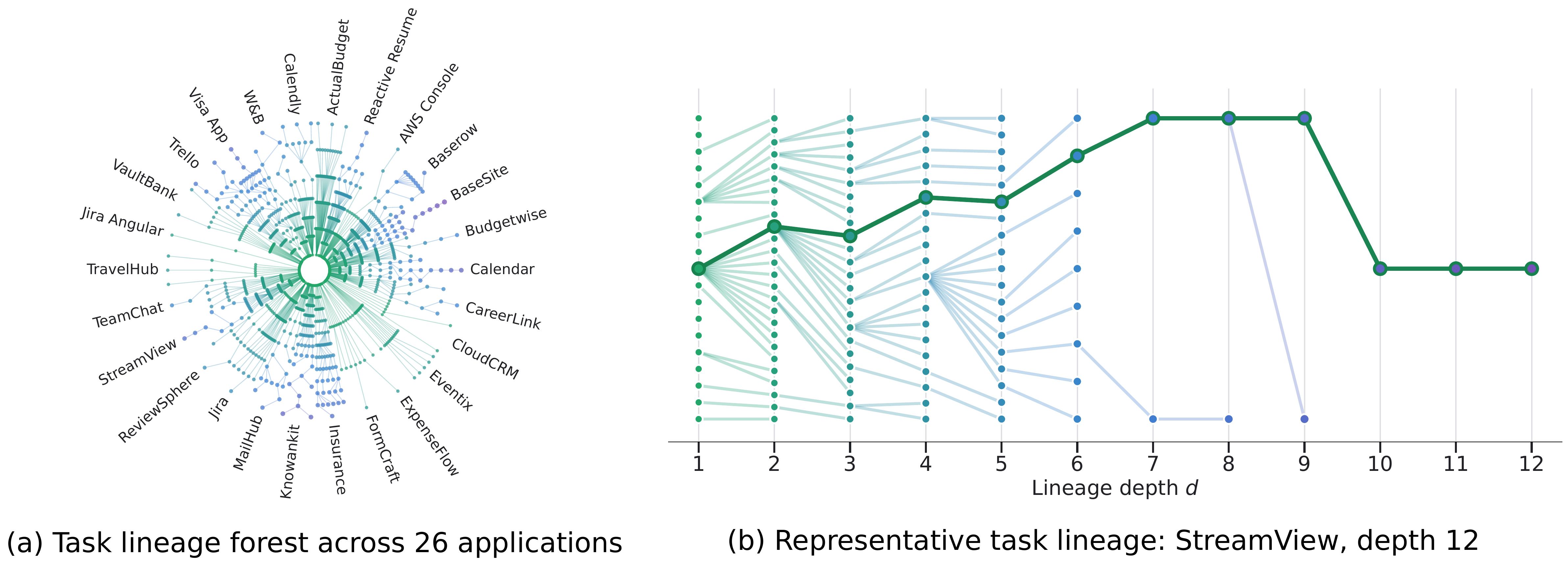}
\caption{\textbf{Prerequisite lineages across applications.}
(a) Mined lineage forests across all 26 applications.
(b) A representative \texttt{StreamView} lineage reaching depth 12, with one
root-to-leaf prerequisite chain highlighted.}
\label{fig:minepatch_summary}
\end{figure}
\fi

\subsection{Deterministic Execution}
\label{app:deterministic}

Reliable replay requires each trace to begin from the same application state and the same application time. Our \texttt{mine-craft-patch} pipeline resets the application before every collection and replay so that each trace starts from the same seeded state. To control time-dependent behavior, each trace also records the application time at which it was collected. On replay, the application clock is shifted so that execution starts from the same time while continuing to advance normally.

The system computes a constant offset between the recorded application time and the host clock. This offset is applied through \texttt{libfaketime} to the application server, interpreter, and database. The browser helper applies the same offset to the page's \texttt{Date} implementation, keeping the frontend and backend synchronized to the same application clock.

Because the offset remains constant, time continues to advance at the normal rate during execution. This preserves relative ordering, durations, and timeout semantics while keeping behavior that depends on the current date or time consistent across replays. For example, a booking interface that renders available dates relative to ``today'' presents the same dates during collection and subsequent replay.

\definecolor{fcframe}{HTML}{00841A}

\lstdefinestyle{fcjson}{
  basicstyle=\ttfamily\scriptsize,
  breaklines=true,
  breakatwhitespace=false,
  columns=fullflexible,
  showstringspaces=false,
  keepspaces=true,
}

\newpage
\section{Browser Helper Implementation Details}
\label{app:browser-helper}

ProgramDistill uses a shared browser helper across mining, replay verification,
and repair evaluation. The helper executes actions in a real Chromium instance
and returns structured observations of the resulting application state.
Table~\ref{tab:helper-actions} summarizes the browser actions supported by the
helper. Reliable replay additionally requires two properties: recorded actions
must refer to stable interaction targets, and observations must avoid transient
intermediate states. We describe these mechanisms below.
The solver-facing view of these observations is filtered as described
in Appendix~\ref{app:harness}.

\subsection{Action Vocabulary}

All stages use the same underlying browser helper, although the actions exposed
to each stage differ slightly. Common actions support observation, navigation,
and interaction with visible elements. Repair additionally supports joint
observation of the reference and editable applications, while mining exposes
actions used for trace construction. Table~\ref{tab:helper-actions} lists the
complete action vocabulary.

\begin{table}[h!]
\centering
\small
\caption{\textbf{Browser-helper action vocabulary.}
Common actions are available during both mining and repair.
Repair additionally exposes side-by-side observation of the reference and editable
applications, while mining provides several actions used for trace construction.}
\label{tab:helper-actions}

\newlength{\helpercmdcol}\newlength{\helperargcol}
\ifnum\workshopbuild=1\small
  \setlength{\helpercmdcol}{0.185\linewidth}\setlength{\helperargcol}{0.325\linewidth}
\else
  \setlength{\helpercmdcol}{0.17\linewidth}\setlength{\helperargcol}{0.27\linewidth}
\fi
\newlength{\helpertabwidth}
\ifnum\workshopbuild=1\setlength{\helpertabwidth}{0.97\linewidth}\else\setlength{\helpertabwidth}{\linewidth}\fi
\begin{tabularx}{\helpertabwidth}{@{}p{\helpercmdcol} p{\helperargcol} Y@{}}
\toprule
\textbf{Command} & \textbf{Arguments} & \textbf{Effect} \\
\midrule

\multicolumn{3}{@{}l}{\textit{Common actions}}\\
\addlinespace[2pt]

\texttt{observe}
& \texttt{page}
& Return a settled observation containing the current URL, visible text, and
interactable elements. \\

\texttt{click}
& \texttt{page}, \texttt{elementId}
& Click an element from the latest observation. \\

\texttt{type}
& \texttt{page}, \texttt{elementId}, \texttt{text}
& Enter text into an interactable field. \\

\texttt{hover}
& \texttt{page}, \texttt{elementId}
& Hover over an element. \\

\texttt{focus}
& \texttt{page}, \texttt{elementId}\,$\vert$\,\texttt{selector}
& Move browser focus to a target element. \\

\texttt{press}
& \texttt{page}, \texttt{key}
& Send a keyboard input to the page. \\

\texttt{scroll}
& \texttt{page}, \texttt{direction}, [\texttt{elementId}]
& Scroll the page, or a specified scrollable element, upward or downward. \\

\texttt{wait}
& \texttt{page}, \texttt{ms}
& Wait for a bounded interval and return the resulting settled observation. \\

\texttt{drag}
& \texttt{page}, \texttt{elementId}+\texttt{targetElementId}
& Drag one observed element to another and record stable selectors for both
endpoints. \\

\texttt{back} / \texttt{forward}
& \texttt{page}
& Navigate backward or forward through browser history. \\

\texttt{upload}
& \texttt{page}, \texttt{elementId}, file
& Upload a file through an observed file input. During mining this may reuse an
artifact produced by a prerequisite trace. \\

\midrule
\multicolumn{3}{@{}l}{\textit{Repair-only actions}}\\
\addlinespace[2pt]

\texttt{observe-both}
& none
& Observe the editable and reference applications together for behavioral
comparison. \\

\texttt{reset-reference}
& none
& Restore the clean seeded reference state after exploratory mutations. \\

\midrule
\multicolumn{3}{@{}l}{\textit{Mining-only actions}}\\
\addlinespace[2pt]

\texttt{navigate}
& \texttt{path}
& Load a specified in-application path. \\

\texttt{waitForSelector}
& \texttt{selector}
& Wait until a specified element appears before continuing trace collection. \\

\bottomrule
\end{tabularx}
\end{table}

\subsection{Replay-Stable Element Addressing}

Frontend-generated DOM identifiers and class names can change across
application builds, making them unreliable for replay. ProgramDistill therefore
separates the temporary identifiers used during live interaction from the
selectors stored in recorded traces.

Each observation assigns a transient numeric \texttt{elementId} to every visible
interactable element and exposes browser-observable attributes such as
accessibility roles and names, labels, placeholders, visible text, links, and
test IDs. For each element, the helper attempts to construct a selector from
these attributes. Actions are issued using \texttt{elementId}, but the recorded
trace stores the corresponding selector instead.

\begin{figure}[t!]
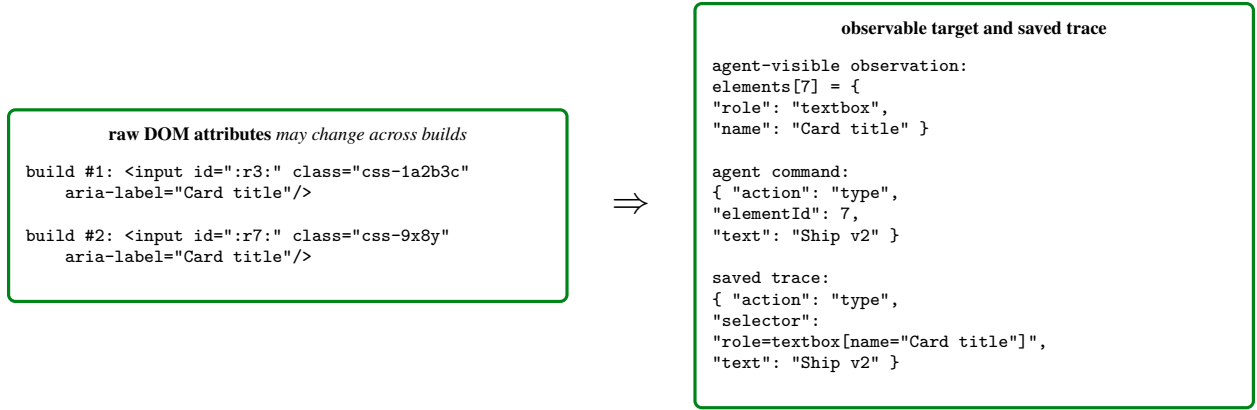

\centering

\begin{minipage}[c]{0.45\linewidth}
\begin{tcolorbox}[
enhanced,
colback=white,
colframe=fcframe,
boxrule=1.2pt,
arc=2pt,
left=3pt,
right=3pt,
top=3pt,
bottom=3pt
]
\centering\scriptsize
\textbf{raw DOM attributes}\
\emph{may change across builds}

\begin{lstlisting}[style=fcjson]
build #1: <input id=":r3:" class="css-1a2b3c"
    aria-label="Card title"/>

build #2: <input id=":r7:" class="css-9x8y"
    aria-label="Card title"/>
\end{lstlisting}
\end{tcolorbox}
\end{minipage}
\hfill
\begin{minipage}[c]{0.06\linewidth}
\centering
\Large$\Rightarrow$
\end{minipage}
\hfill
\begin{minipage}[c]{0.45\linewidth}
\begin{tcolorbox}[
enhanced,
colback=white,
colframe=fcframe,
boxrule=1.2pt,
arc=2pt,
left=3pt,
right=3pt,
top=3pt,
bottom=3pt
]
\centering\scriptsize
\textbf{observable target and saved trace}

\begin{lstlisting}[style=fcjson]
agent-visible observation:
elements[7] = {
"role": "textbox",
"name": "Card title" }

agent command:
{ "action": "type",
"elementId": 7,
"text": "Ship v2" }

saved trace:
{ "action": "type",
"selector":
"role=textbox[name="Card title"]",
"text": "Ship v2" }
\end{lstlisting}
\end{tcolorbox}
\end{minipage}

\caption{\textbf{Replay-stable element addressing.}
Recorded actions use browser-observable descriptors rather than
frontend-generated DOM identifiers, which may change across builds.}
\label{fig:symbolic-element-addressing}
\end{figure}

As illustrated in Figure~\ref{fig:symbolic-element-addressing}, this allows
recorded actions to locate the same logical element even when generated DOM
attributes change across builds. If the observable attributes do not uniquely
identify an element, no selector is produced and actions targeting that element
are refused. Thus, every element-targeted action retained in a trace has a
selector that can be resolved during replay.

\subsection{Settled Observations}

Browser actions can trigger asynchronous updates, so recording state
immediately after an action may capture a transient intermediate state.
ProgramDistill therefore waits for browser activity to settle before recording
each observation.

After an action, the helper applies a short delay and monitors both network
activity and DOM mutations. By default, an observation is considered settled
when no tracked requests remain in flight and the DOM has remained unchanged
for 400,ms. The procedure is capped at 10,s to avoid blocking indefinitely on
applications with continuous updates. DOM mutations include changes to
elements, attributes, and text, allowing the helper to detect updates caused by
timers, WebSockets, and server-sent events.

\begin{figure}[htbp]
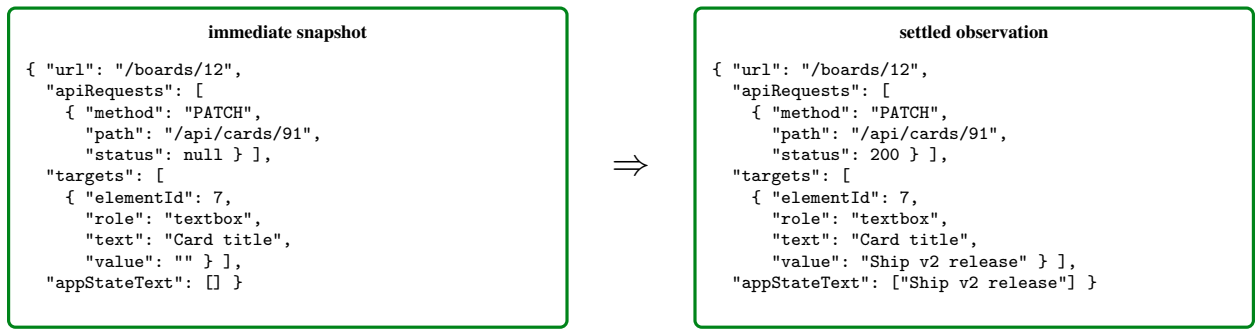

\centering

\begin{minipage}[c]{0.45\linewidth}
\begin{tcolorbox}[
  enhanced,
  colback=white,
  colframe=fcframe,
  boxrule=1.2pt,
  arc=2pt,
  left=3pt,
  right=3pt,
  top=3pt,
  bottom=3pt
]
\centering\scriptsize
\textbf{immediate snapshot}

\begin{lstlisting}[style=fcjson]
{ "url": "/boards/12",
  "apiRequests": [
    { "method": "PATCH",
      "path": "/api/cards/91",
      "status": null } ],
  "targets": [
    { "elementId": 7,
      "role": "textbox",
      "text": "Card title",
      "value": "" } ],
  "appStateText": [] }
\end{lstlisting}
\end{tcolorbox}
\end{minipage}
\hfill
\begin{minipage}[c]{0.06\linewidth}
\centering
\Large$\Rightarrow$
\end{minipage}
\hfill
\begin{minipage}[c]{0.45\linewidth}
\begin{tcolorbox}[
  enhanced,
  colback=white,
  colframe=fcframe,
  boxrule=1.2pt,
  arc=2pt,
  left=3pt,
  right=3pt,
  top=3pt,
  bottom=3pt
]
\centering\scriptsize
\textbf{settled observation}

\begin{lstlisting}[style=fcjson]
{ "url": "/boards/12",
  "apiRequests": [
    { "method": "PATCH",
      "path": "/api/cards/91",
      "status": 200 } ],
  "targets": [
    { "elementId": 7,
      "role": "textbox",
      "text": "Card title",
      "value": "Ship v2 release" } ],
  "appStateText": ["Ship v2 release"] }
\end{lstlisting}
\end{tcolorbox}
\end{minipage}

\caption{\textbf{Settled observations.} Observations are recorded after bounded
browser and network activity has stabilized, reducing the chance that traces encode
transient intermediate states.}
\label{fig:settled-observations}
\end{figure}

Once settled, the helper records the current URL, interactable elements,
application-state signals, observed API statuses, and an interaction
screenshot. Figure~\ref{fig:settled-observations} illustrates how this procedure
avoids recording transient states. Traces that fail to reach a settled
observation within the configured limit are treated as non-reproducible and
excluded from the verified trace bank.

Screenshots are retained for analysis but are not used uniformly as model
inputs. During mining, the Relabeler receives an animated GIF of the cleaned
trace together with the executed actions and outcome evidence. The repair and
full-application reconstruction agents, in contrast, receive only structured
browser observations. Replay verification likewise uses recorded actions and
expected-signal assertions rather than LLM-based visual judgment.

\newpage
\section{Agent Harness and Execution Settings}
\label{app:harness}

The agent harness builds on R2E-Gym~\citep{r2egym}, retaining its
coding-tool interface and adding live browser interaction through the
\texttt{browser} CLI. This section describes runtime boundaries, permitted
communication, observation handling, conversation-history management,
and execution settings. Details of
browser actions and element addressing are provided in
Appendix~\ref{app:browser-helper}, while the instructions for
partial-application and full-application reconstruction
appear in Appendix~\ref{app:solver-instruction}.

\input{archive/section/harness_architecture}

\subsection{Observations and Execution Settings}

\paragraph{Agent-visible information.}
The agent receives text-based tool outputs. These include shell
stdout and stderr, displayed file contents and editing feedback, search
results, and structured browser observations containing the URL, visible
text, element IDs, roles, accessible names, and input values. Full
reconstruction additionally exposes selectors for observed targets.
Screenshots, screenshot paths, raw HTML, API-request logs, and internal
artifact paths are excluded from agent-facing responses.

\paragraph{Recoverable tool observations.}
The harness preserves solver-visible tool outputs separately from the
model context under \texttt{/workspace/programdistill/observations/},
outside the submitted source directory.
Outputs up to 20,000 characters are shown directly, while longer outputs
are paginated into an initial 20,000-character segment followed by
15,000-character segments that can be read on demand.
This lets agents revisit earlier observations even after the corresponding
turns leave the model context, without repeating state-changing browser
actions such as clicks or form submissions.

\paragraph{Conversation-history management.}
When the retained history exceeds the token limit, the harness removes
older complete turns while preserving the task instructions, the first
assistant turn, and recent context. Tool calls are always kept together
with their corresponding results, preventing invalid partial histories.
The resulting message set is checked against the model's token limit
before each call.

\paragraph{Execution settings.}
Table~\ref{tab:harness-settings} summarizes the harness configuration for
partial-application and full-application reconstruction. The history cutoff determines when older
turns are removed, while the token stop threshold limits the size of the
active conversation rather than cumulative API token usage. None of our
runs reached the token stop threshold. The maximum agent-step limit was
reached only once, by Opus~5 in partial-application reconstruction.

\begin{table}[htbp]
\centering
\small
\caption{\textbf{Agent-harness execution settings.}}
\label{tab:harness-settings}
\begin{tabular}{@{}lrr@{}}
\toprule
\textbf{Setting} & \textbf{Partial} & \textbf{Full} \\
\midrule
Maximum agent steps & 1,000 & 10,000 \\
Retained-history cutoff (tokens) & 150,000 & 500,000 \\
Token stop threshold (tokens) & 1,000,000 & 5,000,000 \\
Reasoning effort & high & max \\
Per-tool timeout (seconds) & 90 & 90 \\
Aggregate solver time limit & None & None \\
\bottomrule
\end{tabular}
\end{table}

\newpage
\section{Mining Agent Prompts and Trace Representation}
\label{app:prompts}

This appendix provides the mining prompts and trace representation. It presents
the four mining-role prompts in execution order, the schema of an admitted
trace, and the evolution of a trace across mining stages. The Collector handles both
discovery and clean re-collection, with the latter using a guided prompt variant
and additional context describing the established parent state or a previously
successful route.

\subsection{Mining-Agent Orchestration}
\label{app:mining-orchestration}

\begin{figure*}[htbp]
   \centering

   \begin{minipage}[c]{0.46\textwidth}
       \centering
       \small

       \captionof{table}{Specialized sub-agents within the mining agent.}
       \label{tab:discovery-roles}

       \begin{tabularx}{\linewidth}{@{}lY@{}}
           \toprule
           \textbf{Sub-agent} & \textbf{Role} \\
           \midrule

           Planner
           & Proposes behavior goals and route hints grounded in
             application evidence. \\

           Collector
           & Pursues goals, records interactions, and selects
             observed success signals. \\

           Relabeler
           & Assigns a name and description to the behavior
             actually achieved. \\

           Reflector
           & Summarizes outcomes to guide later rounds. \\

           \bottomrule
       \end{tabularx}
   \end{minipage}
   \hfill
   \begin{minipage}[c]{0.50\textwidth}
       \centering

       \includegraphics[width=\linewidth]{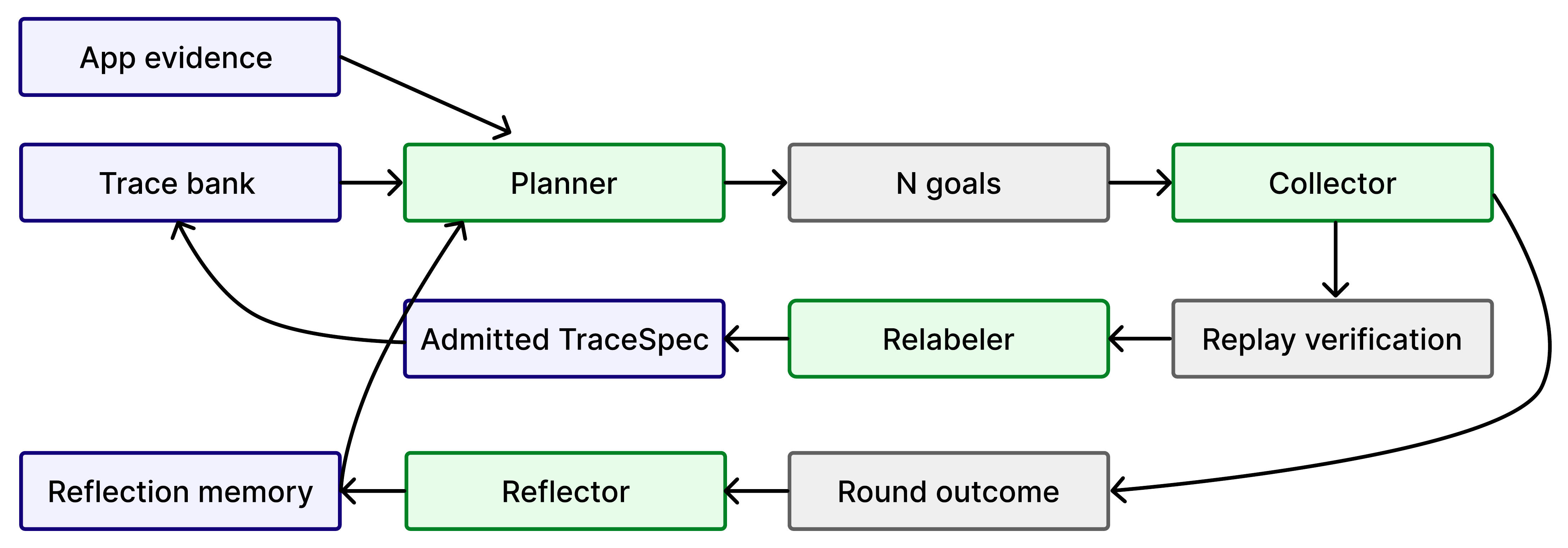}

       \vspace{4pt}

       \captionof{figure}{Mining-agent control flow.
       Each round proposes goals, collects and refines candidate
       traces, admits only replay-verified behaviors, and uses
       reflection to guide the next round.}
       \label{fig:mining-agent-flow-appendix}
   \end{minipage}
\end{figure*}

Mining runs in rounds over the current verified trace bank
\(\mathcal{B}^{(r)}\). Planners propose grounded goals and optional parent
traces. Collectors discover successful behaviors and then re-collect them
from a clean state, while the replay verifier rejects unstable traces.
Relabelers describe the behavior actually achieved, and the Reflector
summarizes coverage and failures to guide the next round. Accepted traces
form \(\mathcal{B}^{(r+1)}\). Figure~\ref{fig:mining-agent-flow-appendix}
and Table~\ref{tab:discovery-roles} show this control flow. The main text
uses the resulting verified traces and parent links as inputs to crafting.

\subsection{Planner}
\label{app:prompt-planner}
At the beginning of each exploration round, the Planner receives the goal
budget together with the current coverage, known failures, application
capabilities, available parent traces, and reflections from prior rounds.
It has read-only access to both the live application and the source code.

\begin{promptbox}{Planner system prompt}
\begin{lstlisting}[style=prompt]
You are TraceDiscoveryPlanner for ProgramDistill. Use 'browser_action' (observe/click only) plus the read-only 'search' and 'file_editor' (view only) tools to understand the running app and its source code, then call 'propose_goals' with a DIVERSE set of grounded discovery goals.

Evidence requirements for EACH goal:
- Ground the behavior and 'trace_hint' in source code you actually searched/read and UI/routes you actually observed; do not invent controls, transitions, or outcomes.
- Write 'trace_hint' as a concise natural-language route for the downstream browser agent, not selectors or action JSON.
- The exact source files you actually searched/read in 'trace_hint_code_paths'.
- The exact UI text, accessible names, titles, or URL paths you actually observed in 'trace_hint_observed_ui'. Unsupported evidence is rejected.

Requirements:
- Prefer uncovered CORE capabilities that directly accomplish the product's purpose. Also reserve goals for meaningful SECONDARY or source-discovered behaviors that add breadth or replace blocked, redundant, or non-verifiable core outcomes. The capability map is guidance, not an allowlist.
- Spread goals across distinct observed surfaces and varied workflow depths: focused state transitions, medium workflows, and grounded end-to-end workflows. Every goal must produce an independently valuable, browser-verifiable product outcome. Do not propose goals whose only novelty is opening a page, performing a routine UI gesture, toggling transient UI, or inspecting debug or implementation state. Prioritize state-changing, verifiable outcomes; do not pad goals. Split only at a persistent checkpoint that has standalone value and enables a deeper follow-up.
- Deprioritize login errors, minor validation paths, generic navigation, tiny gestures, and implementation details unless they are necessary prerequisites for a meaningful deeper workflow.
- Propose behaviors that are semantically distinct from every trace-ledger entry and from other goals in the same round. Treat workflows as equivalent when they exercise the same product capability, state transition, UI path, and verification pattern, even if they use different recipients, records, names, text values, or other inputs. Do not propose parameter-only variants, such as sending the same kind of message to a different person or repeating the same operation on another record, unless the changed entity causes materially different product logic, UI, or outcome.
- Do not re-propose the same or an equivalent failed workflow when its required state or control was unavailable, unless current source/UI evidence shows that the root cause has changed.
- Each goal must be independently attemptable in the browser either from a cold start or after replaying its declared parent chain.
- Actively use the deepest relevant 'parent_eligible=true' trace when its end-state enables the new behavior or avoids repeating meaningful setup. The full chain replays automatically, so describe only the novel behavior from the parent's end-state. Prefer extending that state into a deeper or later outcome; distribute goals across relevant parents where useful.
- Never use the goal itself, a descendant, an unavailable trace, or an unrelated trace as parent. Use a root goal only for genuine cold-start behavior that does not repeat available setup.
- After proposing goals, call 'finish'.
\end{lstlisting}
\end{promptbox}

Each proposed goal has the following structure.
\begin{lstlisting}[style=fcjson]
{
  "name": "...",
  "description": "...",
  "trace_hint": "...",
  "trace_hint_code_paths": ["..."],
  "trace_hint_observed_ui": ["..."],
  "novelty_reason": "...",
  "parent_trace": null
}
\end{lstlisting}

The purpose of the evidence fields is to make each proposed goal more faithful
to the source and UI evidence available to the Planner. They record the source
paths and interface values actually observed during that run, but are not otherwise used by the pipeline.

Each available parent-trace entry includes its ordered replay
\texttt{chain}, \texttt{depth}, and boolean indicators
\texttt{is\_leaf} and \texttt{deepest\_leaf}. The Planner also receives the
current number of traces and leaf chains, along with the maximum tree depth,
and is encouraged to extend a relevant leaf when possible.

\subsection{Collector}
\label{app:prompt-collector}
Each goal is handled by a Collector that interacts with the live application
using only live browser observations. As the Collector executes the goal, the
environment records every browser action it takes. After each observation, the
environment also provides a set of candidate success signals derived from the
observed application state, such as changes in the URL, visible text, element
state, or other observable outcomes. Once the Collector reaches a coherent
outcome, it selects the signals that best demonstrate that the behavior was
achieved. The recorded actions and selected signals together form the candidate
trace, which is subsequently replay-verified.

\begin{promptbox}{Collector system prompt}
\begin{lstlisting}[style=prompt]
You are TraceCollectorAgent for ProgramDistill. You receive one discovery goal and an optional natural-language route hint. Attempt the goal in the live web app through 'browser_action', recording the browser actions you actually execute. The hint is guidance, not an action script: work only from live observations. When a coherent outcome is achieved, select observation-derived evidence that proves it and save the trace.

Rules:
- Begin with 'browser_action' action=observe. Every browser action returns a settled observation; use it to choose the next step. If it conclusively proves success, use that response's 'signalCandidates', call 'save_trace' immediately, and do not observe again. Call observe separately only when the action response is genuinely inconclusive.
- Interact only with elements present in the latest observation, using its 'elementId' or an explicit selector grounded in an observed name, label, text, or test id.
- Use only credentials or input values supplied by the app or runtime. Never invent or brute-force them. If authentication blocks the goal and no value is available, finish with a short reason.
- Pursue the assigned goal methodically and follow the instance instructions about whether a partial but coherent outcome may be saved. Never claim the assigned goal succeeded when it did not.
- For export or download goals, treat a non-empty saved file in the observation's 'downloads' array as success; a headless browser may show no separate download UI.
- Select one or more signal candidates that materially prove the achieved outcome. When multiple meaningful signals are available, prefer signals from different evidence types. Exclude weak, unrelated, or duplicative signals, and never include a signal merely to increase the count. Never author or modify signal content. The selected set must collectively prove every material outcome you claim. A candidate with 'source=form_value' proves only what is currently in an editable control; it does not by itself prove that a save, update, filter, or submission took effect. Likewise, a generic heading or label does not by itself prove the claimed result. Prefer concrete rendered results or other outcome-specific evidence. After saving, call 'finish'.
- Stop immediately at the successful state. Do not log out, undo, navigate away, or perform cleanup afterward, because every subsequent browser action becomes part of the replay trace.
\end{lstlisting}
\end{promptbox}

\paragraph{Prerequisite-state context.}
Some goals depend on application state established by an existing parent trace.
Before the Collector begins, the system replays the validated parent trace and
its ancestors to establish that prerequisite state. The Collector is then told
to continue from the resulting state and record only the novel actions needed
for the current goal.   

\begin{promptbox}{Prerequisite-state block}
\begin{lstlisting}[style=prompt]
PARENT STATE ALREADY ESTABLISHED: before this Collector started, the
system replayed the validated parent trace [<parent_trace>] and all of
its ancestors. The browser is already at the parent's end-state. Do NOT
execute the parent actions a second time; begin from the current state
and record ONLY the novel actions that achieve this goal. The parent
chain is replayed automatically whenever this child trace is replayed
and is not duplicated inside the child's action trace.
\end{lstlisting}
\end{promptbox}

\paragraph{Guided clean re-collection.}
Initial discovery may produce long traces because the Collector explores
multiple interaction paths before reaching the goal. Once a behavior has been
successfully discovered, the earlier successful route is provided back to the
Collector as guidance for clean re-collection. Starting again from a clean
state, the Collector uses live observations to reach the same goal through a
more direct path while omitting unnecessary exploratory steps. The following
block provides the earlier successful route as a hint rather than as an action
sequence to reproduce exactly.

\begin{promptbox}{Guided re-run block}
\begin{lstlisting}[style=prompt]
GUIDED RE-RUN: this goal has ALREADY been confirmed reachable. Below is the
earlier successful run's steps. Use them as route evidence, not as instructions
to reproduce in order. From the CURRENT clean start, use live observations to
take a direct route while preserving every step needed to reach and prove the
assigned outcome. Omit only steps whose removal does not affect that outcome.
'{initial_title_part_0}' in a route is symbolic: replace it with the first
segment of the CURRENT page title from the live initial state, never send the
braces literally.
Earlier successful route:
<discovered route>
\end{lstlisting}
\end{promptbox}

\subsection{Relabeler}
\label{app:prompt-relabeler}
Once a candidate passes replay verification, the Relabeler receives the executed
actions, success evidence, and final observation, but not the original goal. It
then assigns a name and description based only on the behavior actually
achieved. Hiding the original goal prevents the label from being biased toward
the Planner's intended outcome when the Collector reached something different.

\begin{promptbox}{Relabeler system prompt}
\begin{lstlisting}[style=prompt]
You are TraceRelabeler for ProgramDistill. You are given the browser actions an agent actually executed against a live web app, the machine-checkable success evidence it observed, and the full animated GIF rendered from every retained step in the cleaned replayable trace. Use the visual sequence to verify what was visibly accomplished and to avoid labels contradicted by the UI. Author a concise identity for THIS behavior, grounded ONLY in the actions and evidence shown. You are deliberately NOT told what the agent originally intended -- do not guess or restate an intention; describe only what the actions and evidence prove was accomplished.

Return STRICT JSON with exactly two keys and no other prose:
{"name": "<short lower_snake_case behavior id, e.g. create_board_and_add_card>", "description": "<one plain sentence stating what was accomplished>"}
Rules: name is lower_snake_case, specific, at most 6 words; description is a single sentence describing the concrete achieved behavior (the outcome, not raw UI mechanics). If the evidence shows only navigation or a trivial read with no state change, name it accordingly (e.g. view_settings_page).
\end{lstlisting}
\end{promptbox}

\subsection{Round Reflector}
\label{app:prompt-reflector}
The Reflector runs after all goals in an exploration round have been processed.
It reviews the application capabilities, the accumulated trace bank, the goals
and outcomes from the completed round, unresolved failures, and reflections from
earlier rounds. It then produces a short summary of the most important coverage
gaps, failure patterns, and opportunities for deeper exploration. This
reflection is appended to the \texttt{reflection\_memory} provided to the
Planner in the next round.

\begin{promptbox}{Round-reflection system prompt}
\begin{lstlisting}[style=prompt]
You are TraceDiscoveryRoundReflector for ProgramDistill. Review one completed discovery round using the app's curated product specification, the cumulative trace bank, the round's proposed goals and concrete outcomes, the recent proposal-to-actual outcome history, prior failures, and recent round reflections. Produce a short advisory reflection for the NEXT Planner. Identify meaningful coverage bias, neglected product surfaces, useful opportunities to go deeper through existing parent traces, repeated failure patterns, and whether discovery appears close to saturation. Explicitly flag recurring near-duplicate proposals even when their names differ by comparing product surface, state transition, parent state, and the actual relabeled outcome. Distinguish useful depth from redundant repetition. Judge coverage from actual outcomes rather than proposed names. Reassess earlier concerns against current evidence and do not repeat concerns that have already been resolved. Focus on the two to four most consequential gaps instead of inventorying every missing capability. Do not propose a numbered goal list, selectors, or action steps. Do not claim a surface exists unless supported by the supplied app specification or trace evidence. Do not blindly repeat stale guidance.

Return STRICT JSON with exactly one key and no other prose:
{"reflection": "<one concise paragraph, at most 180 words>"}
\end{lstlisting}
\end{promptbox}


\subsection{Admitted Trace Representation}
\label{app:trace-spec}

After clean re-collection and replay verification, each admitted behavior is
stored as a \texttt{TraceSpec}. The \texttt{TraceSpec} records the action
sequence from the clean run with replay-stable selectors, the
observation-grounded expected signals, and an optional reference to its
immediate parent trace. It also stores a \texttt{trace\_hint} summarizing the
latest successful route for use as guidance in subsequent collection. Replay
verification ignores this hint and depends only on the recorded actions, parent
structure, and expected signals.

\begin{tcolorbox}[
enhanced,
breakable,
colback=white,
colframe=fcframe,
colbacktitle=green!10,
coltitle=black,
boxrule=1.2pt,
arc=2pt,
left=3pt,
right=3pt,
top=3pt,
bottom=3pt,
fonttitle=\small,
title={An admitted child \texttt{TraceSpec}: renaming a board list (abridged)}
]
\begin{lstlisting}[style=fcjson]
{
  "name": "rename_done_list_to_completed",
  "candidate_id": "vdevired_trello_clone",
  "description": "Renamed the Done list on the Workflow Pipeline board to Completed.",
  "parent_trace": "create_board_with_ordered_lists",
  "action_trace": [
    {
      "action": "click",
      "selector": "role=button[name=\"Rename list Done\"]"
    },
    {
      "action": "type",
      "selector": "[name=\"title\"]",
      "text": "Completed"
    },
    {
      "action": "press",
      "selector": "[name=\"title\"]",
      "key": "Enter"
    }
  ],
  "expected_signals": {
    "all_accessible_names": [
      "Completed Rename list Completed"
    ],
    "reject_visible_text": [
      "Done"
    ]
  }
}
\end{lstlisting}
\vspace{2pt}
\footnotesize
The parent trace creates and opens the \texttt{Workflow Pipeline} board
with the \texttt{Done}, \texttt{Ideas}, and \texttt{In Progress} lists.
After reset and replay of the parent chain, the child performs three
actions to rename \texttt{Done} to \texttt{Completed}. Its success signals
check the renamed list's accessible name and the absence of the old
\texttt{Done} text. All three recorded actions are shown; the route hint,
empty signal fields, and redundant execution metadata are omitted.
\end{tcolorbox}

\subsection{Trace Evolution Across Mining Stages}
\label{app:trace-evolution}

Table~\ref{tab:trace-evolution} summarizes how a proposed behavior is refined
from an exploratory goal into a replay-verified trace.

\begin{table}[htbp]
\centering
\small
\caption{How a candidate trace evolves through the mining pipeline.}
\label{tab:trace-evolution}
\begin{tabularx}{\linewidth}{@{}p{0.16\linewidth}YY@{}}
\toprule
\textbf{Stage} & \textbf{Input} & \textbf{Output} \\
\midrule

Planning &
source and UI evidence, current coverage, known failures, available parent
traces, and prior reflections &
a proposed behavior goal with a grounded route hint and an optional
\texttt{parent\_trace} \\

Discovery &
the proposed goal and the live state established by its parent chain, if any &
an exploratory action sequence, selected success evidence, and a route hint
derived from the successful execution \\

Clean re-collection &
the previously successful route and a clean state with the parent chain replayed &
a more direct action sequence, newly selected success evidence, and a
\texttt{trace\_hint} regenerated from the clean execution \\

Replay verification &
the clean action sequence, parent structure, and expected signals &
a deterministic pass/fail result after reset and replay of the required parent
chain and target actions \\

Relabeling &
the replay-verified actions, success evidence, and final observation, without
the original proposed goal &
a name and description reflecting the behavior actually achieved \\

\bottomrule
\end{tabularx}
\end{table}


\newpage
\section{Crafting Agent Prompts}
\label{app:taskgen-prompts}

Crafting masks the source implementing a verified behavior using two LLM roles, the task generator and a stateless mask-depth critic.
The generator constructs the counterfactual failures by directly editing the source code, while the critic judges whether the resulting diff removes substantive implementation or only disables it superficially.
The system prompts used for the generator and critic are shown below.


\subsection{Task Generator}
\label{app:prompt-task-generator}
The system prompt is assembled from common masking rules and one scope-specific
clause. 
In each task, the generator additionally receives the app description, trace actions, expected signals, and, on a repair round, the concrete verification failure from the
previous attempt.

\begin{promptbox}{TaskGenerator system prompt (abridged)}
\begin{lstlisting}[style=prompt]
You are TaskGeneratorAgent for ProgramDistill. You build feature-restoration coding tasks from a verified golden behavior trace of a real OSS web app.

Your job: delete-mask every piece of source that implements that feature, so the feature is broken, while the application still builds and boots.

Masking rules:
- Remove the implementation rather than disabling the feature. Delete the substantive serializer, generator, query, render, handler, export, or state-update logic that produces the behavior observed in the trace. Restoration must require rewriting real logic.
- Reject shallow masks such as changing a feature flag, permission, configuration, environment value, enum, or allowlist; adding or editing a guard or early return; or disabling a call site while its implementation remains intact. If reverting one switch or line restores the feature, remove the deeper implementation instead.
- Keep the module compilable. Preserve imports, exports, declarations, signatures, and route registrations. Replace only substantive bodies with minimal inert stubs such as an error, 'return null', a 501 response, 'pass', an empty value, or a no-op handler.
- Mask every applicable feature-owned surface, including the frontend component, API endpoint or route, and service or model logic. Leaving any implementation path intact makes the task invalid.
- For full UI + logic restoration, remove the feature-owned rendered markup as well as its handlers. Replace forms, controls, sections, views, or lists owned by the feature with an inert placeholder while preserving the export and route so the app still boots. Leave unrelated layout, navigation, and surrounding UI intact.
- Do not touch unrelated features, shared utilities used broadly, auth/bootstrap needed just to reach the feature, or the build config.

Verification you are designing for (inverse of validate_instances): after your mask_patch is applied in a dev pod, the app still boots, but the golden trace that used to pass must now fail because the feature is gone.

Workflow: explore with search/read_file/list_dir using the trace's source hints and UI steps; delete the real implementation with edit_file (old_str must be unique); review with view_changes; then submit_task with a solver-facing problem_statement; then finish.

problem_statement rules: state only what the feature is -- its user-facing purpose and workflow and which surfaces it spans -- anchored to the feature's own name/description from the trace. Do not state the observable outcomes that prove it works (on-screen text, toast/validation/error messages, counts, generated identifiers/filenames, URLs, or request/response shapes): those are the verification contract, and the solver must discover them empirically by driving the working feature in the reference app through the browser helper. Do not list the masked files/paths, and never include function names, code, or a step-by-step fix. Scale length to the feature: one sentence for a single surface, a brief per-surface list for a multi-surface or asynchronous one.
\end{lstlisting}
\end{promptbox}

\subsection{Mask-Depth Critic}
\label{app:prompt-mask-critic}

Judging feature deletion from behavioral failure alone is insufficient, as a one-line feature-flag change can also make a trace fail by "turning-off" code segments. 
The critic therefore receives the feature description, success signals, solver-facing statement, and production-to-masked diff.
The critic fails the generated diff unless restoration requires rewriting substantive logic.

\begin{promptbox}{MaskDepthCritic system prompt}
\begin{lstlisting}[style=prompt]
You are MaskDepthCritic for ProgramDistill. A task-generator agent was asked to build a feature-restoration coding task by deleting the real implementation of one feature from an app's source, so a solver must rewrite that logic to restore the feature. You are given the feature description, its success signals, and the unified diff the generator produced (prod -> masked: what it removed/changed).

Your only job: decide whether the diff genuinely deletes the feature's implementation, or merely disables it shallowly.

A mask is deep (good) when the substantive logic that produces the feature's observable behavior is gone -- e.g. the serializer/query/render/export builder/state-update/handler body is removed and replaced by an inert stub that returns empty/placeholder data or throws. Restoring the feature would require writing real logic from scratch.

A mask is shallow (bad) when the implementation is left intact and the feature is only switched off, for example: flipping a boolean / permission / capability / feature flag ('return True'->'return False', 'canExport = false'); changing a config / env / enum / allowlist value; adding or editing an early-return, guard, or 'if' condition; or commenting out a single call site while the function it calls still exists. If a developer could restore the feature by reverting a single flag/guard/line without writing new logic, the mask is shallow.

UI features: the intended scope for this mask is full restoration (UI + logic). So a mask is also shallow if it guts only the event handlers / submit / API logic but leaves the feature's rendered markup (the form fields, buttons, view the user interacts with) fully intact. A deep mask must remove the feature's returned JSX/template markup as well as its logic (a still-rendered, still-interactive form with dead handlers is shallow).

Respond with only a JSON object, no prose, no code fences:
{"verdict": "deep" | "shallow", "reason": "<one or two sentences>", "restoration_requires_rewrite": true | false}
\end{lstlisting}
\end{promptbox}

\subsection{Cumulative Mask Merger}
\label{app:prompt-merger}

The cumulative mask merger is invoked when conflicts arise while applying the atomic masks to the source.
Conflicts occur when two atomic masks attempt to edit the same lines.
The merger starts from a partial merge: every non-conflicting mask is applied first, and the conflicting masks are listed for it to resolve.
For each one it is given the lines in conflict, the mask's own diff, its problem statement, and the masked source it produced on its own. 
It reads the source with the same read-only tools the task generator uses, edits the partial merge until every listed mask is applied, and submits one resolution per mask.
If the submission is rejected, the agent sees why and tries again.


\begin{promptbox}{Cumulative merger system prompt}
\begin{lstlisting}[style=prompt]
You are a conservative conflict merge agent for benchmark masking. Large certified inputs live in workspace artifacts: read only what you need. Resolve overlaps in the seeded source with the smallest coherent union of removals. Never invent implementation or alter unrelated code. Success requires submit_merge.
\end{lstlisting}
\end{promptbox}

\begin{promptbox}{Cumulative merger user message}
\begin{lstlisting}[style=prompt]
Resolve the cumulative mask conflict in the seeded in-memory source workspace. Read the artifact manifest first with read_file:
<manifest_path>

The real production baseline is available at the normal source paths. Those paths currently expose a branch-local seed that already preserves the newest and all non-conflicting masks; the manifest identifies the components missing from that seed. Start with priority_component_ids; components outside that set are already certified in the seed and do not need inspection unless validation feedback says otherwise. Inspect only the conflict ranges, authoritative patches, problem statements, and certified variants you need. Edit the normal source paths to add every missing masking intent without restoring any included component or changing unrelated code. Then call submit_merge with every expected component exactly once and one resolution per component. Do not paste complete source files into the response. If validation_feedback is present, correct it.
\end{lstlisting}
\end{promptbox}

\newpage
\section{Partial- and Full-Application Reconstruction Instructions}
\label{app:solver-instruction}

For both partial-application and full-application reconstruction, the agent
receives one system message and one user message at the start of each task. The
system message is fixed, while the user message is instantiated from a fixed
template. Its \texttt{\{problem\_statement\}} slot is replaced by the task
instruction, which combines the accepted problem statement with fixed runtime
guidance. The \texttt{\{tool\_call\_instruction\}} slot selects the single- or
parallel-tool-call variant used for the model under evaluation. Accordingly,
the first two listings below show the system and user messages, while the third
shows the task instruction inserted into the
\texttt{\{problem\_statement\}} slot of the user message.

\subsection{Partial-Application Reconstruction}
\label{app:instruction-repair}

\begin{promptbox}{Partial-application reconstruction system prompt}
\begin{lstlisting}[style=prompt]
You are a software repair agent restoring behavior intentionally removed from a full application repository. You have a masked current app and a live reference app that is the behavioral oracle. Use the reference empirically, make the correct source change, and verify the restored behavior in the current app.
\end{lstlisting}
\end{promptbox}

\begin{promptbox}{Partial-application reconstruction user message}
\begin{lstlisting}[style=prompt]
The full application repository is available at {repo_path}.

<programdistill_task>
{problem_statement}
</programdistill_task>

Restore the removed behavior in non-test application source files only. Do not
modify generated build output, runtime state, logs, browser artifacts,
dependencies, caches, or harness/tooling files.

Work in this order:
1. Explore the repository to locate the source that implements the feature, reading only the surrounding code needed to understand project conventions.
2. Before editing, exercise the feature in the reference app with the 'browser' CLI.
3. Compare the current app where useful, then implement a focused repair in the source you identified.
4. Drive the repaired workflow in the current app and compare its observable behavior with the reference.
5. When the repair is complete, submit it using the 'finish' tool.

Restore the task description and every relevant observable reference behavior
completely and accurately, while avoiding unrelated or out-of-scope changes.

{tool_call_instruction}
\end{lstlisting}
\end{promptbox}

\begin{promptbox}{Partial-application reconstruction task instruction (injected at \texttt{\{problem\_statement\}})}
\begin{lstlisting}[style=prompt]
You are working in the full <repository> repository at <repo_path>. The application is <app_description>.

Restore <feature_description>.

Source-change boundary: modify only application source files needed for this repair. Do not edit generated build output, runtime state, logs, browser artifacts, dependencies, caches, or harness/tooling files. Do not intentionally trigger or include those files as part of the solution. In this application, excluded path prefixes include: <excluded_path_prefixes>.

Use the surrounding source files to infer the full expected behavior and match the existing project conventions.

Reference behavior via the 'browser' command (IMPORTANT):
A live browser with TWO pages is available through 'execute_bash' using the 'browser' command:
- 'reference': the FULL, working app with the target feature intact. Use it as the ground-truth oracle for how the feature should look and behave. You can observe and interact with it, but you CANNOT read its source or shell -- only its rendered behavior.
- 'current': the app in THIS repo (the one you are editing), with the feature removed. Use it to reproduce the broken state and to confirm your fix restores the behavior.

Recommended workflow: (1) 'browser observe reference' to see the correct behavior and the exact UI/text/state it produces; (2) drive the reference through the feature's steps (click/type) to learn the full flow; (3) restore the implementation in the source; (4) 'browser observe current' and re-drive it to verify it now matches the reference. Always re-'observe' a page after every interaction -- element ids are assigned by the latest observation.

Commands (each is one execute_bash call):
- 'browser observe <current|reference>' -- render a page; returns visible text, url, title, and interactable elements with numeric ids.
- 'browser observe-both' -- observe both pages at once (side-by-side).
- 'browser click <page> <elementId>' -- click an element by its id from the latest observe.
- 'browser type <page> <elementId> <text>' -- type into a field.
- 'browser hover <page> <elementId>' / 'browser focus <page> <elementId>' / 'browser press <page> <key>' / 'browser scroll <page> <up|down>' / 'browser wait <page> [ms]' / 'browser back <page>' / 'browser forward <page>'.
- 'browser upload <page> <elementId> <localFile>' -- upload a local file through a file input from the latest observation.
- To exercise a drag-based feature (reordering a list, moving a kanban card), either focus the drag handle and use press (Space to lift, Arrow keys to move, Space to drop), or 'browser drag <page> <elementId> <targetElementId>' to drag one element onto another.
Example: 'browser observe reference', then 'browser click reference 12', then 'browser observe reference'.

Evaluation replays the recorded user workflow against the current app and checks whether the reference app's observable behavior has been restored completely and accurately.

Command tip: each action is executed through a timeout wrapper. For commands that change directories or use shell operators, wrap the command explicitly, e.g. 'bash -lc 'cd /workspace && ...''.
\end{lstlisting}
\end{promptbox}

\subsection{Full-Application Reconstruction}
\label{app:instruction-rebuild}

\begin{promptbox}{Full-application reconstruction system prompt}
\begin{lstlisting}[style=prompt]
You are implementing a complete web application in a minimal starter repository. Use the application specification and the live reference application to understand the expected behavior. Explore the reference as needed, implement the application in the current workspace, and compare the result with the reference to validate your work.
\end{lstlisting}
\end{promptbox}

\begin{promptbox}{Full-application reconstruction user message}
\begin{lstlisting}[style=prompt]
The blank application scaffold is available at {repo_path}. Build the complete
application there without attempting to recover hidden source, build output,
caches, installed artifacts, or harness and tooling files.

<programdistill_task>
{problem_statement}
</programdistill_task>

Implement the application in non-test application source files only. Do not
modify generated build output, runtime state, logs, browser artifacts,
dependencies, caches, or harness/tooling files.

Work in this order:
1. Systematically explore the reference app with the 'browser' CLI, including its routes, controls, state transitions, validation, and secondary functionality.
2. Implement the observed behavior in the blank scaffold.
3. Exercise the same workflows in the current app and compare their observable results with the reference.
4. Repeat exploration and implementation until the complete product specification and all additional discoverable behavior are covered.
5. When the reconstruction is complete, submit it using the 'finish' tool.

Reproduce every relevant observable reference behavior completely and
accurately while keeping harness and evaluator internals out of the
submission.

{tool_call_instruction}
\end{lstlisting}
\end{promptbox}

The task-instruction template uses placeholders for the repository path
and product specification. Tasks requiring sign-in additionally include a
reference-access block with seeded login instructions.

\begin{promptbox}{Full-application reconstruction task instruction (injected at \texttt{\{problem\_statement\}})}
\begin{lstlisting}[style=prompt]
Rebuild the complete application from the provided blank scaffold.

No current application implementation is available. Do not attempt to
recover it from caches, installed artifacts, build outputs, or harness
and tooling files.

## Browser environment

A live browser with two pages is available through the 'browser' command:

- 'reference': the complete, working original application. Use it as the
  ground-truth oracle for the application's observable information and behavior. You may
  observe and interact with it, but you cannot access its source or shell.
- 'current': the application you are implementing. Use it to exercise your
  implementation and compare it with the reference.

Reconstruct the application from the behavior observable through the
reference browser. Faithfully match every observable aspect, including but
not limited to exposed text and element structure, semantic identity (e.g., element type,
role, accessible name, and label), exposed element selectors, navigation,
validation, and resulting state.

## Evaluation and reproduction requirements

Evaluation replays user interactions using selectors and checks the application
state after each action. Similar text or element structure in observations is
not sufficient. Faithfully reproduce selectors and state changes under the same
conditions and action sequence.

Treat exposed selectors as part of the observable interface that must be
reproduced, not merely as reference information. In equivalent observable
states, implement corresponding controls on 'current' so that they expose
the same selectors as 'reference'.

Successful clicking or typing through an element ID does not establish selector
compatibility. A similar final result alone does not establish that intermediate
interactions and state transitions have been implemented identically.

Reproduce intermediate screens, menus and dialogs, input validation, success
and error messages, navigation, data changes, and saved results throughout
the same interaction sequence.

## Exploration, implementation, and comparison

Systematically explore the entire reference application before and during
implementation. Follow its routes and controls, exercise its interactions
and state transitions, and re-observe after every action.

Implement what you observe, then reproduce equivalent starting states and the
same interaction sequence on 'reference' and 'current'. Compare all of:

- Corresponding element types, roles, accessible names, labels, and exposed selectors.
- Screens, menus, dialogs, and navigation paths immediately after each action.
- Input validation and success or error responses.
- Created, updated, or deleted data and its effect on subsequent interactions.
- Saved results, downloads, and other side effects observed in the reference.

If a corresponding control on 'current' exposes a different selector or no
selector, inspect the preceding state, element type, role, label, DOM structure,
and other relevant details, then correct your implementation to match the
reference. If behavior or state changes differ, find and fix the cause and
repeat the same sequence.

Element IDs belong to the latest observation of each page. Do not reuse a
reference element ID on 'current' or reuse IDs from an earlier observation.
Re-observe after interactions and use the latest IDs for the relevant page.

Browser commands:
- 'browser observe <current|reference>'
- 'browser observe-both'
- 'browser recover <current|reference>' (reopen the app entry page without resetting application data)
- 'browser reset-reference' (restore the clean seeded reference state after exploratory mutations)
- 'browser click <page> <elementId>'
- 'browser type <page> <elementId> <text>'
- 'browser hover <page> <elementId>'
- 'browser focus <page> <elementId>'
- 'browser press <page> <key>'
- 'browser scroll <page> <up|down>'
- 'browser wait <page> [ms]'
- 'browser back <page>' / 'browser forward <page>'
- 'browser upload <page> <elementId> <localFile>'
- 'browser drag <page> <elementId> <targetElementId>'

## Product specification

Purpose: <app_purpose>.

Core capabilities:
- <core_capability_1>
- <core_capability_2>
- ...

Secondary capabilities:
- <secondary_capability_1>
- <secondary_capability_2>

Treat this product specification as required but non-exhaustive coverage.
Faithfully implement every item above, but do not limit the reconstruction
to those items. Discover and reproduce all additional functionality and
user-observable behavior available through the reference application.

## Implementation contract

- Work only inside '<repo_path>'.
- 'app.js' must export 'async function handle(req, res, context)'.
- Do not call 'listen()'; the task-owned runtime owns all ports.
- Keep application state in 'context.state' so the verifier can reset it.
- Use 'context.readBody()', 'context.json()', and 'context.html()' as needed.
- The runtime hot-reloads repository modules on every request.
- Faithfully implement the routes, labels, forms, selectors, state transitions,
  downloads, and validation behavior observed in the reference.
\end{lstlisting}
\end{promptbox}

\section{Model Pricing and Execution Time}
\label{app:costs}

\subsection{Model Pricing}
\label{app:pricing}

Table~\ref{tab:pricing} lists list price per model, in \$ per 1M tokens, read
from the LiteLLM \texttt{model\_cost} table under the same bare model
identifiers the agent harness uses (e.g. \texttt{claude-opus-5},
\texttt{gpt-5.6-sol}).

Figures~\ref{fig:eval-repair}(a) and~\ref{fig:eval-repair-chain}(a)
report mean cost per trajectory in USD.
For each trajectory, we multiply the number of uncached input, cached input,
and output tokens by their respective per-token prices from LiteLLM,
then sum the three amounts.

\begin{table}[htbp]
\centering
\small
\caption{\textbf{List price per model.} Input and output price in \$ per 1M
tokens, from LiteLLM's \texttt{model\_cost} table.}
\label{tab:pricing}
\begin{tabular}{lcc}
\toprule
\textbf{Model} & \textbf{Input (\$/1M)} & \textbf{Output (\$/1M)} \\
\midrule
GPT-6 Astra            & 10.00 & 50.00 \\
Claude Opus 5           & 5.00 & 25.00 \\
GPT-5.6 Sol             & 4.00 & 20.00 \\
GPT-5.3 Codex           & 1.75 & 14.00 \\
Gemini 3.1 Pro Preview  & 2.00 & 12.00 \\
Claude Sonnet 5         & 2.00 & 10.00 \\
Grok 4.6                & 2.00 & 6.00  \\
Gemini 3.6 Flash        & 0.75 & 3.75  \\
Gemini 3.7 Flash        & 0.75 & 3.75  \\
\bottomrule
\end{tabular}
\end{table}

\subsection{Execution Time}
\label{app:runtime}

To give a concrete sense of the time required to run the benchmark,
we report one representative evaluation run.
GPT-6 Astra at high reasoning effort completed all 300 tasks in
approximately 557 minutes of wall-clock time (9 hours 17 minutes)
using four CPU servers.
Each server had two Intel Xeon Gold 6238R processors at 2.20\,GHz
(56 physical cores, 112 logical CPUs) and ran up to eight trials
concurrently, for a maximum of 32 concurrent trials.
Each trial was allocated six logical CPUs, split evenly between the
current and reference applications, with verification reusing the
current application's allocation.
The servers hosted the applications, browsers, and verification
processes, while model inference was served through a remote API.

Wall-clock time is measured from the earliest trial start to the final
trial completion and includes per-trial setup, agent execution,
verification, and API waiting time, but excludes one-time environment
provisioning.
Individual trials took 48.2 minutes on average, with a median of
43.1 minutes.
This example is intended to provide a practical estimate of benchmark
runtime rather than a measure of standalone model-inference speed.

\newpage
\section{More Experimental Results}
\label{app:more-results}

\subsection{Results by Mask Scope}
\label{app:mask-scope}

Every task in ProgramDistill-300 carries one of two mask scopes. A logic-only
task removes the behavior behind an interface that remains unchanged, so the
agent restores handlers, endpoints, queries, and state updates behind existing
controls. A logic-and-UI task also removes the UI owned by the feature, requiring
the agent to reconstruct both the observed interface and its underlying behavior.
Table~\ref{tab:mask-scope} compares performance across these two settings using
binary and chain scores.

Astra achieves \RepairAstraScopeBinaryLogic\% binary success on
logic-only tasks and \RepairAstraScopeBinaryUI\% on logic-and-UI tasks,
a \RepairAstraScopeBinaryGap{}-point gap.
The chain score gives partial credit for behaviors recovered before the
first failure in a cumulative task.
Under this metric, Astra achieves \RepairAstraScopeChainLogic\% on
logic-only tasks and \RepairAstraScopeChainUI\% on logic-and-UI tasks,
a \RepairAstraScopeChainGap{}-point gap.
The scope comparison therefore captures both complete task success and
the extent of partial restoration.

\begin{table}[htbp]
\centering
\small
\caption{\textbf{Performance by mask scope on ProgramDistill-300.}
Logic-only tasks mask the behavior behind an unchanged interface,
whereas logic-and-UI tasks also remove the UI owned by the feature.
The suite contains 140 logic-only and 160 logic-and-UI tasks.
$\Delta$ denotes the difference between logic-only and logic-and-UI
performance.}
\label{tab:mask-scope}

\resizebox{\linewidth}{!}{%
\begin{tabular}{@{}lrrrr@{\hspace{14pt}}rrrr@{}}
\toprule
& \multicolumn{4}{c}{\textbf{Binary score}}
& \multicolumn{4}{c}{\textbf{Chain score}} \\
\cmidrule(lr){2-5}
\cmidrule(l){6-9}

\textbf{Model}
& \textbf{Overall}
& \textbf{Logic-only}
& \textbf{Logic-and-UI}
& $\boldsymbol{\Delta}$
& \textbf{Overall}
& \textbf{Logic-only}
& \textbf{Logic-and-UI}
& $\boldsymbol{\Delta}$ \\
\midrule

GPT-6 Astra
& \RepairAstraBinary & \RepairAstraScopeBinaryLogic
& \RepairAstraScopeBinaryUI & \RepairAstraScopeBinaryGap
& \RepairAstraChain & \RepairAstraScopeChainLogic
& \RepairAstraScopeChainUI & \RepairAstraScopeChainGap \\

Claude Opus 5
& 68.7 & 80.0 & 58.8 & 21.2
& 75.2 & 85.7 & 66.0 & 19.7 \\

GPT-5.6 Sol
& 60.7 & 76.4 & 46.9 & 29.6
& 68.2 & 81.5 & 56.5 & 25.0 \\

Grok 4.6
& 48.3 & 64.3 & 34.4 & 29.9
& 57.1 & 72.7 & 43.5 & 29.2 \\

Claude Sonnet 5
& 47.3 & 65.0 & 31.9 & 33.1
& 57.5 & 73.9 & 43.1 & 30.8 \\

GPT-5.3 Codex
& 45.7 & 59.6 & 33.4 & 26.2
& 53.7 & 67.4 & 41.7 & 25.7 \\

Gemini 3.7 Flash
& 45.3 & 61.4 & 31.2 & 30.2
& 53.2 & 69.5 & 38.9 & 30.6 \\

Gemini 3.6 Flash
& 29.7 & 37.9 & 22.5 & 15.4
& 39.6 & 49.6 & 30.9 & 18.7 \\

Gemini 3.1 Pro Preview
& 22.3 & 31.1 & 14.7 & 16.4
& 30.5 & 41.6 & 20.8 & 20.8 \\

\bottomrule
\end{tabular}}
\end{table}

\subsection{Results Under the Chain Score}
\label{app:chain-score}

The binary score gives credit only when all targets in a cumulative task
pass, while the chain score credits the fraction recovered before the first
failure. We compare these scores for the two evaluation settings.

\begin{figure}[htbp]
\centering
\includegraphics[width=\linewidth]{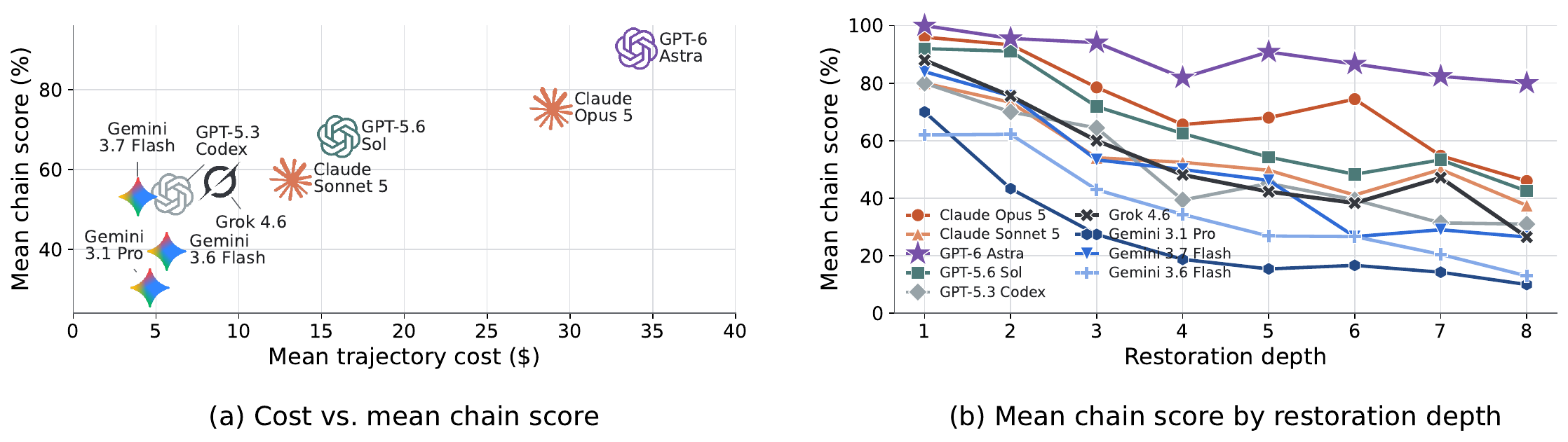}
\caption{\textbf{Partial-application reconstruction performance under the chain
score.} The chain-score counterpart of Figure~\ref{fig:eval-repair}.
Results cover all 300 tasks in \texttt{ProgramDistill-300}, including both
atomic and cumulative tasks.
\textbf{(a)} Mean chain score versus mean trajectory cost (USD),
using the same cost aggregation as Figure~\ref{fig:eval-repair}(a).
\textbf{(b)} Mean chain score by restoration depth $r_L$. The decline with depth is shallower than under the binary score because
partial restoration still earns credit, but the overall depth effect remains.}
\label{fig:eval-repair-chain}
\end{figure}

\paragraph{Partial-Application Reconstruction.}
For the setting in Section~\ref{sec:exp-repair},
Figure~\ref{fig:eval-repair-chain} is the chain-score counterpart of
Figure~\ref{fig:eval-repair}, covering all 300 tasks in
\texttt{ProgramDistill-300}. Table~\ref{tab:partial-chain} compares binary and
chain scores on the 250 cumulative repair tasks.
Astra has the highest scores under both metrics:
\RepairAstraCumulativeBinary\% binary and \RepairAstraCumulativeChain\%
chain score on cumulative tasks.
The gap reflects cases where part of the restoration lineage is recovered
without a complete task pass.
Opus~5 increases from 63.2\% to 71.1\%, and Sol from 54.4\% to 63.4\%.

\begin{table}[t!]
\centering
\small
\caption{\textbf{Partial-application reconstruction under binary and chain scores.}
Mean scores (\%) over the 250 cumulative tasks of \texttt{ProgramDistill-300}.}
\label{tab:partial-chain}
\begin{tabular}{lrr}
\toprule
\textbf{Model} & \textbf{Binary score} & \textbf{Chain score} \\
\midrule
GPT-6 Astra & \RepairAstraCumulativeBinary
& \RepairAstraCumulativeChain \\
Claude Opus 5         & 63.2 & 71.1 \\
GPT-5.6 Sol           & 54.4 & 63.4 \\
Claude Sonnet 5       & 40.8 & 53.0 \\
Grok 4.6              & 40.4 & 50.9 \\
GPT-5.3 Codex         & 38.8 & 48.4 \\
Gemini 3.7 Flash      & 37.6 & 47.0 \\
Gemini 3.6 Flash      & 23.2 & 35.2 \\
Gemini 3.1 Pro Preview & 12.8 & 22.6 \\
\bottomrule
\end{tabular}
\end{table}

\paragraph{Full-Application Reconstruction.}
For the setting in Section~\ref{sec:full-app},
Table~\ref{tab:full-chain-apps} reports chain scores in the same model and
application order as Table~\ref{tab:full-app-reconstruction}.
Mean binary/chain scores are
\CanonicalAstraBinaryWeighted/\CanonicalAstraChainWeighted\% for Astra,
\CanonicalOpusBinaryWeighted/\CanonicalOpusChainWeighted\% for Opus~5,
and \CanonicalSolBinaryWeighted/\CanonicalSolChainWeighted\% for Sol.
The larger gaps between the scores indicate that agents recover workflow
prefixes more often than complete workflows.

\begin{table*}[htbp]
\centering
\caption{\textbf{Full-application reconstruction under the chain score (\%).}
The chain-score counterpart of Table~\ref{tab:full-app-reconstruction}.
The Mean column averages all evaluated cumulative workflows, and bold
indicates the best result among the three models.}
\label{tab:full-chain-apps}
\begingroup
\setlength{\tabcolsep}{3pt}
\newcommand{\appheader}[1]{%
  \textbf{\begin{tabular}[c]{@{}c@{}}#1\end{tabular}}}
\resizebox{\textwidth}{!}{%
\begin{tabular}{lccccccccccccc}
\toprule
\appheader{Model} & \appheader{Reactive\\Resume} & \appheader{AWS\\Console} & \appheader{Baserow} & \appheader{BudgetWise} & \appheader{CareerLink} & \appheader{MailHub} & \appheader{Jira\\(Oldboyxx)} & \appheader{Trello\\(Knowankit)} & \appheader{StreamView} & \appheader{TeamChat} & \appheader{Trello\\(Vdevired)} & \appheader{Weights \&\\Biases} & \appheader{Mean} \\
\midrule
GPT-6 Astra (max) & \textbf{68.5} & \textbf{45.8} & \textbf{36.7} & \textbf{56.0} & \textbf{65.3} & 89.0 & 85.6 & 73.2 & \textbf{91.4} & \textbf{47.0} & \textbf{44.3} & \textbf{80.9} & \textbf{64.58} \\
Opus 5 (max) & 43.7 & 35.4 & 35.3 & 39.8 & 41.4 & \textbf{93.9} & \textbf{93.3} & 73.9 & 28.0 & 42.9 & 29.6 & 35.3 & 47.42 \\
GPT-5.6 Sol (max) & 16.8 & 8.3 & 29.0 & 21.5 & 12.4 & 82.0 & 87.5 & \textbf{76.7} & 32.2 & 35.6 & 42.6 & 9.1 & 34.22 \\
\bottomrule
\end{tabular}}
\endgroup
\end{table*}

\subsection{Does Context Length Explain the Depth Effect?}
\label{app:depth-context}
\input{archive/analysis/generated/repair-context-macros.tex}

Deeper tasks generally expose agents to longer contexts, raising an
alternative explanation for the performance decline in
Section~\ref{sec:exp-repair}. Models may perform worse simply because
they operate over longer contexts rather than because they must compose
more dependent repairs. We examine this possibility in three ways using
mean prompt size per step as a measure of context exposure across
trajectories from all nine models.

First, the growth in context length does not closely track the decline
in performance. The median prompt size per step increases from
\RepairContextPromptDepthOne{}k tokens at restoration depth~1 to
\RepairContextPromptDepthEight{}k at depth~8, a factor of
\RepairContextPromptRatio{}, while the mean binary score falls from
\RepairContextScoreDepthOne\% to \RepairContextScoreDepthEight\%.
Most of the increase in context length occurs by depth~4. Beyond that
point, context length changes little while performance continues to
decline.

\begin{table}[h!]
\centering
\small
\caption{\textbf{Mean binary score by restoration depth among runs with similar
context lengths.} Runs are grouped by their mean prompt size per step, and
columns indicate restoration depth. This diagnostic uses the task outcomes
paired with the analyzed trajectories.}
\label{tab:depth-vs-context}
\begin{tabular}{@{}lrrrrrrrr@{}}
\toprule
& \multicolumn{8}{c}{\textbf{Restoration depth}} \\
\cmidrule(lr){2-9}
\textbf{Mean prompt size / step}
& \textbf{1}
& \textbf{2}
& \textbf{3}
& \textbf{4}
& \textbf{5}
& \textbf{6}
& \textbf{7}
& \textbf{8} \\
\midrule

40--80k & 82.6 & 77.1 & 65.0 & 51.8 & 48.1 & 46.4 & 45.3 & 24.2 \\
80--120k & 77.0 & 58.9 & 43.4 & 39.5 & 32.5 & 30.1 & 30.0 & 22.5 \\
120--200k & 54.5 & 68.8 & 33.9 & 33.3 & 30.0 & 25.4 & 23.2 & 23.1 \\

\bottomrule
\end{tabular}
\end{table}

Second, performance still declines with restoration depth among runs
with similar context lengths. Within each context band in
Table~\ref{tab:depth-vs-context}, binary score shows an overall downward
trend as depth increases. From depth~1 to depth~8, mean binary score
drops by \RepairContextLowBandDrop{} points in the 40--80k band,
\RepairContextMidBandDrop{} points in the 80--120k band, and
\RepairContextHighBandDrop{} points in the 120--200k band.
Thus, deeper tasks remain substantially harder even at comparable
context lengths.

Third, within each restoration depth, prompt size has only a modest
negative Pearson correlation with binary score, ranging from
$\RepairContextCorrelationMin$ to $\RepairContextCorrelationMax$.
After centering both prompt size and binary score within each task to
account for task differences, the pooled correlation is weakly positive
($\RepairContextWithinTaskCorrelation$).

Together, these analyses show that the depth-related performance decline
cannot be explained by longer contexts alone and are consistent with
the increasing difficulty of composing multiple dependent repairs.

\subsection{Agent Interaction Statistics}
\label{app:interaction-stats}

\paragraph{Partial-Application Reconstruction.}
Table~\ref{tab:behavior} provides the full interaction statistics supporting
Figure~\ref{fig:model-behavior} in Section~\ref{sec:analysis}: the minimum,
mean, and maximum counts of agent steps, observation steps, and edit/write
steps for each model. The lower panel groups the same trajectories by
restoration depth and reports mean counts across all nine models.

\begin{table*}[h!]
\centering
\small
\setlength{\tabcolsep}{3.5pt}

\caption{\textbf{Agent interaction behavior.}
The upper panel reports min, mean, and max interaction statistics for each
model on \texttt{ProgramDistill-300}; the lower panel reports per-trajectory
means by restoration depth, pooling all nine models.
An observation step contains a request for application state through an
\texttt{observe} command, a state-returning browser interaction, or a
supported direct browser or API check of the current app. Each agent step
counts at most once, even when it targets both applications.}
\label{tab:behavior}

\resizebox{\linewidth}{!}{%
\begin{tabular}{
    >{\raggedright\arraybackslash}p{3.25cm}
    S[table-format=4.2, table-column-width=1.1cm]
    S[table-format=4.2, table-column-width=1.3cm]
    S[table-format=4.2, table-column-width=1.2cm]
    @{\hspace{7pt}}
    S[table-format=4.2, table-column-width=1.1cm]
    S[table-format=4.2, table-column-width=1.3cm]
    S[table-format=4.2, table-column-width=1.2cm]
    @{\hspace{7pt}}
    S[table-format=4.2, table-column-width=1.1cm]
    S[table-format=4.2, table-column-width=1.3cm]
    S[table-format=4.2, table-column-width=1.2cm]
}
\toprule
& \multicolumn{3}{c}{\textbf{Steps}}
& \multicolumn{3}{c}{\textbf{Observation steps}}
& \multicolumn{3}{c}{\textbf{Edit/write steps}} \\
\cmidrule(lr){2-4}
\cmidrule(lr){5-7}
\cmidrule(lr){8-10}
\textbf{Model}
& {Min} & {Mean} & {Max}
& {Min} & {Mean} & {Max}
& {Min} & {Mean} & {Max} \\
\midrule

GPT-6 Astra & 42 & 229.2 & 757 & 35 & 191.3 & 581 & 1 & 9.9 & 58 \\
Claude Opus 5 & 29 & 209.3 & 1000 & 1 & 95.3 & 329 & 0 & 23.5 & 117 \\
Claude Sonnet 5 & 27 & 222.4 & 522 & 8 & 107.2 & 364 & 1 & 15.7 & 53 \\
GPT-5.6 Sol & 42 & 199.4 & 377 & 15 & 101.2 & 236 & 1 & 25.8 & 85 \\
GPT-5.3 Codex & 58 & 196.4 & 342 & 11 & 86.7 & 195 & 0 & 17.1 & 60 \\
Grok 4.6 & 41 & 160.9 & 370 & 2 & 65.9 & 173 & 1 & 18.5 & 69 \\
Gemini 3.1 Pro & 35 & 159.6 & 858 & 2 & 46.3 & 152 & 1 & 24.2 & 106 \\
Gemini 3.7 Flash & 50 & 206.8 & 626 & 2 & 82.5 & 218 & 1 & 13.5 & 59 \\
Gemini 3.6 Flash & 55 & 180.2 & 386 & 9 & 68.9 & 184 & 0 & 16.8 & 62 \\

\midrule
\textbf{Restoration depth}
& \multicolumn{3}{c}{\textbf{Mean steps}}
& \multicolumn{3}{c}{\textbf{Mean observation steps}}
& \multicolumn{3}{c}{\textbf{Mean edit/write steps}} \\
\cmidrule(lr){2-4}
\cmidrule(lr){5-7}
\cmidrule(l){8-10}
1 & \multicolumn{3}{c}{124.0} & \multicolumn{3}{c}{61.2} & \multicolumn{3}{c}{7.5} \\
2 & \multicolumn{3}{c}{164.0} & \multicolumn{3}{c}{75.9} & \multicolumn{3}{c}{12.4} \\
3 & \multicolumn{3}{c}{187.0} & \multicolumn{3}{c}{92.8} & \multicolumn{3}{c}{17.0} \\
4 & \multicolumn{3}{c}{216.0} & \multicolumn{3}{c}{99.9} & \multicolumn{3}{c}{21.2} \\
5 & \multicolumn{3}{c}{225.9} & \multicolumn{3}{c}{110.5} & \multicolumn{3}{c}{23.0} \\
6 & \multicolumn{3}{c}{230.8} & \multicolumn{3}{c}{110.5} & \multicolumn{3}{c}{24.3} \\
7 & \multicolumn{3}{c}{235.7} & \multicolumn{3}{c}{110.8} & \multicolumn{3}{c}{24.5} \\
8 & \multicolumn{3}{c}{250.6} & \multicolumn{3}{c}{121.0} & \multicolumn{3}{c}{27.2} \\

\bottomrule
\end{tabular}%
}
\end{table*}

\paragraph{Full-Application Reconstruction.}
\label{app:rebuild-steps}

Table~\ref{tab:rebuild-steps} reports agent steps by application for the
full-application runs in Table~\ref{tab:full-app-reconstruction}.
Astra averages \CanonicalAstraStepsMean{} steps, compared with
\CanonicalOpusStepsMean{} for Opus~5 and \CanonicalSolStepsMean{} for Sol.

\begin{table*}[h!]
\centering
\caption{\textbf{Agent steps per full-application reconstruction run.}}
\label{tab:rebuild-steps}
\begingroup
\setlength{\tabcolsep}{3pt}
\newcommand{\appheader}[1]{%
  \textbf{\begin{tabular}[c]{@{}c@{}}#1\end{tabular}}}
\resizebox{\textwidth}{!}{%
\begin{tabular}{lccccccccccccc}
\toprule
\appheader{Model} & \appheader{Reactive\\Resume} & \appheader{AWS\\Console} & \appheader{Baserow} & \appheader{BudgetWise} & \appheader{CareerLink} & \appheader{MailHub} & \appheader{Jira\\(Oldboyxx)} & \appheader{Trello\\(Knowankit)} & \appheader{StreamView} & \appheader{TeamChat} & \appheader{Trello\\(Vdevired)} & \appheader{Weights \&\\Biases} & \appheader{Mean} \\
\midrule
GPT-6 Astra (max) & 853 & 1921 & 923 & 292 & 824 & 421 & 489 & 434 & 895 & 385 & 700 & 496 & 719.4 \\
Opus 5 (max) & 620 & 837 & 572 & 663 & 940 & 417 & 371 & 595 & 601 & 608 & 533 & 773 & 627.5 \\
GPT-5.6 Sol (max) & 1165 & 856 & 860 & 580 & 861 & 348 & 598 & 446 & 990 & 488 & 470 & 966 & 719.0 \\
\bottomrule
\end{tabular}}
\endgroup
\end{table*}

%% file: archive/section/harness_architecture.tex
\subsection{Runtime Architecture and Isolation}
\label{app:harness-architecture}

The harness separates the runtime into three components, the editable
application, the reference-side browser service, and the verifier.
Figure~\ref{fig:harness-architecture} illustrates their roles and the
communication paths used during reference-guided implementation.
The key design principle is that the agent may observe the behavior of the
reference application without receiving its source code or a source checkout
that could be copied directly.

\begin{figure}[htbp]
\centering
\resizebox{\linewidth}{!}{\input{archive/figures/harness_architecture}}
\caption{\textbf{Runtime layout and permitted communication.}
The browser helper runs on the reference side and maintains isolated
contexts for the fixed reference and editable current applications.
The current context communicates with the editable application across the
runtime boundary, while direct agent access to reference-application ports
is blocked. Submitted source is evaluated separately in a fresh verifier.
Arrows denote logical communication paths rather than a shared filesystem.}
\label{fig:harness-architecture}
\end{figure}

\paragraph{Editable current environment.}
The agent's shell and file tools operate only on the current application.
In partial-application reconstruction tasks, the agent starts from a source tree in which
selected implementation details are masked. In full-application reconstruction tasks, it
starts from a minimal executable scaffold. In both cases, the agent may edit
the application source, run the resulting implementation, and inspect its
behavior.

The agent's \texttt{browser} command does not launch a separate browser inside
the editable environment. Instead, it acts as a thin client for the
reference-side browser service, which performs browser interactions on the
agent's behalf.

\paragraph{Reference environment and browser service.}
The reference environment contains the fixed production application and a
Playwright-based browser helper. The helper maintains two isolated browser
contexts, \texttt{reference} and \texttt{current}, with separate cookies,
storage, and page state. The \texttt{reference} context opens the fixed production application locally,
while the \texttt{current} context connects across the runtime boundary to the
editable application. In the latter case, the browser acts only as a client
of the current application. This browser connection does not expose the
agent's source tree to the reference application or browser service.

The helper exposes browser interactions through a constrained API and returns
structured observations to the agent. Browser recordings, temporary state,
and other internal artifacts remain outside the submitted source tree.

\paragraph{Control traffic and application traffic.}
Two kinds of network traffic are involved. The agent sends
\emph{control traffic} to the browser helper on port 7788 to request
browser actions and observations. Separately, the browser generates
\emph{application traffic} when its reference or current context connects
to an application.

Access to the helper does not permit the agent to connect directly to
reference-application ports. The current browser context, however, must
receive responses from the editable application across the runtime
boundary. The network filter therefore allows only reply packets that
connection tracking associates with an existing browser connection.
This preserves communication between the browser and current application
without opening a broader network path from the agent to the reference
application. The same policy is used for both evaluation settings.

\paragraph{Fresh verification and isolation scope.}
After implementation, a trusted controller transfers the submitted source to
a separate verifier. The verifier starts from a clean task baseline, applies
the submission, rebuilds the application, resets persistent state, and replays
the evaluation workflows. It does not reuse the agent's interactive runtime
state.

Private grading information, including grading traces, expected signals, and
the gold patch, remains outside the agent-visible interface.
Success therefore depends on whether the submitted source reproduces the
required behavior under fresh verification, rather than on the state of the
agent's final browser session.

These boundaries separate editable source code, observable reference behavior,
and private grading material. They provide the isolation required for the
intended reference-guided evaluation setting, but are not designed as a
general-purpose sandbox for arbitrary hostile code.

%% file: archive/figures/harness_architecture.tex
\begin{tikzpicture}[
  x=1cm,y=0.85cm,
  font=\sffamily\fontsize{6.3}{7.7}\selectfont,
  box/.style={draw=black!35,rounded corners=3pt,line width=0.6pt},
  heading/.style={font=\sffamily\fontsize{7.2}{8.6}\selectfont\bfseries},
  service/.style={box,fill=none,align=center,text width=4.0cm,
                  minimum height=1.0cm,inner sep=5pt},
  flow/.style={->,>=latex,line width=0.85pt},
  note/.style={align=center,font=\sffamily\fontsize{6}{7.5}\selectfont,fill=none,
               inner sep=2pt}
]
\path[box,fill=blue!3] (0,2.5) rectangle (5.4,7.4);
\path[box,fill=green!4] (8.6,2.5) rectangle (14,7.4);
\node[anchor=north west,heading] at (0.25,7.15)
  {Current environment};
\node[anchor=north west,heading] at (8.85,7.15)
  {Reference environment};

\node[service] (tools) at (2.7,5.9)
  {\textbf{Agent tools}\\Shell and file edits\\Browser CLI};
\node[service] (current) at (2.7,3.55)
  {\textbf{Current application}\\Editable source\\Application state};
\node[service] (helper) at (11.3,5.9)
  {\textbf{Browser helper}\\Playwright and Chromium\\Two browser contexts};
\node[service] (reference) at (11.3,3.55)
  {\textbf{Reference application}\\Fixed production build\\Reference state};

\draw[flow,blue!60!black,<->] (tools.east) -- (helper.west)
  node[midway,above=4pt,note] {Helper API\\commands and observations};
\draw[flow] (tools.south) -- (current.north)
  node[midway,left=3pt,note] {Edit\\and run};
\draw[flow,<->] (helper.south) -- (reference.north)
  node[midway,right=3pt,note] {Reference view\\local traffic};

\draw[flow,green!45!black] (helper.south west) -- (current.north east)
  node[pos=0.43,above=4pt,note] {Open or interact\\with current};
\draw[flow,green!45!black]
  ([yshift=-5pt]current.north east) --
  ([yshift=-5pt]helper.south west)
  node[pos=0.5,below=2pt,note] {Replies on the\\same connection};

\draw[flow,red!65!black,dashed]
  (current.east) -- (reference.west);
\draw[red!65!black,line width=1.1pt]
  (6.83,3.4) -- (7.17,3.7) (6.83,3.7) -- (7.17,3.4);
\node[note,text=red!65!black] at (7,3.05)
  {No new direct connection\\to the reference application};

\path[box,fill=black!3] (0,-0.2) rectangle (14,1.3);
\node[anchor=west,heading] at (0.3,0.92)
  {Fresh verifier};
\node[anchor=west,align=left] at (0.3,0.33)
  {Clean task start + submitted source\\Rebuild and reset the application};
\node[align=center]
  at (11.1,0.56) {\textbf{Private replay checks}\\Held-out workflows and signals\\Submission score};
\draw[flow] (7.15,0.55) -- (8.25,0.55);
\draw[flow,black!65] (current.south) -- (2.7,1.3)
  node[midway,right=4pt,note] {Submitted source\\transferred by the controller};
\end{tikzpicture}

%% file: archive/analysis/generated/repair-context-macros.tex
\newcommand{\RepairContextPromptDepthOne}{47.5}
\newcommand{\RepairContextPromptDepthEight}{107.5}
\newcommand{\RepairContextPromptRatio}{2.3}
\newcommand{\RepairContextScoreDepthOne}{83.4}
\newcommand{\RepairContextScoreDepthEight}{23.1}
\newcommand{\RepairContextCorrelationMin}{-0.20}
\newcommand{\RepairContextCorrelationMax}{-0.05}
\newcommand{\RepairContextWithinTaskCorrelation}{+0.11}
\newcommand{\RepairContextLowBandDrop}{58.4}
\newcommand{\RepairContextMidBandDrop}{54.5}
\newcommand{\RepairContextHighBandDrop}{31.5}